\documentclass[prb,reprint,aps,superscriptaddress,longbibliography]{revtex4-2}
\usepackage[colorlinks=true,linkcolor=blue,citecolor=blue,urlcolor=blue]{hyperref}
\usepackage{graphicx}
\usepackage{bm}
\usepackage{braket}
\usepackage{amsmath}
\usepackage{amssymb}
\usepackage{graphics}
\usepackage{siunitx}
\usepackage[dvipsnames]{xcolor}
\usepackage{enumitem}
\usepackage[english]{babel}
\setlist{nolistsep}
\DeclareSIUnit\angstrom{\text{Å}}

\begin{document}
\title{Theory of bilayer graphene plasmons spectroscopy by quantum twisting microscope}
\title{Theory of plasmon spectroscopy with the quantum twisting microscope}
\author{Nemin Wei}
\affiliation{\mbox{Department of Physics and Yale Quantum Institute, Yale University, New Haven, Connecticut 06520, USA}}
\author{Francisco Guinea}
\affiliation{\mbox{Imdea Nanoscience, Faraday 9, 28015 Madrid, Spain}}
\affiliation{\mbox{Donostia International Physics Center, Paseo Manuel de Lardiz\'{a}bal 4, 20018 San Sebastian, Spain}}
\author{Felix von Oppen}
\affiliation{\mbox{Dahlem Center for Complex Quantum Systems and Fachbereich Physik, Freie Universit\"at Berlin, 14195 Berlin, Germany}}
\author{Leonid I.\ Glazman}
\affiliation{\mbox{Department of Physics and Yale Quantum Institute, Yale University, New Haven, Connecticut 06520, USA}}
\date{\today}

\begin{abstract}
We consider plasmon-assisted electron tunneling in a quantum twisting microscope (QTM). The dependence of the differential conductance on the two control parameters of the QTM -- the twist angle and bias -- reveals the plasmon spectrum as well as the strength of plasmon-electron interactions in the sample. We perform microscopic calculations for twisted bilayer graphene (TBG), to predict the plasmon features in the tunneling spectra of TBG close to the magic angle for different screening environments. Our work establishes a general framework for inelastic tunneling spectroscopy of collective electronic excitations using the quantum twisting microscope.

\end{abstract}

\maketitle

\twocolumngrid

\section{Introduction}
%

The introduction of the quantum twisting microscope (QTM) opened up new ways of studying two-dimensional van der Waals materials \cite{inbar_quantum_2023}. QTM measurements probe the differential conductance $G(V,\theta)$ between twisted van der Waals materials 
as a function of bias voltage $V$ and twist angle $\theta$, providing information about momentum-resolved spectra of excitations as well as their  interactions. Initial experiments at room temperature already exemplified the  effectiveness of the QTM in resolving the electron spectra of monolayer and twisted bilayer graphene (TBG) \cite{inbar_quantum_2023}. Subsequently, the analysis of weaker features in $G(V,\theta)$ at cryogenic temperatures  revealed the phonon dispersions and the strength of electron-phonon coupling in these materials \cite{birkbeck_measuring_2024,xiao_theory_2024}. These measurements  raise the question whether  other collective excitations and their coupling to the electrons can be efficiently resolved by means of the QTM \cite{peri_probing_2024,pichler_probing_2024,carrega2020tunneling}.

Collective oscillations of charge, known as plasmons, are natural candidates for a QTM study. The isolated flat bands of twisted bilayer graphene 
are predicted to  allow for plasmon modes unaffected by Landau damping over a wide range of momenta. These undamped plasmon excitations are expected to have frequencies within the spectral range between the particle-hole continuum associated with the flat bands and the dispersive bands, where there are no low-order decay processes \cite{lewandowski_intrinsically_2019,papaj2023probing}. Interest in the plasmon excitations is further enhanced by suggestions that they are potential candidates for providing the pairing glue in the superconducting state of magic-angle twisted bilayer graphene (MATBG) \cite{lewandowski_pairing_2021,cea_coulomb_2021,sharma2020superconductivity,peng2024theoretical}. Moreover, despite their fundamental importance, there has been a lack of experimental studies of plasmons in moir\'e materials.

In this work, we propose to use inelastic-tunneling spectroscopy with a QTM to explore the plasmon dispersion of moir\'e materials. 
Tunneling between a graphene monolayer placed on the tip of a QTM and a moir\'e graphene sample conserves momentum. Apart from strong elastic tunneling features that map out the band structure of the sample \cite{inbar_quantum_2023,wei2025dirac}, weaker features associated with inelastic tunneling reveal the plasmon dispersion.
The spectroscopic resolution of plasmons is enabled by the mismatch between the energy and momentum of the tunneling electron on the one hand, and of the quasiparticle states in the sample on the other. This mismatch is lifted by the emission or absorption of a plasmon. Thus, the inelastic contributions to the tunneling conductance encode information about the momentum-resolved plasmon dispersion and the electron-plasmon coupling strength. 

According to the current theory of the plasmon spectra of MATBG \cite{stauber2016quasi,lewandowski_intrinsically_2019,cea_coulomb_2021}, the plasmon frequency first exhibits a rapid increase with wave vector, which is followed by a flat part stretching to the Brillouin zone boundary. [We measure plasmon wave vectors from the $\bm\gamma$ point of the moir\'e Brillouin zone (mBZ)]. We find that plasmon-assisted tunneling into the central MATBG bands reveals the flat part of the plasmon spectrum, while tunneling into higher electron bands allows one to trace its momentum dependence.  

It is interesting to compare QTM-based plasmon spectroscopy to the previously studied phonon spectroscopy in twisted graphene-graphene junctions \cite{birkbeck_measuring_2024,xiao_theory_2024}. The mechanical coupling between van der Waals layers modifies the phonon dispersions only weakly, so that the experimentally observed phonon spectra are adequately described within the original graphene Brillouin zone. In contrast, one expects that the interlayer coupling strongly affects the low-energy collective electronic excitations, which should thus be described within the moir\'e Brillouin zone. In particular, the interlayer coupling strongly affects the plasmon dispersions \cite{cavicchi_theory_2024}. This makes plasmon spectroscopy a more generic case study of collective-mode spectroscopy of twisted van der Waals materials. The stronger interlayer coupling also introduces additional challenges. In particular, one has to ensure that tip effects on the plasmons of the sample can be neglected. Even in the limit of weak tip-sample tunneling, the tip interacts with the sample through the long-range Coulomb interaction. We will show that this coupling has a weak effect on the sample's plasmons provided that the Fermi energy of the tip is sufficiently close to the charge-neutrality point.


The paper is organized as follows. In Sec.~\ref{sec:summary}, we give a simplified presentation that dispenses with the details of the single-particle spectrum of twisted bilayer graphene. After introducing the necessary notation for the continuum model of interacting TBG in Sec.~\ref{sec:bm_model}, we describe the formalism for inelastic tunneling spectroscopy of collective excitations in Sec.~\ref{sec:formalism}, as applicable to generic multi-band moir\'e systems. In Sec.~\ref{sec:numerics}, we present numerical results for the spectral function and the tunneling spectra of interacting TBG, both at a twist angle slightly away from the magic angle and at the magic angle. The limitations and potential extensions of our approach are discussed in Sec.~\ref{sec:discussion}.

\section{Qualitative derivation of the main results}\label{sec:summary}
In this section, we illustrate the effect of inelastic electron tunneling assisted by scattering off the collective excitations, on the differential conductance measured by a quantum twisting microscope (QTM). To simplify considerations, we focus on tunneling processes involving a single band in the sample with four-fold spin-valley degeneracy $N_f=4$ and drop the band indices of the sample. More general expressions and the details of the derivations are presented in Sec.~\ref{sec:formalism}.

The differential conductance, $dI/dV$, for momentum- and energy-resolved tunneling in QTM can be expressed in terms of the spectral functions of the tip and sample, 
\begin{align}\label{eq:didv}
    \frac{dI}{dV} = \frac{2\pi e^2N_f}{\hbar} \sum_{\bm k} \sum_{\bm k'}& |\langle \bm k' \text{T}|\hat{H}_{\text{tun}}|\bm k \text{S}\rangle |^2 \notag\\
    &\times A^{\text{T}}\left(\bm k', 0\right) A^{\text{S}}\left(\bm k, eV\right),
\end{align}
where the in-plane wave vectors $\bm k'$ and $\bm k$ are summed over the Brillouin zone of the tip and sample, respectively. 
The interlayer tunneling $\hat{H}_{\text{tun}}$ conserves the wave vector up to reciprocal lattice vectors of the tip and sample. The energy $\epsilon$ in the spectral functions $A^{\text{T/S}}(\bm k,\epsilon)$ of the tip/sample is measured relative to the respective Fermi levels.

We model the tip as a free-electron gas with Fermi wave vector $k_F$ 
and spectral function $A^{\text{T}}(\bm k', \epsilon) = \delta(\epsilon - \xi_{\bm k'}^{\text{T}})$. The spectral function of the sample reflects interactions of the electrons with the collective excitations, $A^{\text{S}} = -\text{Im} [\epsilon-\xi_{\bm k}-\Sigma(\bm k,\epsilon)]^{-1}/\pi$, where $\Sigma$ denotes the retarded self-energy. In the weak-coupling limit, the spectral function $A^{\text{S}}$ exhibits quasiparticle peaks at energy around $\xi_{\bm k}$ which is responsible for elastic tunneling. We are interested in the tunneling current $\delta I$ generated by tip electrons of wave vector $\bm k$ and energy $\epsilon$ (relative to the Fermi level of the sample) which are off-resonant with the quasiparticles in the sample, $|\epsilon-\xi_{\bm k}|\gg |\Sigma(\bm k,\xi_{\bm k})|$. These tunneling processes are enabled by inelastic scattering. In this ($\bm k, \epsilon$) domain,
\begin{equation}\label{eq:A_largeomega}
    A^{\text{S}}(\bm k,\epsilon) \approx-\frac{1}{\pi} \frac{1}{(\epsilon-\xi_{\bm k})^2}\Sigma^{\prime\prime} (\bm k,\epsilon).
\end{equation}
To compute the dissipative part $\Sigma^{\prime\prime}$ of the self-energy, we model the collective excitations of the sample as bosonic modes with spectrum $\omega_{\bm q}$. The Hamiltonian of the sample reads  
\begin{align}
    &H_{\text{S}} = \sum_{\bm k} \xi_{\bm k}c_{\bm k}^{\dagger}c_{\bm k} + \sum_{\bm q} \hbar\omega_{\bm q} b_{\bm q}^{\dagger}b_{\bm q} + H_{\text{e-pl}}, \\
    &H_{\text{e-pl}} = \frac{1}{\sqrt{N}}\sum_{\bm k, \bm q} g(\bm k, \bm q) c_{\bm k}^{\dagger}c_{\bm k-\bm q} b_{\bm q} +\text{H.c.}\,, 
    \label{eq:H_bf}
\end{align}
where $g(\bm k,\bm q)$ denotes the interaction matrix element between the electrons and the bosonic mode in the sample, and $N$ is the number of moir\'e unit cells, each of area $A_M$.
The matrix element of the electron-plasmon interaction is governed by the potential $\phi_{\bm q}$ associated with the plasmon zero-point motion,
\begin{equation}\label{eq:g_phi}
    g(\bm k,\bm q) = \frac{\phi_{\bm q}}{\sqrt{A_M}} \langle u_{\bm k}|u_{\bm k-\bm q}\rangle,\ \ \ \phi_{\bm q} = \sqrt{\frac{\hbar\omega_{\bm q}v_{\bm q}}{2}}.
\end{equation}
Here, $\ket{u_{\bm k}}$ denotes the periodic factor of an electron Bloch state. Note that in the long wavelength limit $\phi_{\bm q}$ is related to the electron-electron interaction $v_{\bm q}$ and the plasmon frequency $\omega_{\bm q}$ \cite{pines1962approach,mahan2013many}, as on average half of the plasmon energy must be stored as electric potential energy, $\frac{1}{2}\times\hbar\omega_{\bm q}/2 = \phi_{\bm q}^2/2 v_{\bm q}$.

One can use the optical theorem to compute $\Sigma^{\prime\prime}$ to second order in the electron-boson coupling. At zero temperature,
%
\begin{align}\label{eq:sigma_im_1}
    \Sigma^{\prime\prime}(\bm k,\epsilon) = -\frac{\pi}{N}&\sum_{\bm q} |g(\bm k,\bm q)|^2 \notag\\
    &\times\left[\Theta(\xi_{\bm k-\bm q})\delta(\epsilon - \hbar\omega_{\bm q} - \xi_{\bm k-\bm q}) \right.\notag\\
    &\ \ \left. + \Theta(-\xi_{\bm k-\bm q})\delta(\epsilon + \hbar\omega_{\bm q} - \xi_{\bm k-\bm q})\right].
\end{align}
The first (second) term in the bracket contributes to the electron (hole) scattering rate with emission of a plasmon at wave vector $\bm q\ (-\bm q)$. Here, we used $\omega_{\bm q}=\omega_{-\bm q}$ and $\phi_{\bm q}=\phi_{-\bm q}^*$ in the time-reversal invariant system.
Plugging Eq.~\eqref{eq:sigma_im_1} into Eq.~\eqref{eq:A_largeomega} and  Eq.~\eqref{eq:didv} yields the inelastic contribution to the differential conductance. This general form can be simplified, as the Fermi wave vector in the tip $k_{F}\rightarrow 0$ (i.e., $\bm k'$ is close to a Dirac point $\bm K_{\theta}$ in the tip),
%
\begin{align}\label{eq:didv_in}
    \frac{d\delta I}{dV} = \frac{e^2\Gamma}{N}&\sum_{\bm q}|\overline{M}(\bm K_{\theta},\bm q)|^2 \notag\\ 
    &\times\left[\Theta(\xi_{\bm K_{\theta}-\bm q})\delta(eV - \xi_{\bm K_{\theta}-\bm q}- \hbar\omega_{\bm q}) \right.\notag\\
    &\ \ \left. + \Theta(-\xi_{\bm K_{\theta}-\bm q})\delta(eV - \xi_{\bm K_{\theta}-\bm q}+ \hbar\omega_{\bm q})\right].
\end{align}
Here, we defined
\begin{equation}
    \Gamma = \frac{2\pi w^2\Omega\nu_{\text{T}} N_b}{\hbar}, 
\end{equation}
with tunneling strength $w$, junction area $\Omega=NA_M$, and total density of states $\nu_{\text{T}}$ at the Fermi level of the tip; $N_{b}=3$ accounts for the three Bragg scattering processes in the graphene superlattice \cite{birkbeck_measuring_2024}. $\Gamma$ represents the rate of elastic tip-sample tunneling in the absence of momentum conservation; we will use it as a convenient scale for the electron tunneling rate. Equation~\eqref{eq:didv_in} contains the normalized rate of inelastic tunneling,
\begin{equation}\label{eq:M_squared}
    |\overline{M}(\bm K_{\theta},\bm q)|^2= \frac{|\overline{T}(\bm K_{\theta})|^2}{w^2} \left|\frac{g(\bm K_{\theta},\bm q)}{eV-\xi_{\bm K_{\theta}}}\right|^2,
\end{equation}
where $|\overline{T}(\bm K_{\theta})|^2$ is obtained by averaging the squared tunneling matrix elements $|\langle \bm k \text{T}| H_{\text{tun}}|\bm k \text{S}\rangle|^2$ over the wave vectors $\bm k$ along the small Fermi line of the tip.

\begin{figure}
    \centering
    \includegraphics[width=1\linewidth]{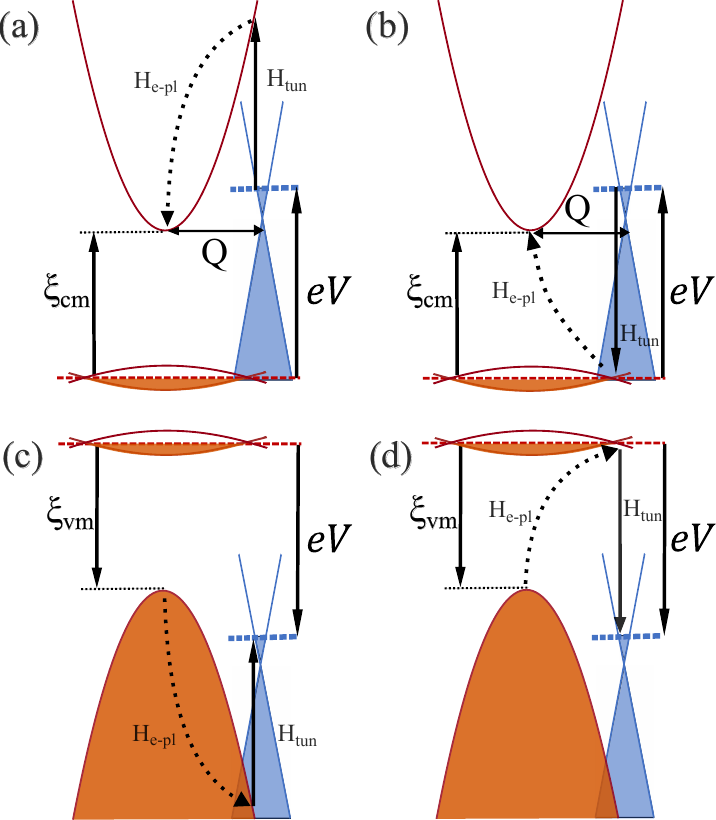}
    \caption{
    Plasmon-assisted inelastic tunneling processes between the tip and dispersive (higher-energy) bands of the sample. The Fermi level of the sample, shown as the red dashed line, is parked in a flat band. The precise position of the Fermi level within the flat band is inconsequential. 
    The Fermi level of the tip is represented by the blue dashed line and is close to the Dirac (charge neutrality) point, $k_{F}\ll |\bm Q|$. (a,b) Two processes for an electron to tunnel from the Fermi level of the tip to the edge of an unoccupied dispersive band of the sample at the threshold bias $eV=\xi_{cm}+\hbar\omega_{\bm Q}$. Each of the processes involves momentum-conserving tunneling (solid line) followed by plasmon emission (dashed line). 
    (c,d) Two complementary processes for an electron to tunnel from the edge of an occupied dispersive band of the sample to the Fermi level of the tip at the threshold bias $eV=\xi_{vm}-\hbar\omega_{\bm Q}$. Processes (b) and (d) involve interband scattering. They are not discussed in Sec.~\ref{sec:summary} but will be included in later sections.
    }
    \label{fig:dispersive_band}
\end{figure}

The first (second) term in the square bracket in Eq.~\eqref{eq:didv_in} corresponds to a two-step tunneling process from the tip (sample) to the sample (tip) when the tip-sample bias is positive (negative) and large enough to emit the bosonic mode, see Fig.~\ref{fig:dispersive_band}(a) and (c) for illustration.

The plasmon-assisted tunneling may add an electron (or a hole) to the flat or dispersive band. To resolve the plasmon spectrum, one can exploit the singularity of the density of states associated with the edge of the dispersive band, or the narrow width of the flat bands. In the following, we assume that the Fermi level in the sample is parked within the flat-band manifold, providing a narrow peak in the energy-resolved density of states. Under this condition, the bias $|eV|$ needed for tunneling into the dispersive bands has to exceed the energy distance $\xi_{cm}$ (or $|\xi_{vm}|$, see Fig.~\ref{fig:dispersive_band}) between the flat-band manifold and the edge of the dispersive band. Similar to the case of phonon spectroscopy \cite{birkbeck_measuring_2024,xiao_theory_2024}, the spectroscopic sensitivity of the differential conductance in Eq.~\eqref{eq:didv_in} requires the electron velocity to be much higher than the plasmon velocity. Given the predicted plasmon spectrum \cite{lewandowski_intrinsically_2019,cea_coulomb_2021,cavicchi_theory_2024}, we then expect the differential conductance to reveal the nearly-flat part of the plasmon spectrum. In contrast, inelastic tunneling into the flat-band manifold is hardly momentum selective because of the low electron velocity. That allows us to relate the differential conductance to the momentum-unresolved plasmon density of states.

\begin{figure}
    \centering
    \includegraphics[width=1\linewidth]{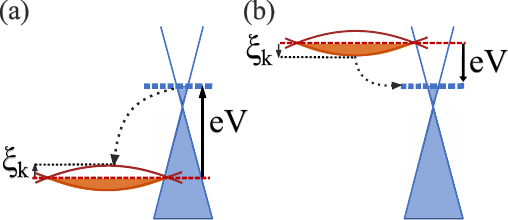}
    \caption{
    Plasmon-assisted tunneling processes between the tip and flat bands of the sample. The Fermi level of the sample is parked in the flat bands. (a) A process for electrons to tunnel from the tip to an empty flat band whose bandwidth is considered to be much smaller than the average plasmon energy. (b) A process for electrons to tunnel from an occupied flat band to the tip. Each process occurs in two steps (not shown), similar to Fig.~\ref{fig:dispersive_band}.}
    \label{fig:flat_band}
\end{figure}

We start by analyzing Eq.~\eqref{eq:didv_in} in the limit of tunneling into a dispersive band and a ``flat'' plasmon, $\omega_{\bm q}\approx \omega_{\bm Q_{\theta}}$. The momentum separation $\bm Q$ between a Dirac point in the tip and an extremum in the sample band can be adjusted by the twist angle $\theta$ between the tip and sample.
For electron tunneling into an empty band with density of states $\nu_{c}(\epsilon)$ and band minimum $\xi_{cm}>0$ [Fig.~\ref{fig:flat_band}(a)], we find
%
\begin{equation}\label{eq:didv_phonon}
    \frac{d\delta I}{dV} \approx e^2\Gamma|\overline{M}(\bm K_{\theta},\bm Q_{\theta})|^2 A_M\nu_{c}(eV - \xi_{cm} -\hbar\omega_{\bm Q_{\theta}}).
\end{equation}
Similarly, for tunneling of a hole into a fully occupied valence band, with density of states $\nu_{v}(\epsilon)$ and band maximum $\xi_{vm}<0$ [Fig.~\ref{fig:flat_band}(c)], Eq.~\eqref{eq:didv_in} yields   
%
\begin{equation}
    \frac{d\delta I}{dV} \approx e^2\Gamma |\overline{M}(\bm K_{\theta},\bm Q_{\theta})|^2 A_M\nu_{v}(eV - \xi_{vm} + \hbar\omega_{\bm Q_{\theta}}). 
\end{equation}
%
For 2D parabolic bands, $\nu_{c/v}(\epsilon) = \frac{m^*}{2\pi\hbar^2}\Theta(\pm\epsilon)$ 
are discontinuous across the band edges, leading to steps in $dI/dV$.
%
The twist-angle dependence of the characteristic bias for these $dI/dV$ steps can trace the energy dispersion of the slow bosonic mode ($\partial_{\bm Q}\omega_{\bm Q}\ll \sqrt{\hbar \omega_{\bm Q}/m^*}$).

We now turn to the case of lower bias voltages, where electron/hole tunneling occurs into the flat-band manifold (see Fig.~\ref{fig:flat_band}). We assume, for simplicity, that the electron-boson coupling is isotropic and depends only on the energy of the bosonic mode. That allows us to make the replacement $g(\bm k, \bm q)\rightarrow g(\hbar\omega_{\bm q})$. In addition, we dispense with the width of the ``flat'' bands, compared to the plasmon energy $\hbar\omega_{\bm q}$; this amounts to taking the limit $\xi_{\bm k}\to 0$. After these simplifications, we find for $\ |eV|\gg \text{max} [|\xi_{\bm k}|]$ that
\begin{align}\label{eq:didv_flatband}
    \frac{d\delta I}{dV}
    \approx e^2\Gamma \frac{|\overline{T}(\bm K_{\theta})|^2}{w^2}\left|\frac{g(|eV|)}{eV}\right|^2 A_M D(|eV|).
\end{align}
%
Here, we use $D(\omega)$ to denote the density of states of the bosonic mode. In this limit, the twist angle affects only the overall magnitude of $dI/dV$ via the tunneling matrix elements, but not the line shape of the $dI/dV$ curve.  

We comment that due to  Eq.~\eqref{eq:A_largeomega}, Eqs.~\eqref{eq:didv_in} and~\eqref{eq:M_squared} are valid only when the bias $eV$ is off-resonant with the single-particle energy $\xi_{\bm K_{\theta}}$, i.e., in the $(V, \theta)$ domain where elastic tunneling is forbidden. Once inelastic features in the tunneling spectra due to long-wavelength plasmons  merge into the quasiparticle peaks, they become challenging to resolve. 

We roughly estimate the magnitude of the electrostatic potential $\phi_{\bm q}$ of the plasmon and the strength of inelastic tunneling for $\bm q$ comparable to the size of the mBZ. The former can be estimated via Eq.~\eqref{eq:g_phi} when $\hbar\omega_{\bm q} \sim \sqrt{N_f\mathcal{B} v_{\bm q}/A_{M}}\gg \mathcal{B}$, where $\mathcal{B}$ stands for the width of the flat-band manifold that hosts the plasmon. This leads to
\begin{equation}
   \frac{\phi_{\bm q}}{\sqrt{A_M}} \sim \sqrt{\frac{(\hbar\omega_{\bm q})^3}{N_f\mathcal{B}}}.
\end{equation}
%
When electrons tunnel to the edge of a dispersive band by emitting a less dispersive plasmon of wave vector $\bm Q_{\theta}$ and energy $\hbar\omega_{\bm Q_{\theta}}\ll|eV-\xi_{\bm K_{\theta}}|\approx \hbar^2\bm Q_{\theta}^2/2m^*$, then
\begin{equation}\label{eq:M_estimate1}
    |\overline{M}(\bm K_{\theta},\bm Q_{\theta})|^2\sim \frac{(\hbar\omega_{\bm Q_{\theta}})^3}{N_f\mathcal{B}\left(\frac{\hbar^2\bm Q_{\theta}^2}{2m^*}\right)^2 }. 
\end{equation}
%
When the electrons, instead, tunnel into a flat band, they can emit plasmons that carry energy
$\hbar\omega_{\bm q}\approx|eV-\xi_{\bm K_{\theta}}|$.
We then find
\begin{equation}\label{eq:M_estimate2}
    |\overline{M}(\bm K_{\theta},\bm q)|^2\sim \frac{\hbar\omega_{\bm q}}{N_f\mathcal{B}}. 
\end{equation}
%
Equations~\eqref{eq:didv_phonon}-\eqref{eq:M_estimate2} show that inelastic tunneling into the dispersive bands generally produces 
weaker features in the tunneling spectrum than tunneling into the flat-band manifold, due to the smaller normalized inelastic tunneling rate $|\overline{M}|^2$ and the smaller density of states of the dispersive bands. At the same time, tunneling into a dispersive band displays a higher sensitivity to the momentum dependence of the plasmon dispersion.



\section{Continuum model of TBG}\label{sec:bm_model}

We define the electron operator \( f_{\bm{p} s l \sigma} \) in twisted bilayer graphene (TBG), where \( s = \uparrow/\downarrow \) labels the spin index, \( l = t/b \) represents the layer index, and \( \sigma = A/B \) is the sublattice index. The wave vector \( \bm{p} \) is measured relative to the \( \Gamma \)-point of the graphene Brillouin zone.
For a small twist angle $\theta_{\text{TBG}}$, we can assign a valley index $\tau=\pm$ to the wave vector $\bm p$ in the $\pm \bm K$ valleys of both graphene layers. The Bistritzer-MacDonald (BM) Hamiltonian $H_0$ for non-interacting TBG reads \cite{bistritzer_moire_2011,lopesdossantos_graphene_2007}
\begin{align}
    H_0 = \sum_{\bm p,s}&\sum_{\sigma,\sigma'}\left(\sum_{l}f_{\bm p sl \sigma}^{\dagger}h_{\bm p}(\theta_l)_{\sigma,\sigma'}f_{\bm p sl\sigma'} \right.\notag\\
 & \left. +\sum_{n=1}^{3} f_{\bm p st \sigma}^{\dagger}(\hat{T}_{\tau n})_{\sigma,\sigma'} f_{\bm p+ \tau\bar{\bm g}_n s b \sigma'} + \text{H.c.}\right).\label{eq:h0_pw}
\end{align}
Here, an individual graphene layer is described by the Dirac Hamiltonian
\begin{align}
   h_{\bm p}(\theta_l) = \hbar v_D(O(\theta_l)\bm p -\tau\bm K)\cdot(\tau\sigma^x, \sigma^y), \label{eq:h}
\end{align}
with $\theta_{t/b}=\pm \theta_{\text{TBG}}/2$ and the rotation matrix
$$
    O(\theta)=\begin{pmatrix}
\cos\theta & \sin\theta\\
-\sin\theta & \cos\theta\\
\end{pmatrix}.
$$
Two layers are coupled by the interlayer tunneling matrix
\begin{equation}\label{eq:tn_general}
    \hat{T}_{\tau n} = w_0 \sigma^0+w_1\sigma^x e^{i\frac{2\pi (n-1)}{3}\tau\sigma^{z}},
\end{equation}
with $w_1>w_0(>0)$ due to the lattice relaxation. We have introduced the moir\'e reciprocal lattice vectors $\bar{\bm g}_1=0$ and $\bar{\bm g}_{2,3}= k_M(\pm\sqrt{3}/2,3/2)^{\text{t}}$ with $k_M = 2|\bm K|\sin (\theta_{\text{TBG}}/2)$.

The Hamiltonian Eq.~\eqref{eq:h0_pw} is diagonal in the Bloch basis labeled by the quasimomentum $\bm k$ restricted to one mBZ for each valley and the energy band index $\lambda$ for individual valley and spin, 
\begin{equation}\label{eq:c}
   c_{\bm k \lambda\tau s}^{\dagger} = \sum_{\bm g}\sum_{l,\sigma}u_{\bm k + \bm g l \sigma}^{\lambda\tau} f_{\bm k+\bm g s l\sigma}^{\dagger}. 
\end{equation}
Here $u_{\bm k+\bm g l\sigma}^{\lambda\tau}$ denotes the normalized reciprocal-space Bloch wave functions, $\bm g$ represents moir\'e reciprocal lattice vectors, 
and $\lambda$ labels the Bloch bands. The Hamiltonian of interacting TBG is given by $H= H_0 + H_{\text{int}}$, with
\begin{align}
    &H_0 = \sum_{\tau,s,\lambda}\sum_{\bm k\in\text{mBZ}}\xi_{\bm k}^{\lambda\tau} c_{\bm k \lambda\tau s}^{\dagger}c_{\bm k \lambda\tau s},\label{eq:h0}\\
    & H_{\text{int}} = \frac{1}{2\Omega}\sum_{\bm q} v_{\bm q}:\rho(\bm q) \rho(-\bm q):.\label{eq:hint}
\end{align}
Here the band energy $\xi_{\bm k}^{\lambda\tau}$ is measured relative to the Fermi level of the sample. `: ... :' means normal ordering, and the density operator reads
\begin{align}
    \rho(\bm q) 
    = \sum_{\tau s} \sum_{\lambda{\rho}}\sum_{\bm k\in\text{mBZ}} \langle u_{\bm k}^{\lambda\tau}| u_{\bm k+\bm q}^{{\rho}\tau}\rangle  c_{\bm k \lambda\tau s}^{\dagger} c_{\bm k+\bm q {\rho}\tau s}, 
\end{align}
with the form factor
\begin{equation}
   \langle u_{\bm k}^{ \lambda\tau}| u_{\bm p}^{{\rho}\tau}\rangle \equiv \sum_{\bm g,l,\sigma}(u_{\bm k + \bm g l\sigma}^{\lambda\tau})^*u_{\bm p + \bm g l\sigma}^{{\rho}\tau}. 
\end{equation}
We use the dual-gate-screened Coulomb potential
\begin{equation}\label{eq:vq}
    v_{\bm q} = \frac{e^2}{2\epsilon_r\epsilon_0} \frac{\tanh |\bm q| d_g}{|\bm q|},
\end{equation}
where $\epsilon_r$ is the relative permittivity of the surrounding dielectrics and $d_g$ represents the sample-gate distance.

\section{Quantitative theory of plasmon-assisted tunneling in QTM}
\label{sec:formalism}

Here we provide a quantitative theory which was sketched in Section~\ref{sec:summary}. We start by accounting for the proper structure of the tunneling matrix elements and electron bands in evaluating the current between the tip and sample (Section~\ref{sec:formalism}A). Next, we relate the electron self-energy in the sample to the particle-hole interaction vertex function (Section~\ref{sec:formalism}B). Lastly, in Section~\ref{sec:formalism}C we single out the plasmon contribution to the vertex, provide concise results for the plasmon-assisted tunneling current, and establish the correspondence to the qualitative considerations of Section~\ref{sec:summary}. 

\subsection{Momentum-conserving tunneling}

In this section, we derive the differential conductance of a QTM tunnel junction between a monolayer-graphene (MLG) tip and a TBG sample, generated by momentum-conserving tunneling. We show that $dI/dV$ as a function of tip-sample bias voltage $V$ and twist angle $\theta$ can measure the momentum-resolved one-particle spectral function of the sample when the tip has a low carrier density and small Fermi wave vector $k_{F}$. As the spectral function exhibits features arising from inelastic scattering of the electrons by collective excitations, $dI/dV$ can reveal information about collective excitations in the sample.

At a small twist angle $\theta$ between the tip and sample, interlayer tunneling preserves the valley number. To simplify the analysis, we consider the system to have spin SU(2) symmetry and time reversal symmetry, so that all four spin-valley flavors ($N_f=4$) contribute equally to the current $I$.
Within the tunneling Hamiltonian formalism \cite{mahan2013many},
\begin{widetext}
\begin{equation}\label{eq:i}
    I = \frac{2\pi e N_f}{\hbar} \int d\epsilon (n_{F}(\epsilon-eV)-n_{F}(\epsilon)) \sum_{\bm k'}\sum_{\bm k\in \text{mBZ}}\text{tr}\left[\hat{H}_{\text{tun}}\hat{A}^{\text{S}}(\bm k, \epsilon) \hat{H}_{\text{tun}}  \hat{A}^{\text{T}}(\bm k', \epsilon - eV)\right], \\
\end{equation}
\end{widetext}
where $n_F(\epsilon)=[\text{exp}(\beta\epsilon) + 1]^{-1}$ denotes the Fermi-Dirac distribution function. The sum over $\bm k'$ extends over wave vectors in the $\tau=+$ valley and the trace is over the energy bands of a single spin and valley flavor (say $s=\uparrow$, $\tau=+$). 

We model the MLG tip as a non-interacting system with a Dirac-delta spectral function in both the conductance and valence bands, $$\hat{A}^{\text{T}}(\bm k',\epsilon) = \sum_{\lambda'}\delta(\epsilon- \xi_{\bm k'}^{\text{T},\lambda'})|\bm k'\lambda'\text{T}\rangle \langle\bm k'\lambda'\text{T}|,$$ but allow the spectral function of the sample to be a positive semidefinite matrix in the Bloch basis,
\begin{equation}\label{eq:A_def}
    \hat{A}^{\text{S}}(\bm k, \epsilon) = \sum_{\lambda{\rho}}A_{\lambda{\rho}}^{\text{S}}(\bm k,\epsilon)|\bm k\lambda\text{S}\rangle \langle\bm k{\rho}\text{S}| = -\frac{\hat{G}^{\prime\prime}(\bm k,\epsilon)}{\pi},
\end{equation}
where $\hat{G}^{\prime\prime}=i(G^{\dagger}-G)/2$ and the retarded Green's function $\hat{G}$ of the sample obeys the Dyson equation
\begin{equation}\label{eq:G}
    \hat{G} = \hat{G}^0 \sum_{n=0}^{\infty}(\hat{\Sigma}\hat{G}^0)^{n}.
\end{equation}
The non-interacting Green's function is diagonal in the Bloch basis, 
\begin{equation}\label{eq:G0}
    \hat{G}^0(\bm k, \epsilon) = \sum_{\lambda}\frac{|\bm k\lambda\text{S}\rangle \langle\bm k\lambda\text{S}|}{\epsilon + i\eta-\xi_{\bm k}^{\lambda}},
\end{equation}
whereas the 
self-energy $\hat{\Sigma}$ and thus the Green's function $\hat{G}$ can in general have off-diagonal matrix elements due to interband scattering.
%
%

The Hamiltonian $\hat{H}_\text{tun}$ describing momentum-conserving tunneling between tip and sample takes the same form as in Eq.~\eqref{eq:h0_pw}:
\begin{equation}
    \hat{H}_{\text{tun}} = \sum_{\bm p,s}\sum_{n=1}^{3} d_{\bm p s}^{\dagger}\hat{T}_{\tau n} f_{\bm p+\tau\delta \bm G_n st}+ \text{H.c.},
\end{equation}
where $d_{\bm p s}^{\dagger}=(d_{\bm p sA}^{\dagger}, d_{\bm p sB}^{\dagger})$ creates an electron on sublattice $A/B$ of the tip with spin $s=\uparrow/\downarrow$ and wave vector $\bm p$ in the $\tau =\pm$ valley. We defined $\delta \bm G_n=\bm G_{n}'-\bm G_n$. Here, $\bm G_1=0$, and $\bm G_{2,3} = \sqrt{3}|\bm K|(\sqrt{3}/2,\pm 1/2)^\text{t}$ represent two primitive reciprocal lattice vectors of the top graphene layer of TBG. The reciprocal lattice vectors of the tip MLG are twisted clockwise by angle $\theta$, i.e., 
$\bm G_n'=
O(\theta)\bm G_{n}$. 
Therefore, $\delta \bm G_1=0$, and $\delta \bm G_{2,3}$ are two primitive reciprocal lattice vectors of the moir\'e superlattice formed by the tip and the top layer of the sample. 

We neglect lattice relaxation and corrugation at the interface of the tip and sample so that the tunneling matrix Eq.~\eqref{eq:tn_general} contains a single parameter $w_0=w_1=w$. We then write the tunneling Hamiltonian in the single-particle Bloch basis in $K$ valley,
\begin{align}\label{eq:h_tun}
    \langle \bm k' \lambda'\text{T} | \hat{H}_{\text{tun}} |\bm k\lambda\text{S}\rangle = &\sum_{n=1}^{3} T_{\lambda'\lambda}(\bm k'+\bm G_{n}')\notag\\
    &\times\sum_{\bm g}\delta_{\bm k'+\bm G_n', \bm k+\bm g + \bm G_n }.
\end{align}
Let us focus on the $n=1$ tunneling matrix elements
\begin{align}\label{eq:matrix_elements}
        &T_{\pm,\lambda}(\bm k') = w\left(\frac{1}{\sqrt{2}}\pm\frac{e^{-i\theta_{\bm k'}}}{\sqrt{2}} \right)\Lambda_{\lambda}(\bm k'), \\
        &\Lambda_{\lambda}(\bm k') = u_{\bm k'tA}^{\lambda}+u_{\bm k' tB}^{\lambda}, \label{eq:average_matrix_elements}
\end{align}
where $\theta_{\bm k'}$ denotes the angle between $\bm k'-\bm K_{\theta}$ and $\bm K_{\theta}$, and $\bm K_{\theta}=O(\theta)(|\bm K|,0)^\text{t}$ represents the tip Dirac point. Note that at generic non-commensurate twist angles between the tip and sample, the three tunneling processes do not interfere because the three Kronecker-deltas in Eq.~\eqref{eq:h_tun} cannot be satisfied simultaneously, and their contributions to the tunneling current $I$ are identical in $C_{3z}-$invariant systems. Therefore, we skip the expressions for $T_{\lambda'\lambda}(\bm k'+\bm G_{2,3})$ here \cite{wei2025dirac}. 
%
%
%

In general, the applied bias $V$  leads to a relative energy shift of the Dirac points of tip and sample. Here, we consider the specific limiting case, in which the density of states associated with the flat-band manifold of the sample is high and the density of states of the tip layer in the vicinity of the Dirac point is negligibly small. In this case, the applied bias $V$
barely changes the electron density in the sample, so that its Dirac point does not shift with respect to the Fermi level. In contrast, screening does not counteract the shift of chemical potential and the associated change of electron density in the tip. As a result, the chemical potential in the tip follows the bias voltage $V$, while the chemical potential in the sample does not move with respect to  
its Dirac point.
Thus, the energies of the single-particle states in the tip remain unchanged relative to the sample's Fermi level, $\partial(\xi_{\bm k'}^{\text{T},\lambda'}+eV)/\partial V\approx 0$.
To compute $dI/dV$ from Eq.~\eqref{eq:i}, we only need to differentiate the first Fermi-Dirac distribution function with respect to the bias. This leads to a simple expression for $dI/dV$ at zero temperature,
\begin{align}\label{eq:didv_full}
    \frac{dI}{dV} &= \frac{2\pi e^2N_bN_f}{\hbar} \sum_{\bm k'}\text{tr}\left[\hat{H}_{\text{tun}}\hat{A}^{\text{S}}\left(\bm k',eV\right)\hat{H}_{\text{tun}}\hat{A}^{\text{T}}(\bm k',0)\right] \notag\\
    &=\frac{e^2\Gamma}{w^2} \sum_{\lambda{\rho}}\oint \frac{d\theta_{k'}}{2\pi} T_{\lambda'{\rho}}^{*}(\bm k')T_{\lambda'\lambda}(\bm k')A_{\lambda{\rho}}^{\text{S}}(\bm k', eV)\notag\\
    &\approx e^2\Gamma \sum_{\lambda{\rho}}\Lambda_{{\rho}}(\bm K_{\theta})^{*}\Lambda_{\lambda}(\bm K_{\theta})A_{\lambda{\rho}}^{\text{S}}(\bm K_{\theta}, eV).
\end{align}
%
%
In the second line of Eq.~\eqref{eq:didv_full}, the wave vector $\bm k'$ is integrated along the Fermi line of the tip, and $\lambda'=\pm 1$ indicates whether the tip Fermi level lies in the conduction ($+$) or valence bands ($-$) of the Dirac cone. To arrive at the final expression, we use Eqs.~\eqref{eq:matrix_elements} and \eqref{eq:average_matrix_elements}, and assume that the tip Fermi momentum is small compared to the momentum uncertainty of the electrons in the sample generated by finite junction size, disorder scattering, or other interaction effects, so that the spectral function $\hat{A}^{\text{S}}(\bm k',\epsilon)$ of the sample is approximately constant along the tip Fermi line centered at the Dirac point $\bm K_{\theta}$. 

Equation~\eqref{eq:didv_full} is the main result of this section. It reduces to Eq.~\eqref{eq:didv}, when restricted to a single band of the sample. It indicates that the differential conductance of the QTM tunnel junction directly measures the momentum-resolved one-particle spectral function of the sample along the trajectory $\bm K_{\theta}$ of the tip Dirac point. The spectral function exhibits not only quasiparticle peaks at energy $\xi_{\bm k}^{\lambda}$ which are responsible for elastic tunneling, but also features of inelastic electron scattering off collective excitations at electron energy $\epsilon$ away from the quasiparticle peaks. At $|\epsilon-\xi_{\bm k}^{\lambda_{1,2}}|\gg |\Sigma_{\lambda_1\lambda_2}(\bm k,\epsilon)|$, by truncating the Dyson equation [Eq.~\eqref{eq:G}] to first order in $\hat{\Sigma}$ , the spectral function takes the form
\begin{equation}\label{eq:A_inelastic}
    A_{\lambda_1\lambda_2}^{\text{S}}(\bm k,\epsilon) \approx - \frac{1}{\pi}\frac{\Sigma_{\lambda_1\lambda_2}^{\prime\prime}(\bm k,\epsilon)}{(\epsilon -\xi_{\bm k}^{\lambda_1})(\epsilon -\xi_{\bm k}^{\lambda_2})}.
\end{equation}
Compared to Eq.~\eqref{eq:A_largeomega}, we allow here for the interband scattering illustrated in Fig.~\ref{fig:dispersive_band}(b) and (d), so that the self-energy is in general a matrix in band space. In the next section, we will derive $\hat{\Sigma}^{\prime\prime}$,  and show that it encodes information about the collective excitations.

\subsection{Off-shell self-energy of electrons}

\begin{figure}
     \centering
     \includegraphics[width=1\linewidth]{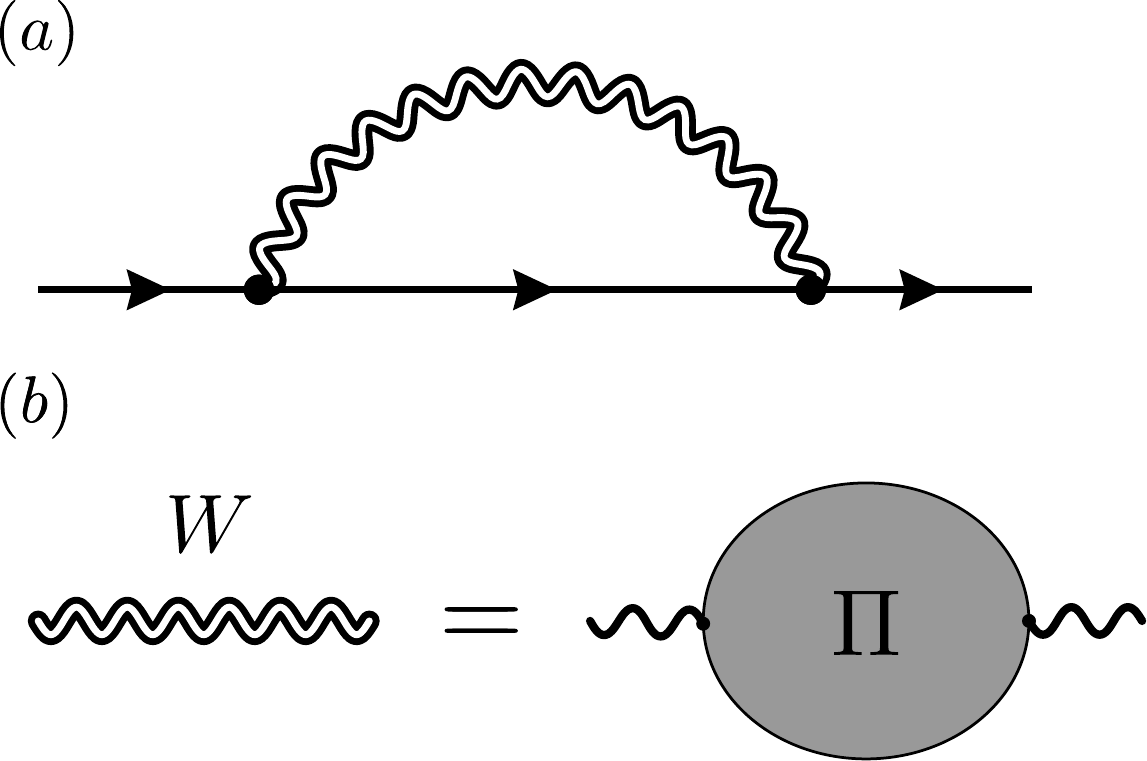}
     \caption{The Feynman diagrams for (a) the electron self-energy $\hat{\Sigma}$ and (b) the dynamical screening potential $\hat{W}$. The solid straight line represents $\hat{G}^{0}$ and the single curly line stands for the bare Coulomb interaction $v_{\bm q}$.}
     \label{fig:selfenergy}
 \end{figure} 
 
We calculate the self-energy $\hat{\Sigma}(\bm k, \epsilon)$ at energies $\epsilon$, which are off-resonant with the Bloch band energies $\xi_{\bm k}^{\lambda}$. We assume that the electron-electron interaction is much smaller than either the widths of the Bloch bands or the energy separation from the narrow bands in cases where narrow bands are present, i.e.,
$$\overline{|\epsilon - \xi_{\bm k}^{\lambda}|}\gg \overline{v_{\bm q}}/A_M,
\ \ \forall\lambda.$$
Here, the overline
stands for averaging over the wave vectors in the first mBZ. Therefore, we consider only Feynman diagrams that contain a single particle-hole interaction vertex $\mathcal{W}_{\bm k_1\lambda_1\rho_1,\bm k_2\lambda_2\rho_2}(\bm q;\tau_1-\tau_2)$, see Fig.~\ref{fig:selfenergy}(a). The notation means that a particle-hole pair ($|\bm k_2\lambda_2\rangle$ and $|\bm k_2-\bm q\rho_2\rangle$) with total wave vector $\bm q$ at imaginary time $\tau_2$ is scattered to another particle-hole pair ($|\bm k_1\lambda_1\rangle$ and $|\bm k_1-\bm q\rho_1\rangle$) at time $\tau_1$. 
Note that we neglect 
the first-order Hartree and Fock diagrams, which merely generate a hermitian self-energy. This renormalizes the quasiparticle band structures but does not contribute to $\hat{\Sigma}^{\prime\prime}$. 
The self-energy at the Matsubara frequencies can be expressed as
\begin{align}
    \Sigma_{\lambda_1\lambda_2}(\bm k,i\omega_n) = \frac{1}{\beta\Omega}\sum_{i\Omega_n}&\sum_{\bm q\in\text{mBZ}}\sum_{\lambda}\mathcal{W}_{\bm k\lambda_1\lambda,\bm k\lambda_2\lambda}(\bm q,i\Omega_n) \notag\\
    &\times G_{\lambda\lambda}^{0}(\bm k-\bm q,i\omega_n-i\Omega_n),\label{eq:Sigma_Matsubara}
\end{align}
The particle-hole vertex admits a spectral representation
\begin{equation}\label{eq:W_spectral}
    \hat{\mathcal{W}}(\bm q,i\Omega_n) = -\int_{-\infty}^{\infty} \frac{d\nu}{\pi}\frac{\hat{W}^{\prime\prime}(\bm q,\nu)}{i\Omega_n-\nu},
\end{equation}
where $\hat{W}^{\prime\prime}(\bm q,\nu)$ is a hermitian matrix, and $i\hat{W}^{\prime\prime}$ is the anti-hermitian part of $\hat{W}(\bm q,\nu) \equiv\hat{\mathcal{W}}(\bm q,\nu+i0^+)$.
After plugging Eq.~\eqref{eq:W_spectral} into Eq.~\eqref{eq:Sigma_Matsubara}, it is straightforward to sum over the Matsubara frequencies and obtain the retarded self-energy via the analytic continuation $i\omega_n\rightarrow \epsilon+i\eta$,
\begin{align}
   \Sigma_{\lambda_1\lambda_2}(\bm k,\epsilon) = \frac{1}{\Omega}\sum_{\bm q\in \text{mBZ}}&\sum_{\lambda}\int \frac{d\nu}{\pi} W_{\bm k\lambda_1\lambda,\bm k\lambda_2\lambda}^{\prime\prime}(\bm q,\nu)\notag\\
   &\ \ \times\frac{n_B(\nu) + n_F(-\xi_{\bm k-\bm q}^{\lambda})}{\nu+\xi_{\bm k-\bm q}^{\lambda}-\epsilon-i\eta}. \label{eq:sigma}
\end{align}
Here, 
$n_B$ is the Bose-Einstein distribution function. 
%
%
%
The anti-hermitian part $i\Sigma^{\prime\prime}$ of the self-energy obeys 
\begin{align}\label{eq:sigma_im}
    \Sigma_{\lambda_1\lambda_2}^{\prime\prime}(\bm k,\epsilon) = \frac{1}{\Omega}&\sum_{\bm q\in \text{mBZ}}\sum_{\lambda} W_{\bm k\lambda_1\lambda,\bm k\lambda_2\lambda}^{\prime\prime}(\bm q, \epsilon-\xi_{\bm k-\bm q}^{ \lambda}) \notag\\
   & \times [n_B(\epsilon-\xi_{\bm k-\bm q}^{\lambda}) + n_F(-\xi_{\bm k-\bm q}^{\lambda})]. 
\end{align}

\subsection{Charge fluctuations and inelastic tunneling}
We have not yet specified the collective excitations. Below we restrict the particle-hole vertex to the screening potential generated by charge fluctuations, as illustrated by the Feynman diagram in Fig.~\ref{fig:selfenergy}(b).
In this case, it is simpler to calculate $\hat{W}^{\prime\prime}$ in reciprocal space first and then transform back to the Bloch basis through the relation
\begin{align}
    & W_{\bm k\lambda_1\rho,\bm k\lambda_2\rho}^{\prime\prime}(\bm q,\nu) \notag\\
    =& \sum_{\bm g_1,\bm g_2} \langle u_{\bm k}^{\lambda_1}|u_{\bm k-\bm q-\bm g_1}^{\rho}\rangle W_{\bm g_1,\bm g_2}^{\prime\prime}(\bm q,\nu) \langle u_{\bm k-\bm q-\bm g_2}^{\rho}|u_{\bm k}^{\lambda_2}\rangle. \label{eq:W_im_band}
\end{align}
Here, $W_{\bm g_1,\bm g_2}^{\prime\prime}=i(W_{\bm g_1,\bm g_2}^{*}-W_{\bm g_2,\bm g_1})/2$ is determined from the screening potential, 
\begin{equation}\label{eq:W}
    W_{\bm g_1,\bm g_2}(\bm q,\nu) = v_{\bm q+\bm g_1}\Pi_{\bm g_1,\bm g_2}(\bm q,\nu)v_{\bm q+\bm g_2},
\end{equation}
with the retarded density-density response function
\begin{equation}\label{eq:Pi}
\Pi_{\bm g_1,\bm g_2}(\bm q,\nu)=-i\int_0^{\infty} dt \langle [\rho(\bm q+\bm g_1;t), \rho(-\bm q-\bm g_2;0)]\rangle e^{i\nu t},
\end{equation}
where $\langle...\rangle$ stands for the thermal average. One can show from Eqs.~\eqref{eq:W} and~\eqref{eq:Pi} that
\begin{equation}\label{eq:W_hermiticity}
    W_{\bm g_1,\bm g_2}(\bm q,\nu) = W_{-\bm g_1,-\bm g_2}(-\bm q,-\nu)^*.
\end{equation}
%
This equation allows us to focus on positive frequencies. 

The poles of $\hat{W}(\bm q,\nu)$ at $\hbar\omega_{\bm q}$ on the positive real frequency axis correspond to undamped collective excitations, plasmons. For $\nu\approx \hbar\omega_{\bm q}>0$,
\begin{equation}
    W_{\bm g_1,\bm g_2}^{\prime\prime}(\bm q,\nu) = -\pi\phi_{\bm q+\bm g_1}\phi_{\bm q+\bm g_2}^*\delta(\nu-\hbar\omega_{\bm q}). \label{eq:W_plasmonpole} 
\end{equation}
Inserting Eq.~\eqref{eq:W_plasmonpole} into Eq.~\eqref{eq:W_spectral} yields an interaction mediated by plasmons that couple to electrons through a local electric potential $\phi(\bm r, t) = \sum_{\bm g} \phi_{\bm q+\bm g}e^{i(\bm q+\bm g)\cdot\bm r-i\omega_{\bm q} t}$. The same screening potential $\hat{W}$ can be deduced from the fermion-boson model, Eq.~\eqref{eq:H_bf}, to the second order in the 
electron-plasmon coupling Hamiltonian
\begin{equation}
    H_{\text{e-pl}} = \frac{1}{\sqrt{N}}\sum_{\bm q,\bm k\in \text{mBZ}}\sum_{\lambda{\rho}} g_{\lambda{\rho}}(\bm k,\bm q) c_{\bm k \lambda}^{\dagger}c_{\bm k-\bm q {\rho}} b_{\bm q} + \text{H.c.} ,\label{eq:H_e_pl}
\end{equation}
with
\begin{equation}
    g_{\lambda{\rho}}(\bm k,\bm q) = \frac{1}{\sqrt{A_M}}\sum_{\bm g}\langle u_{\bm k}^\lambda |u_{\bm k-\bm q-\bm g}^{\rho}\rangle\phi_{\bm q+\bm g}. \label{eq:g}
\end{equation}
Equation~\eqref{eq:g} provides a microscopic expression for the electron-plasmon coupling, where $\phi_{\bm q+\bm g}$ is determined from the screening potential (or the density-density response function) using Eq.~\eqref{eq:W_plasmonpole}. It generalizes Eq.~\eqref{eq:g_phi} to multiband cases and also includes the local field effects in the moir\'e superlattice, which become important at large wave vectors $q\sim k_{M}$.

In the presence of undamped plasmon excitations, we can use the simplified expression Eq.~\eqref{eq:W_plasmonpole} for $W^{\prime\prime}$. Combining it with Eqs.~\eqref{eq:A_inelastic},~\eqref{eq:sigma_im}, and~\eqref{eq:W_im_band}, and plugging the obtained spectral function into Eq.~\eqref{eq:didv_full},
we obtain the zero-temperature differential conductance contributed by the electron-plasmon scattering,
\begin{align}\label{eq:dIdV_plasmon}
    \frac{dI}{dV} = \frac{e^2 \Gamma}{\hbar N} &\sum_{\bm q\in \text{mBZ}}\sum_{\lambda}
    |\overline{M}_{\lambda}(\bm K_{\theta},\bm q)|^2\notag\\
    \times&\left\{\Theta(\xi_{\bm K_{\theta}-\bm q}^{\lambda})\delta(eV -\xi_{\bm K_{\theta}-\bm q}^{\lambda}-\hbar\omega_{\bm q})\right.\notag\\
    &\left.+ \Theta(-\xi_{\bm K_{\theta}-\bm q}^{\lambda})\delta(eV -\xi_{\bm K_{\theta}-\bm q}^{\lambda}+\hbar\omega_{\bm q})\right\}.
\end{align}
where we defined the normalized inelastic tunneling rate
\begin{equation}\label{eq:M_multiband}
    \overline{M}_{\lambda}(\bm K_{\theta},\bm q)= \sum_{{\rho}} \frac{\Lambda_{{\rho}}(\bm K_{\theta}) g_{{\rho}\lambda}(\bm K_{\theta},\bm q)}{eV-\xi_{\bm K_{\theta}}^{\rho}}.
\end{equation}
%
This expression along with Eq.~\eqref{eq:average_matrix_elements} generalizes Eq.~\eqref{eq:didv_in} to cases with multiple intermediate states $|\bm K_{\theta}{\rho}\text{S}\rangle$ in the sample.

In Appendix~\ref{sec:fermi}, we show that the same inelastic tunneling current can be derived from the Fermi's golden rule as well.
The diagrammatic approach has the advantage that it remains applicable when plasmons have a finite decay rate. There Eqs.~\eqref{eq:W_plasmonpole} and~\eqref{eq:dIdV_plasmon} can be generalized by replacing the Dirac-delta with the spectral weight of the plasmons.

{Note that we have not yet considered the Coulomb interactions experienced by the tip electrons. Those interactions, especially the interlayer interactions between the tip and sample, could lead to several effects: First, the tip can screen the Coulomb interactions and potentially damp the charge fluctuations in the sample. This effect can be minimized by maintaining a low doping level in the tip. Second, the interlayer Coulomb interaction renormalizes the tunneling Hamiltonian $\hat{H}_{\text{tun}}$ \cite{vafek2020renormalization} and also introduces an additional channel for the inelastic tunneling through the scattering of the tip electrons by the plasmons in the sample. Although these effects can yield corrections to Eq.~\eqref{eq:i} for tunneling current and Eq.~\eqref{eq:didv_full} for $dI/dV$, we expect the corrections to be quantitatively small thanks to the steep electron dispersion of the tip MLG (see Appendix~\ref{sec:tip_screening} for more details). }

Finally, the formalism discussed in this section could also be applied to the inelastic tunneling arising from other collective bosonic excitations in the electronic system, each associated with different particle-hole vertices $\hat{W}(\bm q,\omega)$.


\section{Tunneling spectrum of interacting TBG}\label{sec:numerics}

In this section, we study the spectral properties of charge-neutral TBG described by the interacting BM model, Eqs.~\eqref{eq:h0}-\eqref{eq:hint}, at small twist angles, and illustrate how to extract information about the plasmons in TBG from the QTM tunneling spectra. We discuss different signatures associated with inelastic tunneling 
into the central flat bands and into the remote dispersive bands, making connections with the qualitative descriptions in Sec.~\ref{sec:summary}. 

Our results are based on the random phase approximation (RPA) for the plasmons and the one-shot $GW$ approximation for the spectral function of TBG. These perturbation theories are most reliable if the Coulomb interaction scale is not larger than the bandwidth of the central flat bands. For this reason, we consider two moderately interacting scenarios: $1.20^{\circ}-$TBG away from the magic angle in Sec.~\ref{sec:non_magic}; and $1.07^{\circ}-$TBG subject to strong gate/dielectric screening in Sec.~\ref{sec:screening}. Throughout the entire section, we fix the model parameters $v_D=10^6\ $\si{\meter\per\second}, $w_1=117$ \si{meV}, $w_0=0.7w_1$, and relative permittivity $\epsilon_r=10$. Additional details of the numerical calculations can be found in Appendix~\ref{sec:numeric_details}. 

\subsection{TBG away from the magic angle}\label{sec:non_magic}

\begin{figure}
    \centering
    \includegraphics[width=1\linewidth]{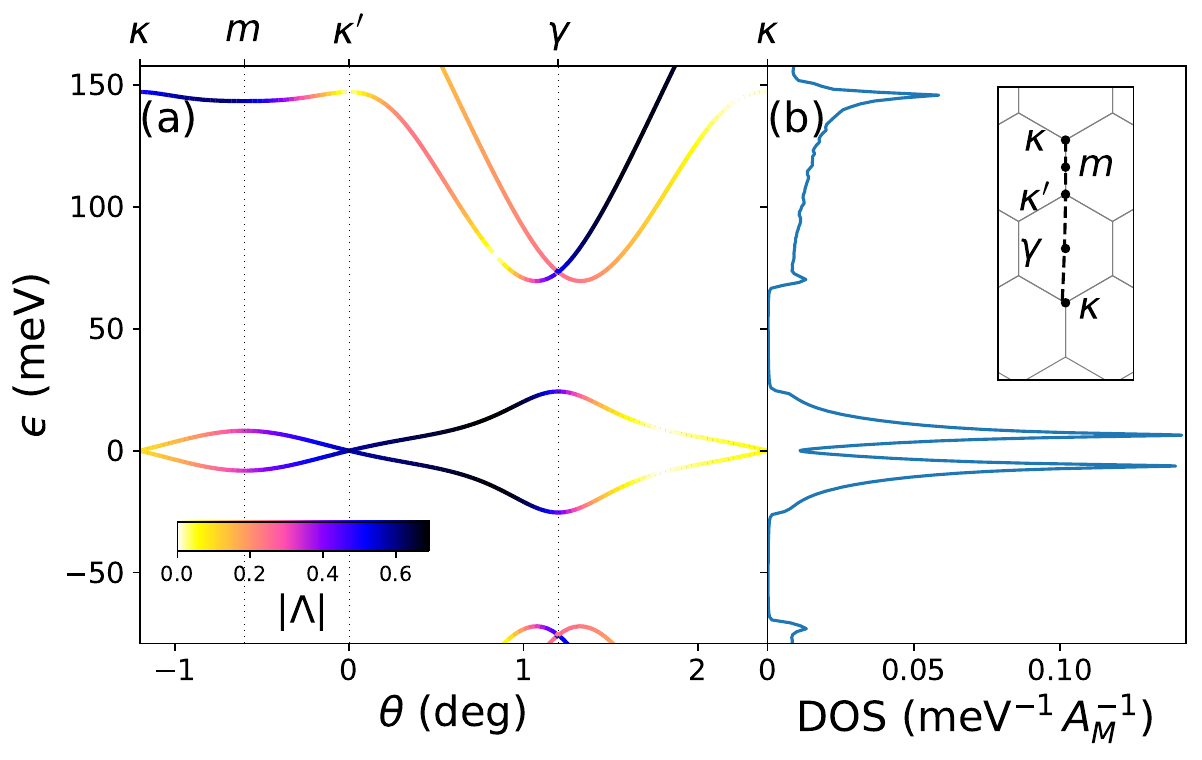}
    \caption{The non-interacting band structures of $1.20^{\circ}-$TBG. (a) The energy dispersion of TBG along the trajectory of the Dirac point of the tip, see the inset in panel (b). $\theta$ denotes the twist angle between the tip and the top layer of TBG. The color encodes the absolute value of the normalized tunneling matrix elements given by Eq.~\eqref{eq:average_matrix_elements}. (b) The tunneling density of states of the non-interacting TBG at charge neutrality, $-\text{Im}\sum_{\bm k} \text{Tr}(G^{0}(\bm k, \epsilon))/\pi\Omega$, with a Lorentzian broadening parameter $\eta=0.4$ \si{meV}.}
    \label{fig:tbg_1.20}
\end{figure}

\begin{figure}
    \centering
    \includegraphics[width=1\linewidth]{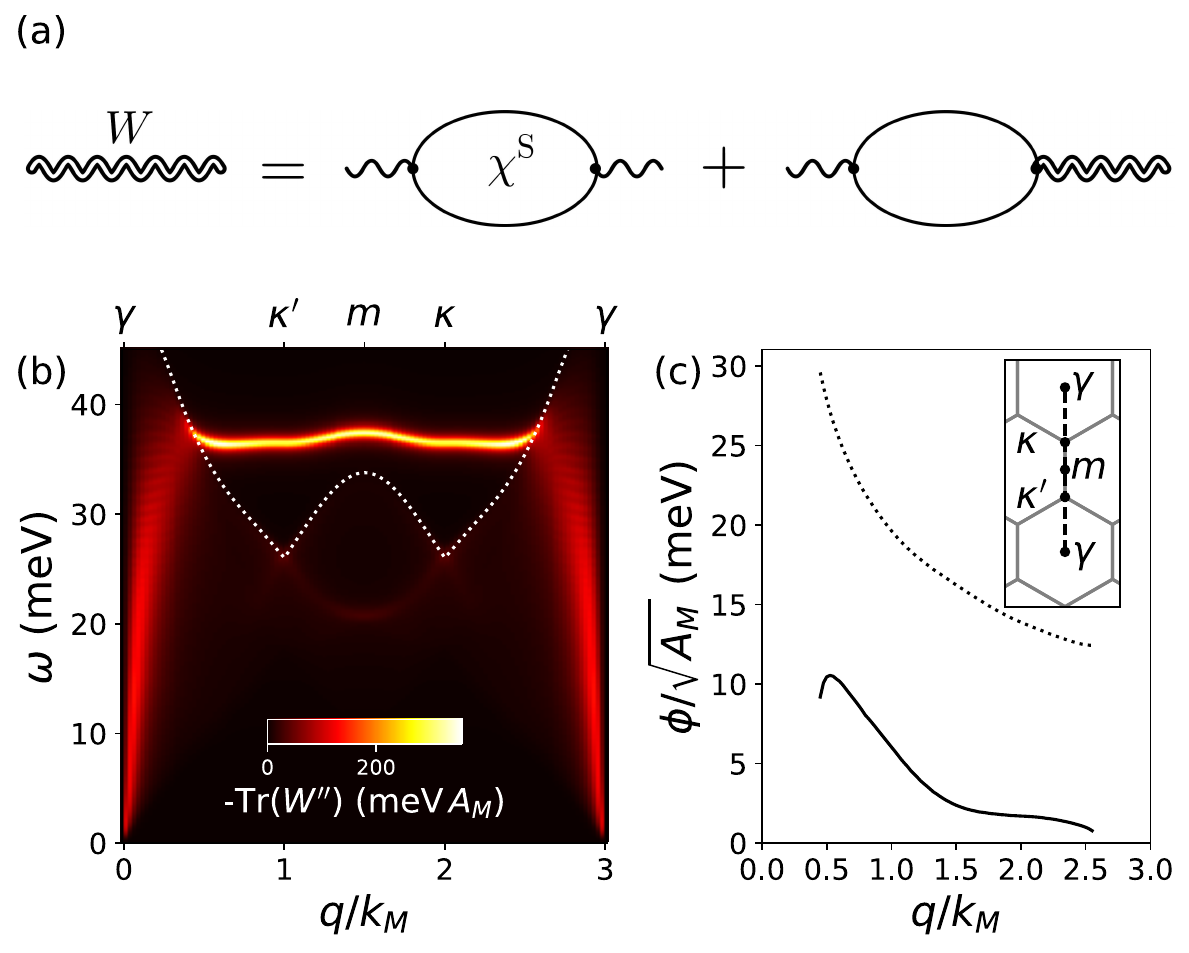}
    \caption{Plasmons in charge neutral $1.20^{\circ}-$TBG within the random phase approximation. (a) Feynman diagrams of the RPA screening potential $\hat{W}$. (b) The dissipative part of the screening potential, $-\text{Tr}(\hat{W}^{\prime\prime}(\bm q,\omega))$, as a function of wave vector $\bm q=q\hat{y}$ and energy $\omega$. The bright curve at energy about $40$ \si{meV} traces the plasmon dispersion $\hbar\omega_{\bm q}$. The broadening of the plasmon spectrum at lower energies (seen at $q\to 0$ and $q\to 3k_M$) comes from the plasmon decay into particle-hole pairs. The white dotted line delineates the boundary of the particle-hole continuum.
    (c) The electrostatic potential $\phi_{\bm q}$ associated with the zero-point motion of plasmons is plotted along a high-symmetry line in the extended mBZ, see the inset. (The black dot corresponds to the origin). In this plot, we exclude the region of $\bm q$ where plasmons enter the particle-hole continuum. The dotted line, $\sqrt{\hbar\omega_{\bm q}v_{\bm q}/2}$ with the bare Coulomb potential $v_{\bm q}$, provides a rough estimate of $\phi$, cf.\ Eq.~(\ref{eq:g_phi}).
    }
    \label{fig:plasmon_1.20}
\end{figure}


For the given parameters, the non-interacting band structure and density of states of a charge-neutral $1.20^{\circ}-$TBG are depicted in Fig.~\ref{fig:tbg_1.20}(a) and (b), respectively.

We compute the screening potential and the plasmons by using the random phase approximation (RPA), as shown in Fig.~\ref{fig:plasmon_1.20}(a) \footnote{In our notation, $\hat{W}$ does not include the first order Coulomb interaction and is different from the conventional notation in the $GW$ approximation.}. Within RPA, the density-density response function of the sample in Eq.~\eqref{eq:Pi} reduces to
\begin{equation}\label{eq:Pi_rpa}
    \Pi_{\bm g_1, \bm g_2}(\bm q, \omega) = \left[\hat{\chi}^{\text{S}}(\bm q, \omega)^{-1} - \hat{V}(\bm q)\right]_{\bm g_1, \bm g_2}^{-1},
\end{equation}
where $V_{\bm g_1, \bm g_2}(\bm q) = v_{\bm q + \bm g_1}\delta_{\bm g_1,\bm g_2}$ denotes the bare Coulomb interaction, and $\hat{\chi}^{\text{S}}$ is the single-particle density-density response function,
%
%
%
\begin{align}\label{eq:chi0}
    \chi_{\bm g_1, \bm g_2}^{\text{S}} (\bm q, \omega) = \frac{2}{\Omega}&\sum_{\tau=\pm}\sum_{\bm k\in \text{mBZ}}\sum_{\lambda,{\rho}} \frac{n_F(\xi_{\bm k}^{\lambda\tau}) - n_F(\xi_{\bm k+\bm q}^{{\rho}\tau})}{\omega + i\eta +\xi_{\bm k}^{\lambda \tau} - \xi_{\bm k+\bm q}^{{\rho}\tau}}  \notag\\
     &\times\langle u_{\bm k}^{\lambda\tau} | u_{\bm k+\bm q+\bm g_1}^{{\rho}\tau}\rangle\langle u_{\bm k+\bm q+\bm g_2}^{{\rho}\tau} |u_{\bm k}^{\lambda\tau}\rangle. 
\end{align}
The response function is block diagonal in reciprocal space because of Umklapp processes. The factor of $2$ accounts for the spin degeneracy. Here, we restore the valley index $\tau$. The two valleys are related by time-reversal symmetry, $\xi_{\bm k}^{\lambda +} = \xi_{-\bm k}^{\lambda -}$ and $u_{\bm k}^{\lambda +} = (u_{-\bm k}^{\lambda -})^{*}$. 

The RPA screening potential $\hat{W}$ can be obtained by inserting the RPA response function, Eq.~\eqref{eq:Pi}, into Eq.~\eqref{eq:W}. We compute its anti-hermitian part $i\hat{W}^{\prime\prime}$ and plot $-\text{Tr}(\hat{W}^{\prime\prime}(q\hat{y},\omega))$ in Fig.~\ref{fig:plasmon_1.20}(b). Sharp peaks of $\hat{W}^{\prime\prime}$ can be observed at energy about $40$ \si{meV} as $q$ approaches the mBZ boundary, see the inset of panel (c). These peaks correspond to the plasmon poles of $\hat{W}$. Here, the plasmon is a collective interband particle-hole excitation because the intraband excitation vanishes at charge neutrality. 
Note that the plasmon energy is smaller than the bandwidth of the central-band manifold ($\sim 50$ \si{meV}), and therefore the particle-hole pair near the $\bm\gamma$ point can damp the plasmons, leading to broad peaks at small $q$. 

Near the zone boundary where the plasmons are undamped, we can determine the electrostatic potential $\phi_{\bm q}$ generated by plasmons according to Eq.~\eqref{eq:W_plasmonpole}. The results are plotted in Fig.~\ref{fig:plasmon_1.20}(c). Overall, $\phi_{\bm q}$ drops rapidly as $\bm q$ increases because, microscopically, these plasmons are charge oscillations within the flat bands and the Bloch wave functions $u_{\bm k+ \bm g}^{\lambda}$ of the flat bands are concentrated in one mBZ. For comparison, the dotted line depicts an estimation of $\phi_{\bm q}$ based on Eq.~\eqref{eq:g_phi} for long-wavelength plasmons. 
The deviations of the estimate from the full numerical result (solid lines) can be attributed to at least two reasons. First, the expression $\phi_{\bm q}=\sqrt{\hbar\omega_{\bm q} v_{\bm q}/2}$ is derived under the assumption that the plasmon energy is well above the particle-hole continuum at the same $\bm q$, i.e., $|\xi_{\bm k+\bm q}^{\lambda}-\xi_{\bm k}^{\rho}|/\hbar\omega_{\bm q}\ll 1$ for the flat bands $\lambda,\rho$. This condition is invalid here. Furthermore, Eq.~\eqref{eq:g_phi} does not consider the local field effects in moir\'e superlattices and therefore cannot capture the rapid decay of $\phi_{\bm q}$ at large $q$. 

\begin{figure}
    \centering
    \includegraphics[width=1\linewidth]{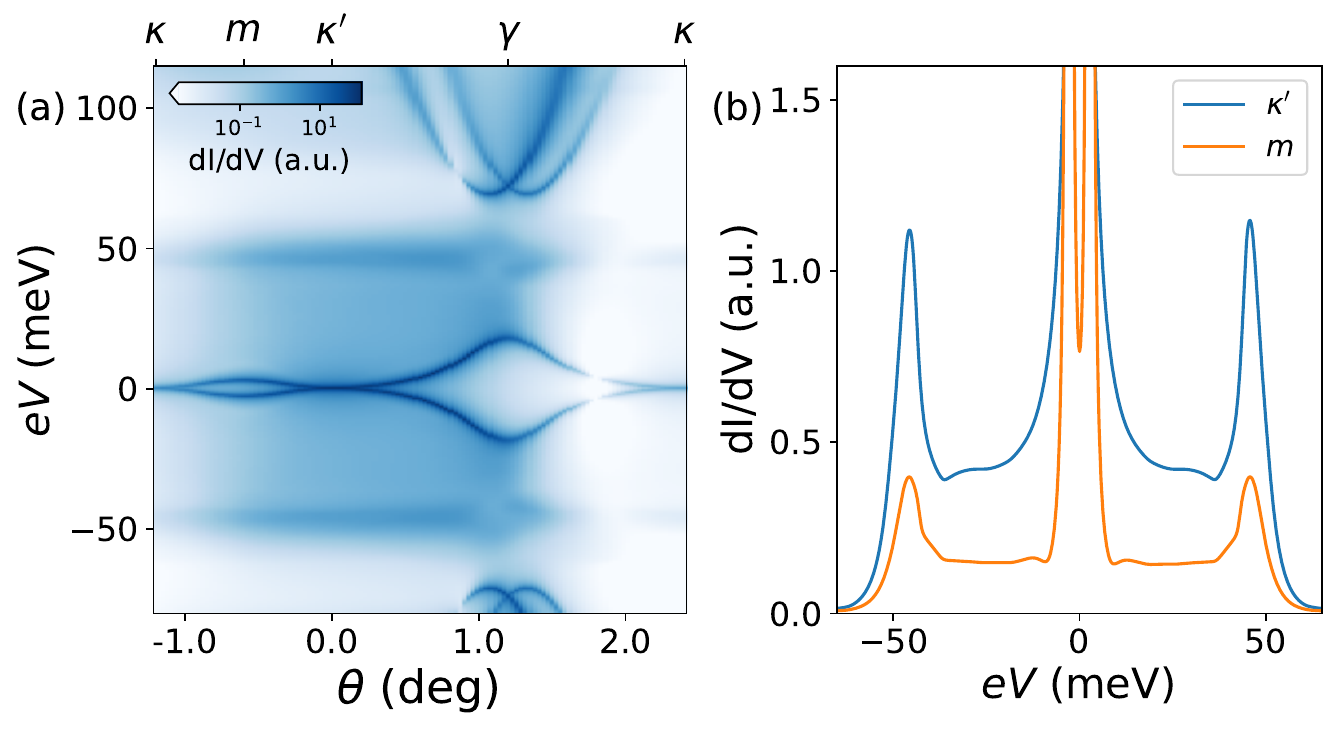}
    \caption{Tunneling spectrum of charge-neutral $1.20^{\circ}-$TBG. (a) $dI/dV$ map as a function of the twist angle $\theta$ and the bias voltage. Sharp lines at low bias represent elastic tunneling into the central bands. Nonlinear conductance at higher bias comes from inelastic tunneling facilitated by emission of damped plasmons and particle-hole excitations. The higher value of $dI/dV$ at $eV\approx 50$ \si{meV} corresponds to the emission of plasmons with nearly-flat spectrum, see Fig.~\ref{fig:plasmon_1.20}(b). (b) Line cuts of (a) at two different twist angles corresponding to when the tip Dirac point passes the $\bm\kappa^{\prime}$ and $\bm m$ points in the mBZ. There are two plasmon satellites, each separated from the strong quasiparticle peaks of the two central bands by approximately the average plasmon energy.}
    \label{fig:dIdV_1.20}
\end{figure}

\begin{figure}
    \centering
    \includegraphics[width=1\linewidth]{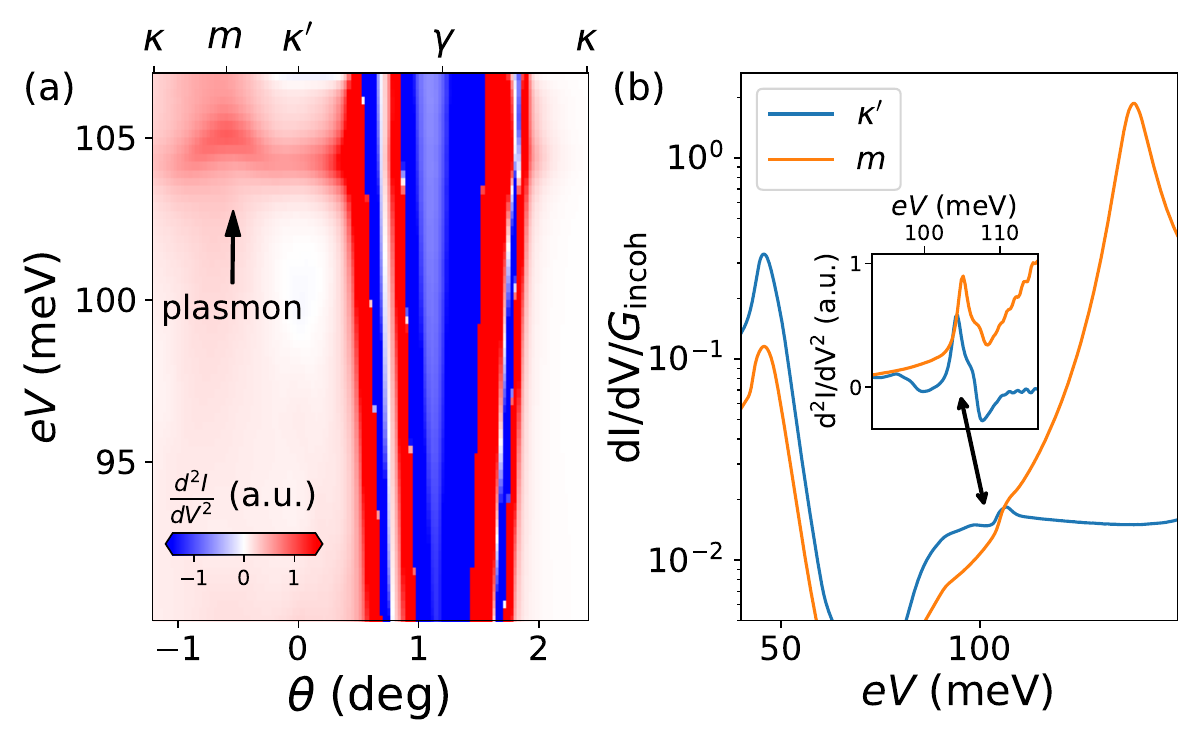}
    \caption{Signatures of plasmons in the tunneling spectrum of charge-neutral $1.20^{\circ}-$TBG. (a) $d^2I/dV^2$ map as a function of the twist angle $\theta$ and the bias $eV$. The broad horizontal feature corresponds to the onset bias allowing the plasmon-assisted tunneling into the remote bands. It traces out the dispersion of the plasmon shown in Fig.~\ref{fig:plasmon_1.20}. Bright almost-vertical lines arise from the elastic tunneling associated with the remote bands of TBG. (b) $dI/dV$ when the tip Dirac point is twisted to the $\bm\kappa^{\prime}$ and $\bm m$ points. The onset of plasmon-assisted tunneling leads to a kink in $dI/dV$ pointed by the arrow. This kink corresponds to a peak in $d^2I/dV^2$, see the inset. 
    The size of the kink is orders-of-magnitude weaker than that of the plasmon satellites at lower bias and the quasiparticle peaks of the nearest remote bands at higher bias (see, e.g., the orange line). At $\bm \kappa^{\prime}$, however, the quasiparticle peak of the lowest remote conduction band vanishes due to the suppression of the tunneling matrix elements for this band. }
\label{fig:d2IdV2_1.20}
\end{figure}

To calculate the tunneling conductance, we first use the above RPA screening potential $\hat{W}$ and Eqs.~\eqref{eq:sigma} and~\eqref{eq:W_im_band} to calculate the self-energy $\hat{\Sigma}(\bm k,\epsilon)$. By plugging this approximated $\hat{\Sigma}(\bm k,\epsilon)$ into the Dyson series, Eq.~\eqref{eq:G}, we obtain the $GW$ approximation for the Green's function
\begin{equation}\label{eq:G_rpa}
    \hat{G}(\bm k,\epsilon) = [\hat{G}^0(\bm k,\epsilon)^{-1}-\hat{\Sigma}(\bm k,\epsilon)]^{-1}.
\end{equation}
We then express the spectral function via Eq.~\eqref{eq:A_def} and calculate the tunneling conductance via Eq.~\eqref{eq:didv_full}. Figure~\ref{fig:dIdV_1.20}(a) depicts the tunneling conductance between the tip and the charge neutral TBG as a function of the twist angle $\theta$ and bias $eV$ between the tip and sample. 
Besides the quasiparticle peaks which map out the band structures of TBG, cf.\ Fig.~\ref{fig:tbg_1.20}(a), we find satellite peaks at biases $eV\sim\pm 50$ \si{meV}, which are nearly twist-angle independent except near the $\bm\gamma$ point. The two satellites arise from inelastic scattering of an electron and a hole off one plasmon, respectively. The energy of the satellite peak on the positive bias side is approximately given by the plasmon energy$~\sim 40$ \si{meV} plus the electron energy corresponding to the Van Hove singularity$~\sim 6$ \si{meV} of the central conduction band, cf.\ Fig.~\ref{fig:tbg_1.20}(b). This peak in $dI/dV$ comes from the convolution of the plasmon density of states $D(\omega)$ and the density of states of the central-conduction-band electrons. 
Compared to the simplified case considered in Section~\ref{sec:summary}, the latter peaks at the energy$~\xi_{\rm VH}$ of the Van Hove singularity, rather than at $\xi=0$. That shifts $eV\to eV-\xi_{\rm VH}$ in Eq.~(\ref{eq:didv_flatband}). The line cuts at $\bm \kappa'$ ($\theta=0$) and $\bm m$ ($\theta=-0.6^{\circ}$) plotted in Fig.~\ref{fig:dIdV_1.20}(b) indicate that the plasmon satellites are much weaker than the quasiparticle peaks at low bias.

The signatures of plasmons can also be found at sufficiently high bias where electrons (holes) have enough energy to tunnel into the remote bands by emitting a plasmon simultaneously. Figure~\ref{fig:d2IdV2_1.20}(a) plots a map of $d^2I/dV^2$ vs.\ $\theta$ and $eV$. Apart from the strong elastic tunneling features near the $\bm\gamma$ point, we find weak $d^2I/dV^2$ peaks at bias $eV^*(\theta)\sim 105$ \si{meV}, marked by the arrow. We have confirmed that $V^*(\theta)$ traces the dispersion of the plasmons, $eV^*(\theta)\approx \xi_{cm} + \hbar\omega_{\bm Q_{\theta}}$ where the band minimum of the upper remote bands is at energy $\xi_{cm}\approx 70$ \si{meV}, cf.\ Fig.~\ref{fig:tbg_1.20}(b), and the wave vector of the emitted plasmon $\bm Q_{\theta} =\bm K_{\theta} - \bm K_{\theta_{\text{TBG}}}$. According to Eq.~\eqref{eq:didv_phonon} and the discussions in Sec.~\ref{sec:summary}, these $d^2I/dV^2$ peaks indicate the onset of the inelastic tunneling into the remote bands assisted by less dispersive bosons (plasmons).  

To gain insight into the intensity of the $d^2I/dV^2$ peaks, we consider an inelastic tunneling process illustrated in Fig.~\ref{fig:dispersive_band}(b) at $\theta=0$. The emitted plasmon carries wave vector $\bm Q_{\theta}=k_M\hat{y}$. 
We plug the electron-plasmon coupling $|g_{c,f}(\bm\gamma, \bm Q_{\theta})|\sim \phi_{\bm Q_{\theta}}/\sqrt{A_M}\sim  6$ \si{meV} (cf.\ Fig.~\ref{fig:plasmon_1.20}(b)), and $eV^{*}\approx 100$ \si{meV} into Eq.~\eqref{eq:M_multiband}, and estimate the normalized inelastic tunneling rate from the tip to the two lowest remote conduction bands, $|\overline{M}_c(\bm K_{\theta},\bm Q_{\theta})|^2\sim |g_{c,f}(\bm \gamma,\bm Q_{\theta})/eV^*|^2\sim 4\times 10^{-3}$, where $c$ and $f$ label the two lowest remote conduction bands and the flat bands, respectively. Based on Eq.~\eqref{eq:didv_phonon}, we estimate that the plasmon-assisted inelastic tunneling increases the differential conductance by $\Delta(dI/dV)\sim 4\times 10^{-3} G_{\text{incoh}}$, where~\footnote{Note that $G_{\text{incoh}}$ is the scale for the features in $dI/dV$ resulting from elastic, momentum-conserving tunneling~\cite{xiao_theory_2024}.}
\begin{equation}\label{eq:G_incoh}
    G_{\text{incoh}}\equiv \Gamma A_M\nu_c(0).
\end{equation}
%
Figure~\ref{fig:d2IdV2_1.20}(b) plots the numerical simulation of $dI/dV$ when the tip Dirac point is twisted to $\bm m$ and $\bm\kappa'$ in the mBZ. The arrow points to kinks in $dI/dV$ associated with the plasmon-assisted inelastic tunneling. 
The size of the kink roughly matches our estimate, indicating that near $\theta=0$, the inelastic tunneling primarily occurs through the virtual electronic states in the flat bands of TBG, see Fig.~\ref{fig:dispersive_band}(b).

We caution that the kinks in $dI/dV$ are relatively weak features in the tunneling spectrum. As shown in Fig.~\ref{fig:d2IdV2_1.20}(b), these features are about two orders of magnitude weaker than the plasmon satellites generated by inelastic tunneling into the central bands at low bias. The difference can be attributed to two factors: the smaller density of states in the remote bands compared to the central bands, and the large energy separation between the remote and central bands. The latter reduces the inelastic tunneling rate, Eq.~\eqref{eq:M_multiband}.

The plasmon-induced kinks in $dI/dV$ are generally much weaker than the quasiparticle peaks associated with the elastic tunneling. The orange curve in Fig.~\ref{fig:d2IdV2_1.20}(b) shows that the quasiparticle peaks of the remote bands at higher bias can be three orders of magnitude stronger than the kinks in $dI/dV$. 
However, interestingly, the quasiparticle peaks are strongly suppressed near $\bm\kappa'$ (blue curve in Fig.~\ref{fig:d2IdV2_1.20}(b)) due to the suppression of the tunneling matrix elements $\Lambda_{r}(\bm K_{\theta=0})$ of the lowest remote conduction band, cf.\ Fig.~\ref{fig:tbg_1.20}(a). The origin of the suppression is a strong polarization of the electron states  
at $\bm\kappa'$ towards the bottom layer of TBG, remote from the tip.  
Consequently, the plasmon-induced inelastic tunneling features exhibit better contrast against the background near $\bm\kappa'$ in Fig.~\ref{fig:d2IdV2_1.20}.

\subsection{MATBG under strong gate/dielectric screening}\label{sec:screening}

In this section, we study possible signatures of the plasmons hosted by the flat bands of MATBG in the QTM tunneling spectrum. To avoid complications arising from band reconstructions in strongly correlated MATBG, one can reduce the Coulomb interaction using metallic gates or high-$\kappa$ dielectrics \cite{gao2024double} (see also \cite{wang2025independently, barrier_coulomb_2024}) that are vertically separated from MATBG by a small distance $d_g\ll a_M$, where $a_M$ is the moir\'e lattice constant. In the following, we use the dual-gate-screened Coulomb potential, Eq.~\eqref{eq:vq}, with $d_g=1$ \si{nm}. The gate screening makes the Coulomb interaction short-ranged. For a $1.07^{\circ}-$TBG, we estimate the bare on-site Coulomb interaction $U$ to be comparable to the bandwidth of the two flat bands $\mathcal{B}\approx U\approx 10$ \si{meV} \cite{cualuguaru2023twisted}.

\begin{figure}
    \centering
    \includegraphics[width=1\linewidth]{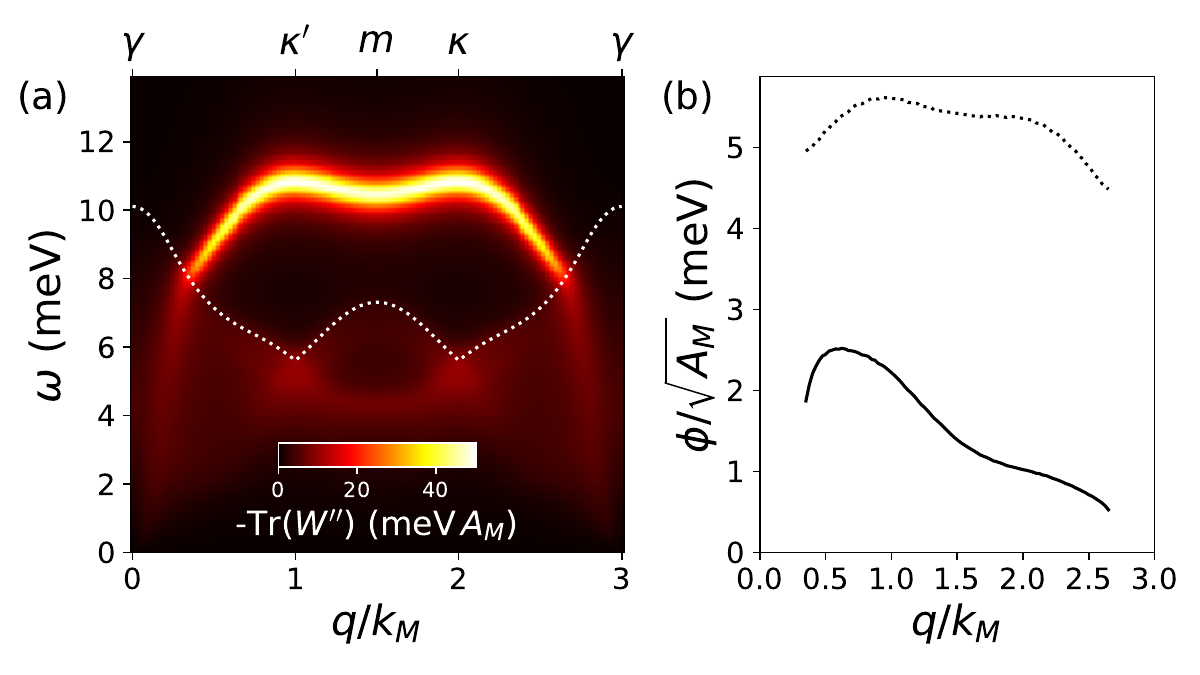}
    \caption{Plasmons of charge neutral $1.07^{\circ}-$TBG situated in the middle of two metallic gates with a sample-gate distance $d_g=1$ \si{nm}. (a) The wave vector and energy dependence of the trace of the dissipative part of the RPA screening potential, $-\text{Tr}(\hat{W}^{\prime\prime}(q\hat{y},\omega))$. The bright curve traces the plasmon dispersion $\hbar\omega_{\bm q}$. The white dots mark the boundary of the particle-hole continuum.
    (b) The same quantities as plotted in Fig.~\ref{fig:plasmon_1.20}(c). The solid line represents the electrostatic potential $\phi_{\bm q}$ associated with the zero-point motion of plasmons, and the dotted line, $\sqrt{\hbar\omega_{\bm q}v_{\bm q}/2}$, provides an order-of-magnitude estimate of $\phi$.}
    \label{fig:plasmon_1.07}
\end{figure}

As gate screening becomes less effective at large wave vectors $q\gtrsim d_g^{-1}\approx 0.3k_M$, there remain well-defined plasmons close to the mBZ boundary, as shown in Fig.~\ref{fig:plasmon_1.07}. In real space, the electric fields generated by these plasmons away from the $\bm \gamma$ point are concentrated in the AA stacking region over a length scale of a few \si{nm} \cite{stauber2016quasi}, where most of the flat-band electrons are localized. 

\begin{figure}
    \centering
    \includegraphics[width=1\linewidth]{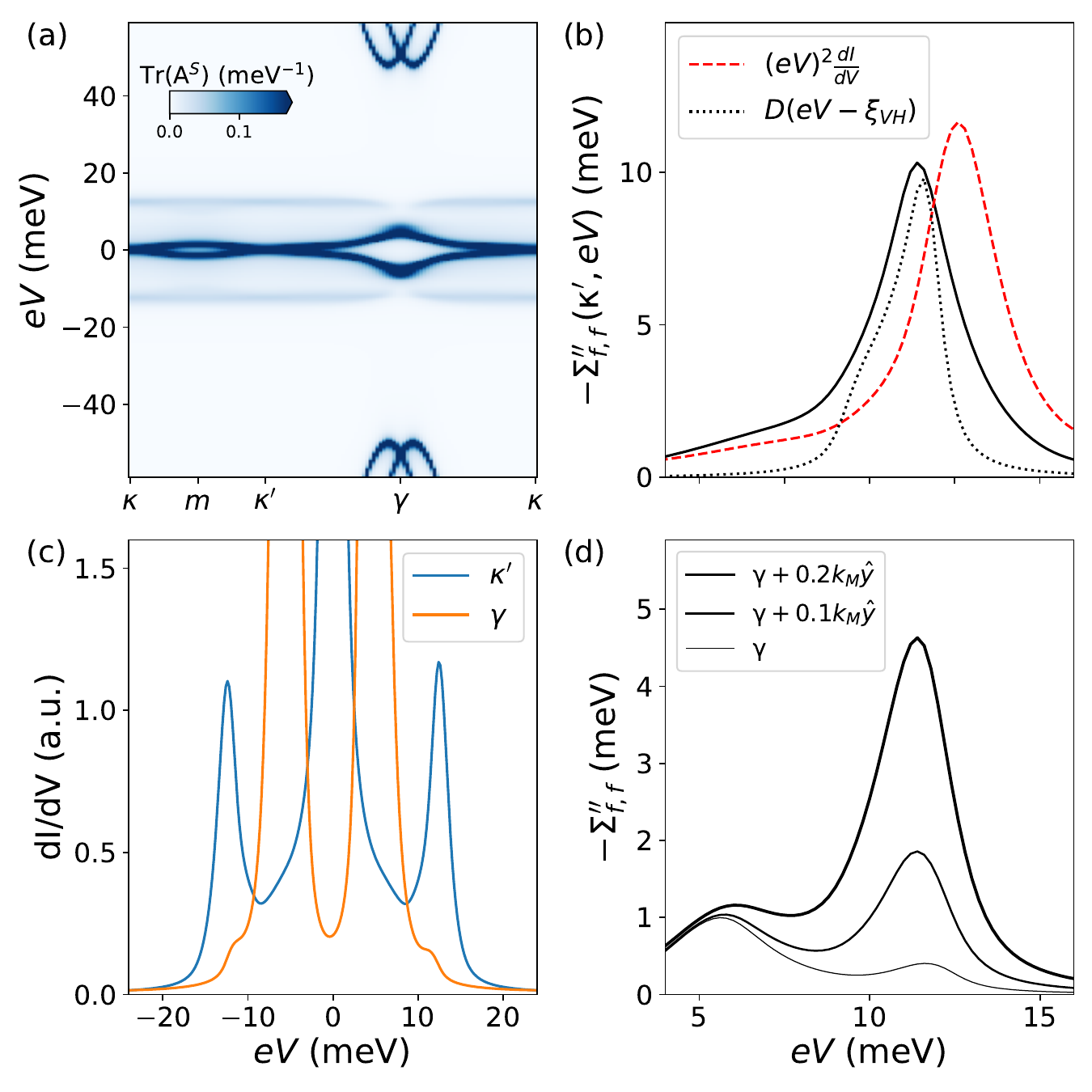}
    \caption{(a) Spectral function $A^{\text{S}}(\bm k, \epsilon)$ of the spin-valley symmetric phase in charge neutral $1.07^{\circ}-$TBG, evaluated in one-shot $GW$ approximation and plotted along a high symmetry line of mBZ traversed by the Dirac point of the MLG tip. In addition to the strong quasiparticle peaks, $A^{\text{S}}$ exhibits two almost dispersionless plasmon satellites centered at energies $\epsilon\sim \pm 10$ \si{meV}. The intensities of the satellites drastically decrease close to the $\bm \gamma$ point. (b) The solid line represents the imaginary part of the self-energy $\Sigma_{f,f}^{\prime\prime}$ of the flat conduction band at $\bm\kappa'$ at energies away from the quasiparticle energy $\xi_{\bm \kappa'}^{f}=0$. The dotted line plots the density of states $D$ (a.u.) of the undamped plasmons shifted horizontally by the Van Hove singularity $\xi_{\text{VH}}=1.2$ \si{meV} of the flat conduction band. It indicates a Van Hove singularity of the plasmon dispersion at $\omega_0\approx 10$ \si{meV}. The red dashed line plots the rescaled tunneling conductance $dI/dV$, which follows the trend of $\Sigma_{f,f}^{\prime\prime}$ and $D$. (c) The simulated tunneling conductance $dI/dV$ as a function of bias $eV$ when the tip Dirac point $\bm K_{\theta}$ is twisted to the $\bm \kappa^{\prime} (\theta=0)$ and $\bm \gamma (\theta=\theta_{\text{TBG}})$ points, respectively. In agreement with the color plot of the spectral function in panel (a), the $dI/dV$ curves show two satellites at energies higher than the quasiparticle peaks, but the satellites nearly vanish at $\bm\gamma$. (d) The imaginary part $\Sigma_{f,f}^{\prime\prime}(\bm k,eV)$  of the self-energy at three different values of wave vector $\bm k$. As $\bm k\rightarrow \bm \gamma$, $\Sigma_{f,f}^{\prime\prime}$ rapidly decreases at energy $eV\approx \omega_0+\xi_{\text{VH}}$, indicating a suppression of the electron-plasmon coupling for electrons near $\bm \gamma$.}
    \label{fig:A_1.07}
\end{figure}

Next, we compute the $GW$ self-energy $\hat{\Sigma}$ and spectral function $A^{\text{S}}(\bm k,\epsilon)$ of MATBG. The spectral function plotted in Fig.~\ref{fig:A_1.07}(a) is qualitatively the same as that of TBG at larger twist angles: We find quasiparticle peaks near the Fermi level as well as almost dispersionless plasmon satellites for both signs of the energy. In Fig.~\ref{fig:A_1.07}(b), the solid line represents the imaginary part $\Sigma^{\prime\prime}$ of the electron self-energy at the $\bm\kappa'$ point of MATBG. (The flat conduction and valence bands are degenerate and have the same self-energy at $\bm\kappa'$.) The red dashed line plots exhibit the rescaled tunneling conductance $(eV)^2dI/dV$ at $\theta=0$ where the tip Dirac point passes $\bm\kappa'$. 
In the limit of weak electron-plasmon coupling where Eq.~\eqref{eq:A_largeomega} holds, $(eV)^2dI/dV$ should match with $\Sigma^{\prime\prime}(\bm\kappa',eV)$, and according to Eq.~\eqref{eq:didv_flatband} they should correlate with the density of states of the undamped plasmon (dotted line), 
\begin{equation}\label{eq:D}
    D(\omega) = -\frac{1}{\pi\Omega}\sum_{\bm q\in \text{mBZ}}\text{Im}\frac{2\hbar\omega_{\bm q}}{(\omega+ i\eta)^2-(\hbar\omega_{\bm q})^2}.
\end{equation}
The numerical results show that the three lines exhibit similar line shapes and peak at nearly the same energies, in qualitative agreement with the above expectations. The quantitative discrepancy between Eq.~\eqref{eq:didv_flatband} and our microscopic calculations of $dI/dV$ based on the $GW$ approximation can be understood as follows. First, the finite bandwidth of the flat bands broadens $\Sigma^{\prime\prime}$ and $A^{\text{S}}$ compared to $D$. Second, Landau damping of the plasmons makes Eqs.~\eqref{eq:didv_flatband} and~\eqref{eq:D} inapplicable at low energy $eV<8$ \si{meV}. Furthermore, neither Eq.~\eqref{eq:didv_flatband} nor $GW$ describes the plasmon satellites in the spectrum accurately when the electron-plasmon coupling (or the self-energy) is not much smaller than the average plasmon energy. It is known that the $GW$ approximation often overestimates the energies and intensities of the plasmon satellites \cite{martin2016interacting,caruso2016gw}.  

Interestingly, the intensity of the satellites drops rapidly near the $\bm\gamma$ point. As shown in Fig.~\ref{fig:A_1.07}(c), the tunneling conductance $dI/dV$ exhibits prominent satellite peaks at $\bm \kappa'$ but only small kinks at $\bm\gamma$. We further elaborate on this trend by plotting 
$\Sigma^{\prime\prime}(\bm k,eV)$ for $\bm k$ in the vicinity of
$\bm\gamma$ in Fig.~\ref{fig:A_1.07}(d). The peak
in $\Sigma^{\prime\prime}$ at $eV\approx\omega_0$ becomes lower
as $\bm k\rightarrow\bm\gamma$. We mention a similar self-energy behavior in the non-magic-angle TBG discussed in Sec.~\ref{sec:non_magic}. 
A common cause of the suppression of the electron-plasmon scattering for $\bm\gamma$-electrons in flat bands is a small overlap between the Bloch wave functions at the $\bm\gamma$-point and in other parts of the mBZ \cite{ledwith_nonlocal_2025}. 
A more drastic effect seen here in Fig.~\ref{fig:A_1.07} is due to the screening of the Coulomb potential by gates.
The gate screening confines the plasmons more strongly within the AA-stacking region of the moiré unit cell and further reduces their coupling with the $\bm\gamma$-electrons in the flat bands, which vanishes at the AA-stacking high symmetry point. 

\begin{figure}
    \centering
    \includegraphics[width=1\linewidth]{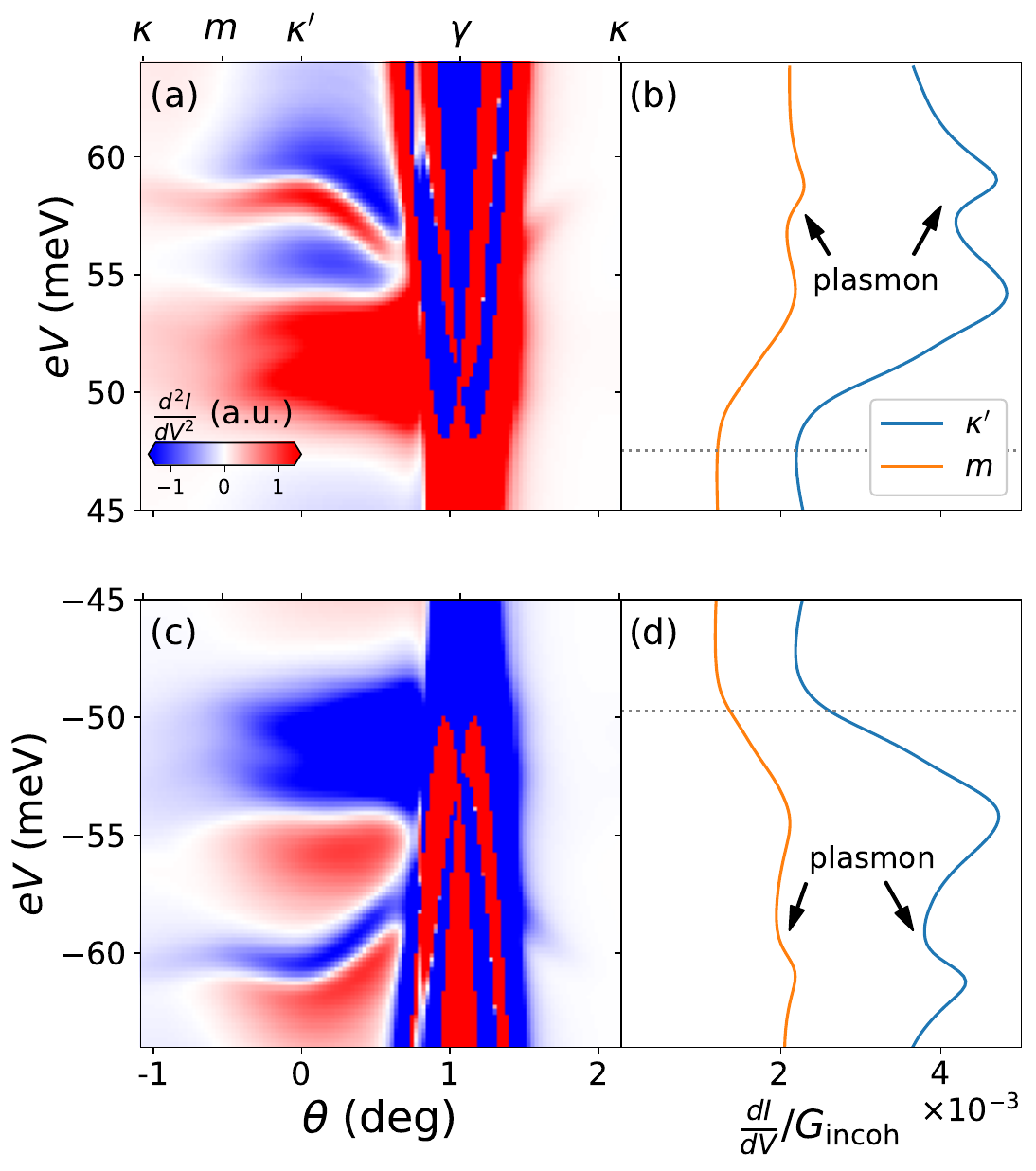}
\caption{ (a) $d^2I/dV^2$ map as a function of the twist angle $\theta$ and the bias voltage.  We choose a bias window allowing the onset of elastic tunneling into the remote dispersive bands and the plasmon satellites. The broader horizontal feature to the left of the $\bm \gamma$ point corresponds to inelastic tunneling into the remote band accompanied by creation of a particle-hole excitation in the flat band (the width of this feature is comparable to the flat bands width). A narrower feature above it comes from the plasmon-assisted tunneling into the edge of the remote band. There is a weaker counterpart of that feature to the right of $\bm \gamma$ point.
(b) $dI/dV$ at two different twist angles corresponding to the line cuts of the $d^2I/dV^2$ color plot, panel (a), at $\bm \kappa^{\prime}$ and $\bm m$. Each line shows an increase of $dI/dV$ associated with creating a particle-hole excitation at biases above the elastic tunneling threshold. Arrows point to the plasmon feature. 
(c) $d^2I/dV^2$ map in the bias range close to the maximum of the lower remote bands. As in (a), the plasmon dispersion can be revealed from $d^2I/dV^2$ dips on the two sides of the strong elastic tunneling feature around $\bm \gamma$. (d) The same as (b) except at negative bias. The dashed lines in (b) and (d) mark $\xi_{cm}$ and $\xi_{vm}$, respectively, and indicate the elastic tunneling thresholds. }
    \label{fig:d2Id2V_1.07}
\end{figure}

We next analyze the tunneling spectrum in the large-bias regime where electrons (holes) in the tip can tunnel into the upper (lower) remote bands. Figure~\ref{fig:d2Id2V_1.07}(a) depicts a map of $d^2I(\theta, eV)/dV^2$ at bias voltages close to the minimum of the remote conduction bands, $\xi_{cm}$. As the tip Dirac point is twisted away from the $\bm \gamma$ point, the features of elastic tunneling into the remote bands can be avoided, and at a fixed twist angle, $d^2I/dV^2$ exhibits at least two peaks as the bias increases. They correspond to two step-like increments of $dI/dV$ atop a smooth background, as shown in Fig.~\ref{fig:d2Id2V_1.07}(b), and mark the onset of inelastic scattering off two different types of particle-hole excitations.

The $dI/dV$-step at larger bias pointed at by the arrow is associated with the emission of plasmons. Figure~\ref{fig:d2Id2V_1.07}(a) show that the bias, $eV^*(\theta)$, of the $d^2I/dV^2$ peak traces out the momentum-resolved energy dispersion of the plasmon, cf.\ Fig.~\ref{fig:plasmon_1.07}(a). This is again qualitatively consistent with Eq.~\eqref{eq:didv_phonon} for plasmon-assisted inelastic tunneling into the dispersive bands. 

The rapid increase of $dI/dV$ at biases close to the edge of the remote bands (edge is marked by dashed line in Fig.~\ref{fig:d2Id2V_1.07}(b)) can be attributed to inelastic tunneling into the flat bands when the tip electrons have sufficient energy to scatter a quasiparticle 
from the flat/remote valence band to the remote/flat conduction band. Due to the narrow bandwidth 
of the flat bands,
these interband scattering can lead to stronger $d^2I/dV^2$ peaks than the plasmon-induced signals, as shown in panel(a).
Both types of inelastic tunneling involve virtual electronic states primarily in the flat bands. Therefore, as $\theta$ decreases and the tip Dirac point is twisted towards the bottom-layer Dirac point $(\bm\kappa)$ of the sample, the intensities of the $d^2I/dV^2$ peaks drop rapidly with the tunneling matrix elements of the flat bands \cite{wei2025dirac}.

Similar phenomena are found in the negative-bias regime in Fig.~\ref{fig:d2Id2V_1.07}(c) and (d). When the Fermi level of the tip is below
the edge of the lower remote bands, electrons in the remote bands can tunnel into the tip by scattering off plasmons; cf.\ Fig.~\ref{fig:dispersive_band}(c) and (d), leading to a dip of $d^2I/dV^2$ at bias $V^{*}(\theta)$ which traces out the momentum-resolved energy dispersion of plasmons.

\section{Discussion}\label{sec:discussion}

In this work, we develop a theory for the inelastic momentum-conserving 
tunneling spectroscopy with the quantum twisting microscope (QTM), and propose methods to access 
the dispersion of a bosonic collective mode, the plasmon.
We specifically study the lowest band of moir\'e plasmons in twisted bilayer graphene (TBG) with a twist angle close to its magic value. By calculating the spectral function of the interacting Bistritzer-MacDonald model of TBG within the $GW$ approximation, we predict two types of signatures associated with plasmons  in the electron tunneling spectrum. 

The first signature consists of nearly dispersionless plasmon satellites in $dI/dV$ at low bias. These satellite peaks arise from inelastic tunneling 
into the central bands of TBG. Peaks in  $dI/dV$ originate from the prominent maximum in the plasmon density of states, cf.~Fig.\ \ref{fig:A_1.07}(b), coming from the Van Hove singularities at the Brillouin zone boundaries \cite{stauber2016quasi,lewandowski_pairing_2021}.

The other signature of plasmons is a peak in $d^2I(V,\theta)/dV^2$ at high bias due to the onset of inelastic tunneling into the dispersive bands. It shows much lower intensities, but the dependence of the  
peak bias $V$ on the twist angle $\theta$ between the tip and sample can trace out the momentum-resolved plasmon dispersion. 

Unlike previous theories of the inelastic tunneling spectroscopy with the QTM \cite{peri_probing_2024,pichler_probing_2024,xiao_theory_2024}, which assumed finite Fermi surfaces 
on both sides of the tunnel junction, we focus on charge neutral TBG.
In our case, the inelastic tunneling features are induced by singularities in the density of states of the sample.
The advantage of using singular points in reciprocal space is that it enables a sharp selection of electron momenta and removes the Fermi-surface averaging of the transferred momenta. Such averaging is inconsequential for studying the phonon spectra at wave vectors comparable to the reciprocal lattice period of {\it monolayer} graphene, but would be detrimental in studying the plasmon spectrum {\it varying within the moir\'e Brillouin zone} (mBZ).

Doping of TBG would also generate a Hartree potential that increases the bandwidth of the flat bands \cite{guinea_electrostatic_2018}. This effect broadens the particle-hole continuum and enhances Landau damping of the plasmons near the mBZ boundary \cite{cea_coulomb_2021}. On the other hand, the Fermi surface in doped TBG ensures conventional plasmons with $\sqrt{q}$-dispersion in the long wavelength limit (for an unscreened Coulomb potential) \cite{cavicchi_theory_2024,novelli2020optical}. These significant changes in the band structure and the characteristics of the plasmons might lead to a  nontrivial evolution of the QTM tunneling spectra with varying doping levels, which we leave for future study.   

We comment on the domain of validity and limitations of our theory. Firstly, the random phase approximation of the screened Coulomb potential is formally exact in the large flavor limit, $N_f\rightarrow \infty, v_{\bm q}\rightarrow 0$ with $N_f v_{\bm q}$ fixed. In this limit, the system can host well-defined plasmons outside the particle-hole continuum while being weakly interacting. In TBG, $N_f=4$ is given, and our calculations suggest that there exist undamped plasmons near the mBZ boundary even when the Coulomb interaction is suppressed by metallic gates as close to the sample as $1$\si{nm}. Based on the RPA, the $GW$ approximation perturbatively accounts for dynamical screening of the long-range Coulomb interactions. It was previously applied to investigate the thermodynamic properties, e.g., the filling dependence of the ground-state energy and chemical potential \cite{zhu2024weak,peng2025many} of MATBG, but has not been used for the spectral properties of TBG until now. There have been alternative approaches to study the spectral function of TBG in the strong coupling regime, such as quantum Monte Carlo \cite{hofmann2022fermionic,huang2025angle}, DMFT \cite{datta_heavy_2023,rai_dynamical_2024,zhou2024kondo,calderón2025cascades,georges_dynamical_1996}, and approximate analytical solutions \cite{ledwith_nonlocal_2025,hu_projected_2025,ledwith2025exotic}. These nonperturbative methods consistently predict that the flat bands of MATBG split into Hubbard bands well above the ordering temperature of the spin and pseudospin and that the Hubbard bands become dispersive near the $\bm \gamma$ point due to the nontrivial topology of the TBG flat bands \cite{ledwith_nonlocal_2025,song_magic-angle_2022}. The $GW$ approximation fails to describe these Mott states and Hubbard bands. Nevertheless, the Mott states with finite-width Hubbard bands 
could, in principle, support plasmon excitations \cite{vanloon2014plasmons}. Inelastic tunneling of the electrons (holes) into the upper (lower) Hubbard band by emitting a plasmon would give rise to 
satellites in the tunneling spectrum at biases above the elastic tunneling peaks that correspond to the Hubbard bands. 


Finally, the formalism in Sec.~\ref{sec:formalism} can be applied to other collective modes. One natural extension of this work would be to incorporate the hybridization between the plasmons and phonons in the calculations \cite{cea_coulomb_2021}.
Another interesting direction for future work is to investigate signatures of the Goldstone modes in the tunneling spectra of various symmetry-broken phases in MATBG \cite{wu2020collective,vafek2020renormalization,kumar2021lattice,bernevig2021twisted_excitations,khalaf2020soft,bultinck2020ground,xie2020nature}. More broadly, our study establishes a theoretical framework for probing collective electronic excitations in correlated moir\'e materials with the QTM.

\acknowledgements{
We thank Elena Bascones, Andrei Bernevig, John Birkbeck, Dumitru Călugăru, Shahal Ilani, Patrick Ledwith, and Jiewen Xiao for insightful discussions and comments. Calculations were conducted at the Yale Center for Research Computing. N.W.\ acknowledges support through the Yale Prize Postdoctoral Fellowship in Condensed Matter Theory. Research at Yale was supported by NSF Grant No.\ DMR-2410182 and by the Office of Naval Research (ONR) under Award No.\ N00014-22-1-2764. Research at Freie Universit\"at Berlin and Yale was supported by Deutsche Forschungsgemeinschaft through CRC 183 (project C02 and a Mercator Fellowship).   Research at Freie Universit\"{a}t Berlin was further supported by Deutsche Forschungsgemeinschaft through a joint ANR-DFG project (TWISTGRAPH). F. G. acknowledges support from NOVMOMAT, project PID2022-142162NB-I00 funded by MICIU/AEI/10.13039/501100011033 and by FEDER, UE as well as financial support through the (MAD2D-CM)-MRR MATERIALES AVANZADOS-IMDEA-NC. This research was supported in part by grant NSF PHY-2309135 to the Kavli Institute for Theoretical Physics (KITP).
}


\newpage
\appendix
\onecolumngrid

\section{Derivation of the inelastic contribution to differential conductance using Fermi's golden rule}\label{sec:fermi}

In this section, we show that when the sample can be described by an interacting electron-boson model (e.g., Eq.~\eqref{eq:H_bf}), the inelastic tunneling current derived from Fermi's golden rule is consistent with Eq.~\eqref{eq:i} in the main text, to the leading order in the electron-boson coupling strength.

Consider the inelastic tunneling of electrons (or holes) from the sample to the tip. This process involves the emission or absorption of a plasmon in the sample, followed by momentum-conserving interlayer tunneling. The $\mathcal{T}-$matrix elements from an initial state of a single electron $|\bm p-\bm q\lambda\text{S}\rangle$ and a plasmon $|\bm q\rangle$ at wave vector $\bm q$ in the sample to a final state $|\bm p\lambda'\text{T}\rangle$ in the tip,
\begin{align}\label{eq:T_in_layer}
    \langle \bm p\lambda' \text{T} |\mathcal{T}|\bm p-\bm q \lambda\text{S}; \bm q\rangle &= \langle\bm p \lambda'\text{T}|H_{\text{tun}}\frac{1}{\epsilon\mathcal{I}-H_0}H_{\text{e-pl}}|\bm p-\bm q \lambda\text{S}; \bm q\rangle \notag\\
    &=\sum_{{\rho}}\frac{T_{\lambda'{\rho}}(\bm p) g_{{\rho}\lambda}(\bm p,\bm q)}{eV+ \xi_{\bm p}^{\text{T},\lambda'}-\xi_{\bm p}^{\rho}} 
\end{align}
%
%
In the energy denominator, $\epsilon = eV+\xi_{\bm p}^{\text{T},\lambda'}$ denotes the energy of the tunneling electron relative to the Fermi level of the sample. We also defined $\mathcal{I} \equiv \sum_{\bm k\lambda}c_{\bm k\lambda}^{\dagger}c_{\bm k\lambda}$.

Likewise, we can write down the $\mathcal{T}-$matrix elements from an initial state of a single hole $|\overline{\bm p-\bm q\lambda\text{S}}\rangle$ and a plasmon $|-\bm q\rangle$ in the sample to a final hole state $|\overline{\bm p\lambda'\text{T}}\rangle$ in the tip,
\begin{align}\label{eq:T_in_layer_hole}
    \langle \overline{\bm p\lambda' \text{T}} |\mathcal{T}| \overline{\bm p-\bm q \lambda\text{S}}; -\bm q\rangle &= \langle \overline{\bm p \lambda'\text{T}}|H_{\text{tun}}\frac{1}{\epsilon\mathcal{I}-H_0}H_{\text{e-pl}}|\overline{\bm p-\bm q \lambda\text{S}}; -\bm q\rangle \notag\\
    &=\sum_{{\rho}}\frac{T_{\lambda'{\rho}}^*(\bm p) g_{{\rho}\lambda}(\bm p-\bm q,-\bm q)}{eV+ \xi_{\bm p}^{\text{T},\lambda'}-\xi_{\bm p}^{\rho}} \notag\\
    &= \langle \bm p\lambda' \text{T} |\mathcal{T}|\bm p-\bm q \lambda\text{S}; \bm q\rangle^{*}.
\end{align}
In the last line, we used time-reversal ($\mathcal{T}$) symmetry,
\begin{equation}\label{eq:TRS}
    \phi_{\bm q+\bm g}=\phi_{-\bm q-\bm g}^{*},\quad g_{\lambda\rho}^*(\bm k-\bm q,-\bm q)=g_{\lambda\rho}(\bm k,\bm q). 
\end{equation}
Using the Fermi's golden rule, it is straightforward to write down the tunneling current at finite temperature
\begin{align}\label{eq:I_inelastic}
    \delta I \approx \frac{2\pi N_bN_f e}{\hbar\Omega} &\sum_{\bm q\in \text{mBZ}}\sum_{\bm p,\lambda'}\sum_{\lambda}\Big|\langle \bm p\lambda' \text{T} |\mathcal{T}_{\text{inel}}|\bm p-\bm q \lambda\text{S}; \bm q\rangle\Big|^2 \delta(eV + \xi_{\bm p}^{\text{T},\lambda'}-\xi_{\bm p-\bm q}^{\lambda}-\hbar\omega_{\bm q}) \notag\\
    &\times\left\{[1-n_{F}(\xi_{\bm p }^{\text{T},\lambda'})]n_{F}(\xi_{\bm p-\bm q}^{\lambda})n_{B}(\hbar\omega_{\bm q}) - n_{F}(\xi_{\bm p }^{\text{T},\lambda'})[1-n_{F}(\xi_{\bm p-\bm q}^{\lambda})][1+n_{B}(\hbar\omega_{\bm q})]\right\} \notag\\
    & - (\omega_{\bm q}\rightarrow -\omega_{\bm q}).
\end{align}
The Dirac-delta enforces the energy conservation. In the last line, we used $\omega_{\bm q}=\omega_{-\bm q}$. 
%
%

Alternatively, we can derive the inelastic contribution to the tunneling current from the linear response theory, see Eq.~\eqref{eq:i}. We write down the dissipative part of the screening potential mediated by the plasmon, according to Eqs.~\eqref{eq:W_im_band},~\eqref{eq:W_hermiticity}, and~\eqref{eq:W_plasmonpole},
\begin{equation}
    W_{\bm k\lambda_1\rho,\bm k\lambda_2\rho}^{\prime\prime}(\bm q,\nu) = -\pi g_{\lambda_1\rho}(\bm k,\bm q)g_{\lambda_2\rho}^*(\bm k, \bm q) \delta(\nu-\hbar\omega_{\bm q}) + \pi g_{\lambda_1\rho}^*(\bm k-\bm q,-\bm q)g_{\lambda_2\rho}(\bm k-\bm q,-\bm q)\delta(\nu +\hbar\omega_{-\bm q}).
\end{equation}
By simplifying the above expression using the $\mathcal{T}$ symmetry, see Eq.~\eqref{eq:TRS}, we obtain the following self-energy from Eq.~\eqref{eq:sigma_im},
\begin{align}
    \Sigma_{\lambda_1\lambda_2}^{\prime\prime}(\bm k,\epsilon) = \sum_{\bm q\in \text{mBZ}} \sum_{\rho}&g_{\lambda_1\rho}(\bm k,\bm q)g_{\lambda_2\rho}^*(\bm k, \bm q) \left(\delta(\epsilon-\xi_{\bm k-\bm q}^{\rho}-\hbar\omega_{\bm q}) - \delta(\epsilon-\xi_{\bm k-\bm q}^{\rho} +\hbar\omega_{\bm q})\right)\notag\\
    &\times [n_{B}(\epsilon-\xi_{\bm k-\bm q}^{\rho}) + n_{F}(-\xi_{\bm k-\bm q}^{\rho})].
\end{align}
Inserting the self-energy into Eq.~\eqref{eq:A_inelastic} yields the spectral function $\hat{A}^{\text{S}}(\bm k,\epsilon)$ at energy $\epsilon$ away from the quasiparticle peaks. We then plug $\hat{A}^{\text{S}}$ and $A_{\lambda'}^{\text{T}}(\bm k,\epsilon)=\delta(\epsilon-\xi_{\bm k'}^{\text{T},\lambda'})$ into Eq.~\eqref{eq:i}, and simplify the results by using the following identity,
\begin{align}
     &[n_{F}(\epsilon) - n_{F}(\xi^{\text{T}})][n_{B}(\epsilon-\xi) + n_{F}(-\xi)] \notag\\
    = &[1-n_{F}(\xi^{\text{T}})]n_{F}(\xi)n_{B}(\epsilon-\xi) \notag\\
    &- n_{F}(\xi^{\text{T}})[1-n_{F}(\xi)][1+n_{B}(\epsilon-\xi)].
\end{align}
After some algebra, one can reproduce Eq.~\eqref{eq:I_inelastic}.

\section{Effects of interlayer Coulomb interactions between the tip and sample on inelastic tunneling}\label{sec:tip_screening}
In this work, we focus on the charge fluctuations in the sample to illustrate the inelastic tunneling spectroscopy for collective excitations with QTM. However, in contrast to other types of collective excitations, the charge fluctuations are coupled with electrons not only in the sample but also in the tip via the long-range Coulomb interactions. 
Therefore, when applied to charge fluctuations, the formalism in Sec.~\ref{sec:formalism} can be improved by including two missing factors: (1) the dynamical screening effects of the tip on the plasmon excitations in TBG; (2) inelastic scattering between the electrons and plasmon on different sides of the tunnel junction. We will analyze these two effects separately in the following two subsections, and argue that they do not lead to qualitative changes on the tunneling spectrum of the MLG-TBG tunnel junctions studied in the main text.

\subsection{Dynamical screening from the tip}

For simplicity, we neglect the finite thickness of TBG by assuming that its charges reside in the center of two graphene layers due to strong layer hybridization of the flat-band wave functions. These charges are separated from the monolayer graphene tip by a vertical distance $d_{\text{T}}$. It will be convenient to introduce a layer index $l=0,1$ to label the TBG and MLG, respectively. Electrons in one layer can interact with electrons in the same or different layers. The bare Coulomb interaction vertex is diagonal in the layer subspace but the intra-and inter-layer interactions can have different strength. We denote the interaction between the layer $l$ and $l'$ as $v_{\bm q}^{l,l'}$. Provided that the layer separation is much smaller than the distance from the graphene layers to gates, $d_{\text{T}}\ll d_g$,
\begin{equation}\label{eq:v_interlayer}
    v_{\bm q}^{l, l'} \approx v_{\bm q}e^{-|l-l'||\bm q|d_{\text{T}}}.
\end{equation}
Let us first write down the non-interacting response function of the full system, which is a diagonal matrix in the layer subspace:
\begin{align}
    \hat{\chi} = \begin{pmatrix}
        \hat{\chi}^{\text{S}} & 0\\
        0& \hat{\chi}^{\text{T}}
    \end{pmatrix},
\end{align}
where $\hat{\chi}^{\text{T/S}}$ represent the non-interacting density-density response functions of the tip and sample, respectively. The RPA density-density response function is also a $2\times 2$ matrix in the layer space,
\begin{equation}\label{eq:W_qtm}
    \hat{W}(\bm q,\omega) = 
    \begin{pmatrix}
        \hat{W}^{0,0} & \hat{W}^{0,1}\\
        \hat{W}^{1,0} & \hat{W}^{1,1}\\
    \end{pmatrix}_{\bm q,\omega}  = 
    \begin{pmatrix}
        \hat{1} - \hat{V}^{0,0}\hat{\chi}^{\text{S}} & - \hat{V}^{0,1}\hat{\chi}^{\text{T}}\\
         - \hat{V}^{1,0}\hat{\chi}^{\text{S}} & \hat{1} - \hat{V}^{1,1}\hat{\chi}^{\text{T}}\\
    \end{pmatrix}_{\bm q,\omega}^{-1}
    \begin{pmatrix}
        \hat{V}^{0,0} & \hat{V}^{0,1}\\
        \hat{V}^{1,0} & \hat{V}^{1,1}\\
    \end{pmatrix}_{\bm q} -\begin{pmatrix}
        \hat{V}^{0,0} & \hat{V}^{0,1}\\
        \hat{V}^{1,0} & \hat{V}^{1,1}\\
    \end{pmatrix}_{\bm q},
\end{equation}
with the bare interaction matrix $V_{\bm g, \bm g'}^{l,l'}(\bm q) = v_{\bm q + \bm g}^{l, l'}\delta_{\bm g, \bm g'}$. Note that in our convention, we have subtracted the first-order Coulomb interactions from the screening potential.

Combining Eqs.~\eqref{eq:v_interlayer}-\eqref{eq:W_qtm} and using the following formula for bockwise inversion,
\begin{equation}
   \begin{pmatrix}
       A & B\\
       C& D\\
   \end{pmatrix}^{-1} =     
   \begin{pmatrix}
       (A - BD^{-1}C)^{-1} & 0\\
       0 & (D-CA^{-1}B)^{-1}\\
   \end{pmatrix}
    \begin{pmatrix}
       I & -BD^{-1}\\
       -CA^{-1} & I\\
   \end{pmatrix},
\end{equation}
we find that the expression for the screened interactions between two electrons in the sample is analogous to Eq.~\eqref{eq:W}, 
\begin{align}\label{eq:W_00}
    \hat{W}^{0,0}(\bm q,\omega) &= \hat{\epsilon}^{\text{eff}}(\bm q,\omega)^{-1}\hat{V}^{\text{eff}}(\bm q,\omega) - \hat{V}^{0,0}(\bm q),\\
    \hat{\epsilon}^{\text{eff}}(\bm q,\omega) &= \hat{1} - \hat{V}^{\text{eff}}(\bm q,\omega)\hat{\chi}^{\text{S}}(\bm q,\omega),
\end{align}
with the bare Coulomb interaction $\hat{V}(\bm q)$ in Eq.~\eqref{eq:W} replaced by a frequency-dependent effective interaction that takes into account the tip screening,
\begin{align}
    \hat{V}^{\text{eff}}(\bm q,\omega) &= \hat{V}^{0,0}(\bm q) + \hat{V}^{0,1}(\bm q) \left(\hat{\chi}^{\text{T}}(\bm q,\omega)^{-1}-\hat{V}^{1,1}(\bm q)\right)^{-1}\hat{V}^{1,0}(\bm q) \\
    &\approx \left(\hat{1}-\hat{V}(\bm q)\hat{\chi}^{\text{T}}(\bm q,\omega)\right)^{-1}\hat{V}(\bm q). \label{eq:V_d=0}
\end{align}
The last equation assumes the tip-sample distance $d_{\text{T}}\ll q^{-1}$, and $\hat{V}^{l,l'}\approx \hat{V}$.

For a monolayer graphene tip, we may neglect the band structure renormalization induced by the periodic Hartree-potential generated by the TBG and approximate $\hat{\chi}^{\text{T}}$ by its value in a decoupled MLG \cite{wunsch2006dynamical} at zero temperature, $\chi_{\bm g_1,\bm g_2}^{\text{T}}(\bm q,\omega)\approx \chi^{\text{G}}(|\bm q+\bm g|,\omega)\delta_{\bm g_1,\bm g_2}$, and
\begin{align}\label{eq:chi_g}
    \chi^{\text{G}}(q,\omega) = -\frac{N_f \mu_{\text{T}}}{2 \pi \hbar^2 v_{D}^2}+\frac{F(q, \omega)}{\hbar^2 v_{D}^2}\left[G\left(\frac{2 \mu_{\text{T}}+ \omega}{\hbar v_{D} q}\right)-G\left(\frac{2 \mu_{\text{T}}-\omega}{\hbar v_{D} q}\right)\right],
\end{align}
with
\begin{align}
    F(q, \omega)=\frac{N_f}{16 \pi} \frac{\hbar^2 v_{D}^2 q^2}{\sqrt{\omega^2-\hbar^2v_{D}^2 q^2}}, \quad G(x)=x \sqrt{x^2-1}-\ln \left(x+\sqrt{x^2-1}\right).
\end{align}
Here, $\sqrt{x^2-1}$ has a branch cut on the interval $(-1,1)$ and the frequency $\omega$ approaches the real axis from above. For charge neutral MLG with four spin-valley flavors ($N_f=4$), the chemical potential $\mu_{\text{T}}=0$, and Eq.~\eqref{eq:chi_g} reduces to $\chi^{\text{G}}(q, \omega) = -i q^2/4\sqrt{\omega^2-\hbar^2v_D^2q^2}$. For $q\gg \omega/\hbar v_D$, 
\begin{align}
    &\chi^{\text{G}}(q,\omega)\approx -\frac{q}{4\hbar v_D}, \\
    &\epsilon^{\text{G}}(q,\omega) = 1- v_{q} \chi^{\text{G}}(q,\omega) \approx 1 + \frac{\pi}{2}\frac{\alpha_{\text{G}}}{\epsilon_r},\label{eq:epsilon_g_largeq}
\end{align}
with the fine structure constant of MLG, $\alpha_{\text{G}} = e^2/4\pi\epsilon_0\hbar v_D\approx 2.2$. These two equations hold for finite $\mu_{\text{T}}=\hbar v_D k_F$ as long as $q\gg 2k_{F}$. 

Equation~\eqref{eq:epsilon_g_largeq} indicates that MLG screens the short-wavelength charge potential perturbation as if it were a conventional dielectric, with a relative permittivity of $(1 + \pi \alpha_{\text{G}}/2)$. By plugging Eq.~\eqref{eq:epsilon_g_largeq} into Eq.~\eqref{eq:V_d=0}, we find that the tip renormalize relative permittivity of the dielectric environment,
\begin{equation}
    \epsilon_r \rightarrow \epsilon_r+ \pi\alpha_\text{G}/2.
\end{equation}

\begin{figure}
    \centering
    \includegraphics[width=1\linewidth]{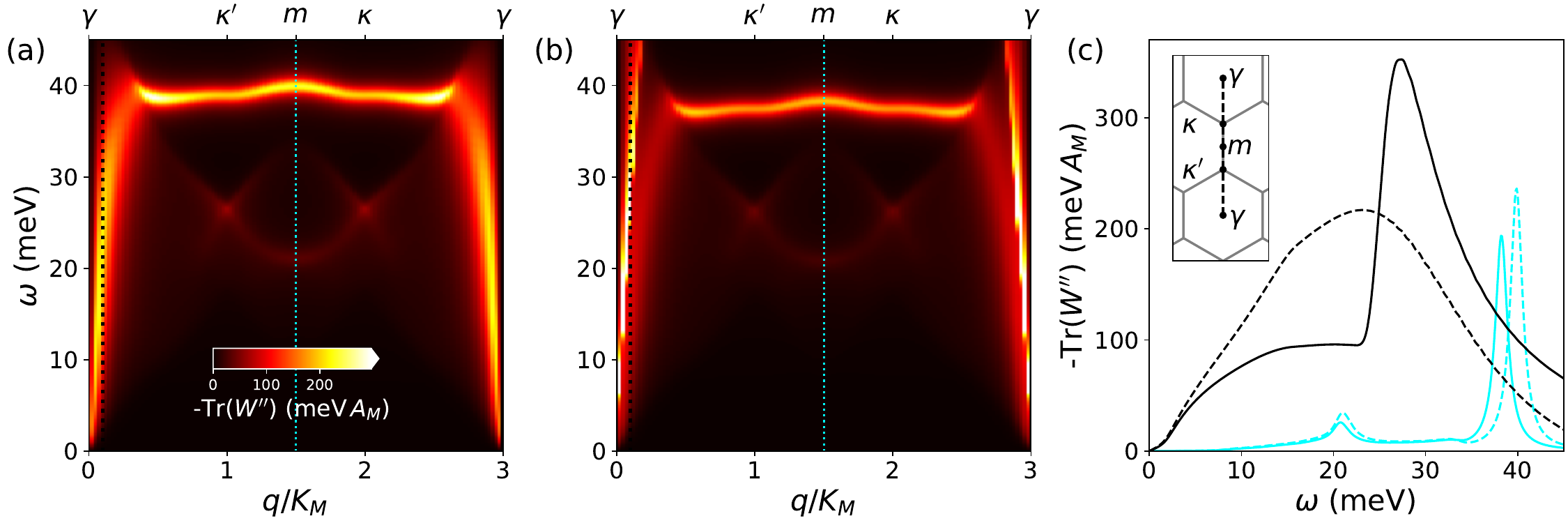}
    \caption{Comparison between the dissipative part of the in-layer RPA screening potential $-\text{Tr}(W^{0,0}(q\hat{y},\omega)^{\prime\prime})$ of TBG without (a) and with (b) dynamical screening from the tip. We consider a charge neutral $1.20^{\circ}-$TBG and an electron-doped MLG tip with chemical potential $\mu_{\text{T}}=20$ \si{meV}, separated vertically by a distance $d_{\text{T}}=5$\si{\angstrom}.  (a) The nearly dispersionless peaks at energy $\omega\approx 40$ \si{meV} arise from the plasmons in the sample. At small $q$, the plasmon is damped by the interband particle-hole transitions and leads to a broad peak of the screening potential. (b) In addition to the features in (a), there are fast dispersing peaks on the left and right of the map induced by the plasmon of the MLG tip. (c) The dashed lines are line cuts of (a) at $q=0.1k_M$(black) and $1.5k_M$(blue), while the solid lines are line cuts of (b) at the same wave vectors. The differences between the two blue lines indicate that the tip screening reduces the plasmon energy and the interaction mediated by the plasmons in the sample. The screening effect of tip becomes more significant at small $q$, as shown by the black lines. The tip also gives rise to a sharp peak in the black solid line, which can be understood as the interaction mediated by the plasmons in the tip. The inset depicts a high symmetry line used in (a) and (b). For these plots, we take the relative permittivity $\epsilon_r=6.6$ and broadening parameter $\eta=0.8$ \si{meV} for both the tip and sample. Other parameters are given in Appendix~\ref{sec:numeric_details}.}
    \label{fig:plasmon_0_qtm}
\end{figure}

Because the plasmon energy in the flat bands of TBG is much smaller than $\hbar v_D k_M \sim 200$ \si{meV}, the plasmons near the zone boundary readily satisfy the condition $q\gg \omega_{\bm q}/\hbar v_D, k_{F}$. As a result, these modes are largely insensitive to screening effects from the tip. We have verified this conclusion by numerically computing the in-layer screening potential, Eq.~\eqref{eq:W_00}, with a tip chemical potential $\mu_{\text{T}}=20$ \si{meV} (Fermi wave vector $k_{F} \approx 0.1k_M$) and relative permittivity $\epsilon_r=6.6$, see Fig.~\ref{fig:plasmon_0_qtm}(a). Compared to the case without the tip screening in panel (b), we see that away from the $\bm\gamma$ point the plasmon peaks are qualitatively unchanged. Screening from the tip becomes more pronounced at smaller $q$, as illustrated by the difference between the dashed and solid black lines in panel (c), which represent line cuts of panels (a) and (b) at $q = 0.1k_M$, respectively. Additionally, in the regime $q \lesssim 2k_F$, the plasmon mode in the doped MLG tip mediates electron-electron interactions in TBG via the interlayer Coulomb interaction, giving rise to the plasmon peak in the solid black line. However, this mode contributes only a weak and smooth background to the differential conductance due to its low density of states when $k_F \ll k_M$. This background can be further suppressed experimentally by minimizing tip doping and vanish completely in the $\mu_{\text{T}} = 0$ limit. Consequently, we neglect the plasmon mode of the MLG tip in the main text.
%
%


\subsection{Interlayer electron-plasmon scattering}
\begin{figure}
    \centering
    \includegraphics[width=1\linewidth]{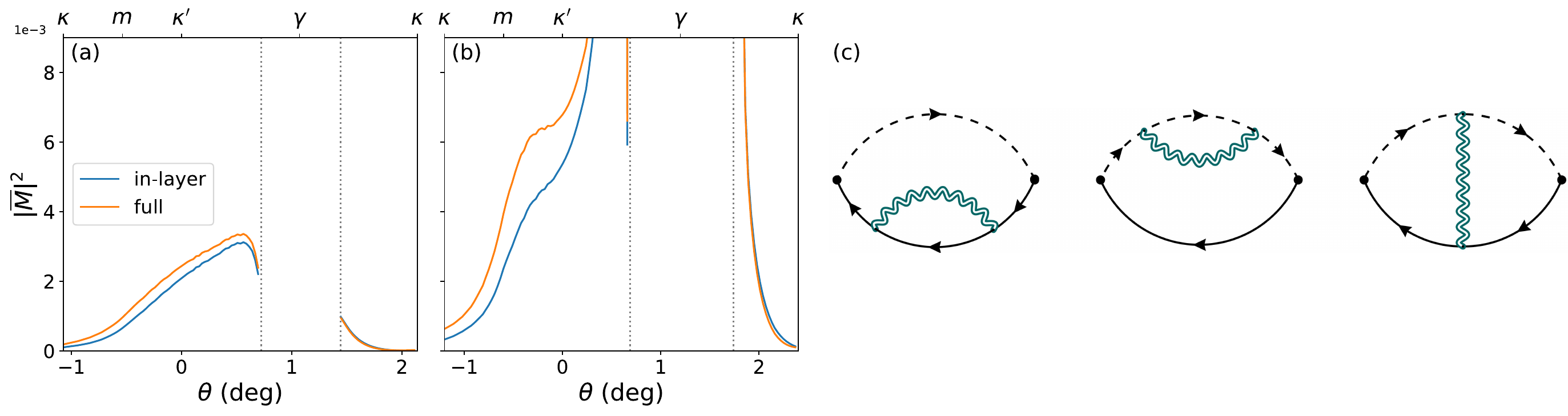}
    \caption{The absolute square of the normalized rate of inelastic tunneling between the tip Fermi line and the $\bm\gamma$ point of the lowest remote conduction bands, $|\overline{M}(\bm K_{\theta}, \bm Q_{\theta})|^2$, as a function of the twist angle $\theta$ between the tip and sample. The sample is (a) a $1.07^{\circ}-$TBG at charge neutrality with the sample-gate distance $d_g=1$\si{nm}; (b) $1.20^{\circ}-$TBG at charge neutrality with $d_g=40$\si{nm}. The orange lines take into account both in-layer and interlayer electron-plasmon interactions, while the blue lines are computed from Eq.~\eqref{eq:M_multiband} in the main text, which include only the electron-plasmon scattering in the TBG. The twist-angle regime between the two dashed lines is excluded. In this regime, the tip Dirac point $\bm K_{\theta}$ is twisted close to the $\bm \gamma$ point, and the small-$\bm Q_{\theta}$ plasmons required for inelastic tunneling are damped by the particle-hole continuum. For these plots, we use the tip-sample distance $d_{\text{T}}= 0.5$\si{nm}. (c) Examples of Feynman diagrams that contribute to the tunneling current. The dashed and solid lines represent the non-interacting electron propagtors in the tip and sample, respectively. The curly line corresponds to the screening potential $\hat{W}$. The left and right vertices stand for the bare tunneling Hamiltonian, $\hat{H}_{\text{tun}}$. The last diagram exemplifies the vertex correction due to the interlayer Coulomb interactions, which is not included in Eq.~\eqref{eq:i}, and accounts for the main differences between the orange and blue lines in both panel (a) and panel (b).}
    \label{fig:T}
\end{figure}

The plasmons in the sample generate an electric field that extends out of the sample plane $(z=0)$ and interact with the electrons in the tip. In this section, we compute the corrections to the inelastic tunneling conductance due to the interlayer electron-plasmon coupling. 
The full electron-plasmon coupling Hamiltonian can be expressed in the real space as follows:
\begin{equation}\label{eq:H_e_pl_full}
    H_{\text{e-pl}} = \int d^2\bm r\  \hat{\phi}(\bm r;z=0)\rho(\bm r) + \hat{\phi}(\bm r; z=d_{\text{T}})\rho_{\text{T}}(\bm r),
\end{equation}
where $\rho_{\text{T}}(\bm r) = \sum_{\bm p,\bm q} \sum_{s\sigma}d_{\bm p s\sigma}^{\dagger}d_{\bm p+\bm q s\sigma} e^{i\bm q\cdot \bm r}$ is the density operator on the tip. We assume that the sample-gate distance $d_g$ is much larger than the sample-to-tip distance $d_{\text{T}}$ ($d_g\gg d_{\text{T}}$). The electric potential $\hat{\phi}(\bm r;z)$ generated by the plasmon reads
\begin{equation}
    \hat{\phi}(\bm r;z) = \sum_{\bm q\in\text{mBZ}}e^{i\bm q\cdot\bm r}b_{\bm q}\left(\sum_{\bm g}\phi_{\bm q+\bm g}e^{-|\bm q+\bm g||z|+i\bm g\cdot\bm r} \right) + \text{H.c.}, \quad |z|\ll d_g. \label{eq:phi_r}
\end{equation}
%
%
Note that the first term in Eq.~\eqref{eq:H_e_pl_full} is equivalent to Eq.~\eqref{eq:H_e_pl} in the main text.

For the electron Hamiltonian, the non-interacting Bistritzer-MacDonald model of the TBG sample is denoted as $H_0$ (see Eq.~\eqref{eq:h0}), while the tip Dirac Hamiltonian is given by (see Eq.~\eqref{eq:h}) 
\begin{equation}
    H_{\text{T}} = \sum_{\bm k'}\sum_{s} d_{\bm k' s}^{\dagger}h_{\bm k'}(\theta +\theta_{\text{TBG}}/2)d_{\bm k' s},
\end{equation}
where $\theta$ denotes the twist angle between the tip and the top layer of the TBG. The electrons in the tip have the band dispersion $\xi_{\bm k'}^{\text{T},\lambda'=\pm} = \lambda'\hbar v_D|\bm k'-\bm K_{\theta}|-\mu_{\text{T}}$.

We compute the $\mathcal{T}-$matrix of the inelastic tunneling from the sample to the tip by absorbing a plasmon. The electrons in the sample can either abosrb a plasmon before tunneling into the tip or tunnel into the tip first and then absorb a plasmon. The $\mathcal{T}-$matrix should therefore include a coherent superposition of these two processes:

\begin{align}
    \langle \bm k'\lambda'\text{T}|\mathcal{T}_{\text{in}}| \bm k'-\bm q\lambda\text{S}; \bm q\rangle &= \langle \bm k'\lambda'\text{T}; \bm q| H_{\text{tun}}\frac{1}{\epsilon\mathcal{I}-H_0}H_{\text{e-pl}} + H_{\text{e-pl}}\frac{1}{(\epsilon + \mu_{\text{T}} - eV )\mathcal{I} - H_{\text{T}}}H_{\text{tun}} |\bm k'-\bm q\lambda\text{S}; \bm q\rangle \notag\\
    & = \sum_{\rho} \frac{T_{\lambda'\rho}(\bm k')g_{\rho\lambda}(\bm k', \bm q)}{eV+\xi_{\bm k'}^{\text{T},\lambda'}-\xi_{\bm k'}^{\rho}} + \sum_{\rho^{\prime}}\sum_{\bm g} \frac{g_{\lambda'\rho'}^{\text{inter}}(\bm k',\bm q+\bm g)T_{\rho'\lambda}\left(\bm k^{\prime}-\bm q-\bm g\right)}{\xi_{\bm k'}^{\text{T},\lambda'} - \xi_{\bm k^{\prime}-\bm q- \bm g}^{\text{T},\rho'}}.  \label{eq:T}
\end{align}
Here, $\epsilon=eV+ \xi_{\bm k'}^{\text{T},\lambda'}$ denotes the energy of the tunneling electron relative to the Fermi level of the sample. We defined the interlayer electron-boson coupling strength
\begin{equation}
    g_{\lambda'\rho'}^{\text{inter}}(\bm k',\bm p) = \langle \bm k'\lambda'\text{T}| \phi_{\bm p}e^{-|\bm p|d_{\text{T}}}e^{i\bm p\cdot\bm r}|\bm k'-\bm p \rho'\text{T}\rangle. 
\end{equation}
As shown in Sec.~\ref{sec:formalism}, the zero-temperature $dI/dV$ is determined by the tunneling electrons on the Fermi level of the tip, $\xi_{\bm k'}^{\text{T},\lambda'}=0$. As we assume the Fermi wave vector $k_F$ of the tip to be negligible throughout this work, we can make the approximation $\bm k'\approx K_{\theta}$ in the energy denominator of Eq.~\eqref{eq:T}. The steep energy dispersion of the MLG tip then ensures that for most of $\bm q$ in the mBZ, $|\xi_{\bm K_{\theta}-\bm q}^{\text{T},\lambda'}|= \hbar v_D|\bm q| \gtrsim \text{min}_{\rho}\{|eV-\xi_{\bm K_{\theta}}^{\rho}|\}$. It suppresses the second term in Eq.~\eqref{eq:T}, thereby reducing the effects of the interlayer electron-plasmon coupling. 

This is particularly the case for the dispersionless plasmon satellites observed at $|eV|\sim \hbar\omega_{\bm q}\ll \hbar v_D |\bm q|$. The effects of the interlayer electron-plasmon coupling on the tunneling conductance can be neglected at such a small bias. 

For inelastic tunneling into remote bands at relatively high bias, let us consider that electrons on the Fermi level of the tip are scattered to the $\bm \gamma$ point by a plasmon with wave vector $\bm q = \bm K_{\theta} - \bm K_{\theta_{\text{TBG}}}\equiv \bm Q_{\theta}$, where $\bm K_{\theta}$ is the tip Dirac point, which coincides with the $\bm \gamma$ point at the twist angle $\theta=\theta_{\text{TBG}}$. We average the absolute square of the $\mathcal{T}-$matrix elements in Eq.~\eqref{eq:T} over the tip Fermi line. The results normalized by the tunneling strength squared, $w^2$, read
\begin{align}
    |\overline{M}_{\lambda}(\bm K_{\theta},\bm Q_{\theta})|^2 &\equiv \frac{1}{w^2\nu_{\text{T}}} \int \frac{d^2 k^{\prime}}{(2 \pi)^2} \left|\langle \bm k'\lambda'\text{T}|\mathcal{T}_{\text{in}} | \bm k'-\bm Q_{\theta}\lambda\text{S}; \bm Q_{\theta}\rangle \right|^2\delta(\xi_{\bm k^{\prime}}^{\text{T}, \lambda^{\prime}}) \notag\\
    & \approx \frac{1}{2}\sum_{\sigma=\pm} \left|\sum_{\rho}\frac{\Lambda_{\rho}(\bm K_{\theta})g_{\rho\lambda}(\bm K_{\theta}, \bm Q_{\theta})}{eV-\xi_{\bm K_{\theta}}^{\rho}} + \sum_{\bm g}\frac{\phi_{\bm Q_{\theta}+\bm g}e^{-|\bm Q_{\theta}+\bm g|d_{\text{T}}}\Lambda_{\lambda}(\bm K_{\theta_{\text{TBG}}}-\bm g)}{\mu_{\text{T}}-\hbar v_D [K_{\theta_{\text{TBG}}}^x- g^x-i\sigma(K_{\theta_{\text{TBG}}}^{y}- g^y)]} \right|^2. \label{eq:M_full}
\end{align}
Here, $\sigma=\pm$ correspond to the A and B sublattices of graphene. The notations $g^{x,y}$ represent the $x$ and $y$ components of the vector $\bm g$, respectively. Note that we can recover Eq.~\eqref{eq:M_multiband} when the second term on the right-hand side of Eq.~\eqref{eq:M_full} is negligible. In TBG, because the two lowest remote conduction bands $c_{1,2}$ are degenerate at the $\bm\gamma$ point, we define the root mean square of the normalized inelastic tunneling rates of these two bands, $|\overline{M}|=[(|\overline{M}_{c_1}|^2+|\overline{M}_{c_2}|^2)/2]^{\frac{1}{2}}$.
Physically, $|\overline{M}(\bm K_{\theta},\bm Q_{\theta})|^2$ is proportional to the increase in $dI/dV$ at the onset of plasmon-assisted inelastic tunneling into remote bands of TBG.

We compute $|\overline{M}|^2$ numerically for the two examples in the main text: (a) $1.07^{\circ}-$TBG subject to strong gate screening ($d_g=1$\si{nm}); (b) $1.20^{\circ}-$TBG with $d_g=40$\si{nm}. In the calculation, we assume negligible tip Fermi energy, $\mu_{\text{T}}=0$, and a small tip-sample distance $d_{\text{T}}=0.5$\si{nm}. The results are plotted as the orange lines in Fig.~\ref{fig:T}. Comparing them to the blue lines obtained from Eq.~\eqref{eq:M_multiband} in the main text, we observe noticeable corrections generated by the interlayer electron-plasmon coupling, but the corrections are overall smaller than the in-layer contributions to $|\overline{M}|^2$. The corrections are more pronounced for TBG at larger twist angle because there the onset feature occurs at higher bias. 

The analysis in this section reveals one limitation of the formalism in the main text. According to linear response theory, the tunneling current is given by \cite{mahan2013many},
\begin{equation}\label{eq:i_basic}
    I(V) = \frac{2e}{\hbar^2}\text{Im}\left\{-i\int_{0}^{\infty}dt'\ e^{-\frac{ieVt'}{\hbar}}\left\langle \left[\sum_{\bm p,s}\sum_{n=1}^{3} f_{\bm p+\tau\delta \bm G_n st}^{\dagger}(t')\hat{T}_{\tau n}^{\dagger}d_{\bm p s}(t'), \sum_{\bm p,s}\sum_{n=1}^{3} d_{\bm p s}^{\dagger}(0)\hat{T}_{\tau n} f_{\bm p+\tau\delta \bm G_n st}(0)\right] \right\rangle\right\}.
\end{equation}
This leads to Eq.~\eqref{eq:i} in the main text if and only if the electrons in the tip and sample are uncorrelated in the absence of interlayer tunneling \cite{bistritzer2010transport}. Figure~\ref{fig:T}(c) depicts the Feynman diagrams for the response function in Eq.~\eqref{eq:i_basic} that contain a single interaction vertex $\hat{W}$. The first and second diagrams are associated with the correlations between the electrons in the same layer and, in principle, can be included in Eq.~\eqref{eq:i} by using the spectral functions $A^{\text{T,S}}$ of the interacting electrons. In contrast, the last diagram arises from the interlayer Coulomb interactions and is neglected in Eq.~\eqref{eq:i}. This diagram corresponds to the crossing terms in Eq.~\eqref{eq:M_full}, which constitute the primary corrections distinguishing the orange curves from the blue ones in Fig.~\ref{fig:T}(a) and (b).

Nonetheless, at least within the second-order perturbation theory and at low bias, the scattering between the tip electrons and the TBG plasmons gives rise to quantitatively small corrections to the $\mathcal{T}-$matrix elements, and does not introduce new inelastic tunneling features in the spectrum. This, together with the analysis in the previous subsection, justifies neglecting the interlayer electron-plasmon coupling in the main text.
\\
\twocolumngrid

\section{Numerical details}\label{sec:numeric_details}

In this section, we present the details of the numerical calculations. Let us consider the following Hamiltonian
\begin{equation}
    H = H_0 + H_{\text{int}} + H_c,
\end{equation}
where the single-particle Bistritzer-MacDonald(BM) Hamiltonian $H_0$ and the Coulomb interaction $H_{\text{int}}$ are given by Eqs.~\eqref{eq:h0} and~\eqref{eq:hint}, respectively. In addition, a counterterm $H_c$ is added to regularize the divergent single-particle self-energy contributed by the Dirac sea. However, the choice of the counterterm is not unique \cite{xie2020nature, bultinck2020ground, liu2021nematic, bernevig2021twisted_symmetries}.
We follow Ref.~\cite{liu2021nematic} to use
\begin{equation} \label{eq:Hc}
    H_c = -H_{HF}^{\nu=0},
\end{equation}
where $H_{HF}^{\nu=0}$ is the Hartree-Fock mean field potential generated by the ground state of non-interacting Hamiltonian $H_0$ at charge neutrality. In the following one-shot $GW$ calculations, Eq.~\eqref{eq:Hc} automatically cancels with the first-order Hartree-Fock self-energy. 

Assuming that the band structure of the non-interacting BM Hamiltonian $H_0$ provides a good approximation of the quasiparticle spectrum of $H$, we use the unperturbed Green's function $G^0(\bm k, \epsilon)$ in Eq.~\eqref{eq:G0} as a starting point for the perturbative calculation.



We first calculate the RPA screening potential $\hat{W}$ via Eqs.~\eqref{eq:W} and~\eqref{eq:Pi_rpa}. To numerically compute the non-interacting density-density response function $\chi_{\bm g_1,\bm g_2}^0(\bm q,\omega)$ of TBG, Eq.~\eqref{eq:chi0}, we included the 10 Bloch bands closest to the charge-neutrality points. We kept the 19 shortest reciprocal lattice vectors $\bm g_{1,2}$'s and used a grid of wave vectors with $120\times120$ points in one moir\'e Brillouin zone, $\bm q = \frac{i}{N_q} \bar{\bm g}_2  + \frac{j}{N_q}\bar{\bm g}_{3}\ (i,j\in \mathbb{Z}, 0\leq i,j < N_2)$, where $\bar{\bm g}_{2,3} = k_M(\pm\sqrt{3}/2, 3/2)^{\text{t}}$ and $N_q=120$. 
For $1.07^{\circ}-$TBG, we used a regular mesh of 1300 real-axis frequency points in the interval $[-130, 130)$ \si{meV}. For $1.20^{\circ}-$TBG, we used a regular mesh of 1200 real-axis frequency points in the interval $[-150, 150)$ \si{meV}. Unless otherwise mentioned, we chose a broadening parameter $\eta=0.4$ \si{meV} and smearing temperature $T=1$\si{\kelvin}. The calculations can be simplified by taking advantage of $C_{2z}$ and $\mathcal{T}$ symmetries of TBG,
\begin{align}
    \hat{\chi}^{0}(\bm q,\omega) &\overset{C_{2z}}{=} \hat{\chi}^{0}(\bm q,-\omega)^* \overset{\mathcal{T}}{=} \hat{\chi}^{0}(\bm q,\omega)^{\text{t}},
\end{align}
%
as well as the $C_{3z}$ symmetry,
\begin{align}
    \chi_{\bm g_1, \bm g_2}^0(\bm q, \omega) = \chi_{C_{3z}\bm g_1, C_{3z}\bm g_2}^0(C_{3z}\bm q, \omega).
\end{align}

The self-energy $\Sigma_{\lambda_1\lambda_2}(\bm k,\epsilon)$ is computed using Eq.~\eqref{eq:sigma}, with the frequency integral evaluated via the fast Fourier transform. Although, strictly speaking, Eq.~\eqref{eq:sigma} is derived only for $\eta=0^+$, we used a finite broadening parameter $\eta=0.4$ \si{meV} to smooth the numerical results. We then construct the $GW$ Green's function via Eq.~\eqref{eq:G_rpa} and the spectral function $A^{\text{S}} = -G^{\prime\prime}/\pi$.

Finally, we obtain the tunneling conductance from the spectral function $A_{\lambda_1\lambda_2}^{\text{S}}(\bm k,\epsilon)$ of the sample according to the approximate formula, Eq.~\eqref{eq:didv_full}. To reach the convergency of the weak $dI/dV$ features in Figs.~\ref{fig:d2IdV2_1.20} and~\ref{fig:d2Id2V_1.07}, we kept the 16 lowest Bloch bands $\lambda_{1,2}$ when calculating the self-energy and spectral function. Note that the small-$k_F$ assumption in Eq.~\eqref{eq:didv_full} greatly simplifies the calculation by allowing us to compute $A^{\text{S}}$ only along the trajectory of the tip Dirac point $\bm K_{\theta}$, see the dashed line in the inset of Fig.~\ref{fig:tbg_1.20}(b). In practice, we evaluate $A^{\text{S}}$ along the vertical line passing through the high-symmetry points marked in Fig.~\ref{fig:tbg_1.20}(b), which closely approximates the trajectory of $\bm K_\theta$ for small $\theta$ and $\theta_{\text{TBG}}$.    



\bibliography{references}

\begin{thebibliography}{54}%
\makeatletter
\providecommand \@ifxundefined [1]{%
 \@ifx{#1\undefined}
}%
\providecommand \@ifnum [1]{%
 \ifnum #1\expandafter \@firstoftwo
 \else \expandafter \@secondoftwo
 \fi
}%
\providecommand \@ifx [1]{%
 \ifx #1\expandafter \@firstoftwo
 \else \expandafter \@secondoftwo
 \fi
}%
\providecommand \natexlab [1]{#1}%
\providecommand \enquote  [1]{``#1''}%
\providecommand \bibnamefont  [1]{#1}%
\providecommand \bibfnamefont [1]{#1}%
\providecommand \citenamefont [1]{#1}%
\providecommand \href@noop [0]{\@secondoftwo}%
\providecommand \href [0]{\begingroup \@sanitize@url \@href}%
\providecommand \@href[1]{\@@startlink{#1}\@@href}%
\providecommand \@@href[1]{\endgroup#1\@@endlink}%
\providecommand \@sanitize@url [0]{\catcode `\\12\catcode `\$12\catcode `\&12\catcode `\#12\catcode `\^12\catcode `\_12\catcode `\%12\relax}%
\providecommand \@@startlink[1]{}%
\providecommand \@@endlink[0]{}%
\providecommand \url  [0]{\begingroup\@sanitize@url \@url }%
\providecommand \@url [1]{\endgroup\@href {#1}{\urlprefix }}%
\providecommand \urlprefix  [0]{URL }%
\providecommand \Eprint [0]{\href }%
\providecommand \doibase [0]{https://doi.org/}%
\providecommand \selectlanguage [0]{\@gobble}%
\providecommand \bibinfo  [0]{\@secondoftwo}%
\providecommand \bibfield  [0]{\@secondoftwo}%
\providecommand \translation [1]{[#1]}%
\providecommand \BibitemOpen [0]{}%
\providecommand \bibitemStop [0]{}%
\providecommand \bibitemNoStop [0]{.\EOS\space}%
\providecommand \EOS [0]{\spacefactor3000\relax}%
\providecommand \BibitemShut  [1]{\csname bibitem#1\endcsname}%
\let\auto@bib@innerbib\@empty
\bibitem [{\citenamefont {Inbar}\ \emph {et~al.}(2023)\citenamefont {Inbar}, \citenamefont {Birkbeck}, \citenamefont {Xiao}, \citenamefont {Taniguchi}, \citenamefont {Watanabe}, \citenamefont {Yan}, \citenamefont {Oreg}, \citenamefont {Stern}, \citenamefont {Berg},\ and\ \citenamefont {Ilani}}]{inbar_quantum_2023}%
  \BibitemOpen
  \bibfield  {author} {\bibinfo {author} {\bibfnamefont {A.}~\bibnamefont {Inbar}}, \bibinfo {author} {\bibfnamefont {J.}~\bibnamefont {Birkbeck}}, \bibinfo {author} {\bibfnamefont {J.}~\bibnamefont {Xiao}}, \bibinfo {author} {\bibfnamefont {T.}~\bibnamefont {Taniguchi}}, \bibinfo {author} {\bibfnamefont {K.}~\bibnamefont {Watanabe}}, \bibinfo {author} {\bibfnamefont {B.}~\bibnamefont {Yan}}, \bibinfo {author} {\bibfnamefont {Y.}~\bibnamefont {Oreg}}, \bibinfo {author} {\bibfnamefont {A.}~\bibnamefont {Stern}}, \bibinfo {author} {\bibfnamefont {E.}~\bibnamefont {Berg}},\ and\ \bibinfo {author} {\bibfnamefont {S.}~\bibnamefont {Ilani}},\ }\bibfield  {title} {{\selectlanguage {english}\bibinfo {title} {The quantum twisting microscope}},\ }\href {https://doi.org/10.1038/s41586-022-05685-y} {\bibfield  {journal} {\bibinfo  {journal} {Nature}\ }\textbf {\bibinfo {volume} {614}},\ \bibinfo {pages} {682} (\bibinfo {year} {2023})}\BibitemShut {NoStop}%
\bibitem [{\citenamefont {Birkbeck}\ \emph {et~al.}(2024)\citenamefont {Birkbeck}, \citenamefont {Xiao}, \citenamefont {Inbar}, \citenamefont {Taniguchi}, \citenamefont {Watanabe}, \citenamefont {Berg}, \citenamefont {Glazman}, \citenamefont {Guinea}, \citenamefont {von Oppen},\ and\ \citenamefont {Ilani}}]{birkbeck_measuring_2024}%
  \BibitemOpen
  \bibfield  {author} {\bibinfo {author} {\bibfnamefont {J.}~\bibnamefont {Birkbeck}}, \bibinfo {author} {\bibfnamefont {J.}~\bibnamefont {Xiao}}, \bibinfo {author} {\bibfnamefont {A.}~\bibnamefont {Inbar}}, \bibinfo {author} {\bibfnamefont {T.}~\bibnamefont {Taniguchi}}, \bibinfo {author} {\bibfnamefont {K.}~\bibnamefont {Watanabe}}, \bibinfo {author} {\bibfnamefont {E.}~\bibnamefont {Berg}}, \bibinfo {author} {\bibfnamefont {L.}~\bibnamefont {Glazman}}, \bibinfo {author} {\bibfnamefont {F.}~\bibnamefont {Guinea}}, \bibinfo {author} {\bibfnamefont {F.}~\bibnamefont {von Oppen}},\ and\ \bibinfo {author} {\bibfnamefont {S.}~\bibnamefont {Ilani}},\ }\href {http://arxiv.org/abs/2407.13404} {\bibinfo {title} {Measuring {Phonon} {Dispersion} and {Electron}-{Phason} {Coupling} in {Twisted} {Bilayer} {Graphene} with a {Cryogenic} {Quantum} {Twisting} {Microscope}}} (\bibinfo {year} {2024}),\ \bibinfo {note} {arXiv:2407.13404 [cond-mat]}\BibitemShut {NoStop}%
\bibitem [{\citenamefont {Xiao}\ \emph {et~al.}(2024)\citenamefont {Xiao}, \citenamefont {Berg}, \citenamefont {Glazman}, \citenamefont {Guinea}, \citenamefont {Ilani},\ and\ \citenamefont {von Oppen}}]{xiao_theory_2024}%
  \BibitemOpen
  \bibfield  {author} {\bibinfo {author} {\bibfnamefont {J.}~\bibnamefont {Xiao}}, \bibinfo {author} {\bibfnamefont {E.}~\bibnamefont {Berg}}, \bibinfo {author} {\bibfnamefont {L.~I.}\ \bibnamefont {Glazman}}, \bibinfo {author} {\bibfnamefont {F.}~\bibnamefont {Guinea}}, \bibinfo {author} {\bibfnamefont {S.}~\bibnamefont {Ilani}},\ and\ \bibinfo {author} {\bibfnamefont {F.}~\bibnamefont {von Oppen}},\ }\href {http://arxiv.org/abs/2407.12092} {\bibinfo {title} {Theory of phonon spectroscopy with the quantum twisting microscope}} (\bibinfo {year} {2024}),\ \bibinfo {note} {arXiv:2407.12092 [cond-mat]}\BibitemShut {NoStop}%
\bibitem [{\citenamefont {Peri}\ \emph {et~al.}(2024)\citenamefont {Peri}, \citenamefont {Ilani}, \citenamefont {Lee},\ and\ \citenamefont {Refael}}]{peri_probing_2024}%
  \BibitemOpen
  \bibfield  {author} {\bibinfo {author} {\bibfnamefont {V.}~\bibnamefont {Peri}}, \bibinfo {author} {\bibfnamefont {S.}~\bibnamefont {Ilani}}, \bibinfo {author} {\bibfnamefont {P.~A.}\ \bibnamefont {Lee}},\ and\ \bibinfo {author} {\bibfnamefont {G.}~\bibnamefont {Refael}},\ }\bibfield  {title} {{\selectlanguage {english}\bibinfo {title} {Probing quantum spin liquids with a quantum twisting microscope}},\ }\href {https://doi.org/10.1103/PhysRevB.109.035127} {\bibfield  {journal} {\bibinfo  {journal} {Physical Review B}\ }\textbf {\bibinfo {volume} {109}},\ \bibinfo {pages} {035127} (\bibinfo {year} {2024})}\BibitemShut {NoStop}%
\bibitem [{\citenamefont {Pichler}\ \emph {et~al.}(2024)\citenamefont {Pichler}, \citenamefont {Kadow}, \citenamefont {Kuhlenkamp},\ and\ \citenamefont {Knap}}]{pichler_probing_2024}%
  \BibitemOpen
  \bibfield  {author} {\bibinfo {author} {\bibfnamefont {F.}~\bibnamefont {Pichler}}, \bibinfo {author} {\bibfnamefont {W.}~\bibnamefont {Kadow}}, \bibinfo {author} {\bibfnamefont {C.}~\bibnamefont {Kuhlenkamp}},\ and\ \bibinfo {author} {\bibfnamefont {M.}~\bibnamefont {Knap}},\ }\bibfield  {title} {{\selectlanguage {english}\bibinfo {title} {Probing magnetism in moiré heterostructures with quantum twisting microscopes}},\ }\href {https://doi.org/10.1103/PhysRevB.110.045116} {\bibfield  {journal} {\bibinfo  {journal} {Physical Review B}\ }\textbf {\bibinfo {volume} {110}},\ \bibinfo {pages} {045116} (\bibinfo {year} {2024})}\BibitemShut {NoStop}%
\bibitem [{\citenamefont {Carrega}\ \emph {et~al.}(2020)\citenamefont {Carrega}, \citenamefont {Vera-Marun},\ and\ \citenamefont {Principi}}]{carrega2020tunneling}%
  \BibitemOpen
  \bibfield  {author} {\bibinfo {author} {\bibfnamefont {M.}~\bibnamefont {Carrega}}, \bibinfo {author} {\bibfnamefont {I.~J.}\ \bibnamefont {Vera-Marun}},\ and\ \bibinfo {author} {\bibfnamefont {A.}~\bibnamefont {Principi}},\ }\bibfield  {title} {\bibinfo {title} {Tunneling spectroscopy as a probe of fractionalization in two-dimensional magnetic heterostructures},\ }\href {https://doi.org/10.1103/PhysRevB.102.085412} {\bibfield  {journal} {\bibinfo  {journal} {Phys. Rev. B}\ }\textbf {\bibinfo {volume} {102}},\ \bibinfo {pages} {085412} (\bibinfo {year} {2020})}\BibitemShut {NoStop}%
\bibitem [{\citenamefont {Lewandowski}\ and\ \citenamefont {Levitov}(2019)}]{lewandowski_intrinsically_2019}%
  \BibitemOpen
  \bibfield  {author} {\bibinfo {author} {\bibfnamefont {C.}~\bibnamefont {Lewandowski}}\ and\ \bibinfo {author} {\bibfnamefont {L.}~\bibnamefont {Levitov}},\ }\bibfield  {title} {{\selectlanguage {english}\bibinfo {title} {Intrinsically undamped plasmon modes in narrow electron bands}},\ }\href {https://doi.org/10.1073/pnas.1909069116} {\bibfield  {journal} {\bibinfo  {journal} {Proceedings of the National Academy of Sciences}\ }\textbf {\bibinfo {volume} {116}},\ \bibinfo {pages} {20869} (\bibinfo {year} {2019})}\BibitemShut {NoStop}%
\bibitem [{\citenamefont {Papaj}\ and\ \citenamefont {Lewandowski}(2023)}]{papaj2023probing}%
  \BibitemOpen
  \bibfield  {author} {\bibinfo {author} {\bibfnamefont {M.}~\bibnamefont {Papaj}}\ and\ \bibinfo {author} {\bibfnamefont {C.}~\bibnamefont {Lewandowski}},\ }\bibfield  {title} {\bibinfo {title} {Probing correlated states with plasmons},\ }\href@noop {} {\bibfield  {journal} {\bibinfo  {journal} {Science Advances}\ }\textbf {\bibinfo {volume} {9}},\ \bibinfo {pages} {eadg3262} (\bibinfo {year} {2023})}\BibitemShut {NoStop}%
\bibitem [{\citenamefont {Lewandowski}\ \emph {et~al.}(2021)\citenamefont {Lewandowski}, \citenamefont {Chowdhury},\ and\ \citenamefont {Ruhman}}]{lewandowski_pairing_2021}%
  \BibitemOpen
  \bibfield  {author} {\bibinfo {author} {\bibfnamefont {C.}~\bibnamefont {Lewandowski}}, \bibinfo {author} {\bibfnamefont {D.}~\bibnamefont {Chowdhury}},\ and\ \bibinfo {author} {\bibfnamefont {J.}~\bibnamefont {Ruhman}},\ }\bibfield  {title} {{\selectlanguage {english}\bibinfo {title} {Pairing in magic-angle twisted bilayer graphene: {Role} of phonon and plasmon umklapp}},\ }\href {https://doi.org/10.1103/PhysRevB.103.235401} {\bibfield  {journal} {\bibinfo  {journal} {Physical Review B}\ }\textbf {\bibinfo {volume} {103}},\ \bibinfo {pages} {235401} (\bibinfo {year} {2021})}\BibitemShut {NoStop}%
\bibitem [{\citenamefont {Cea}\ and\ \citenamefont {Guinea}(2021)}]{cea_coulomb_2021}%
  \BibitemOpen
  \bibfield  {author} {\bibinfo {author} {\bibfnamefont {T.}~\bibnamefont {Cea}}\ and\ \bibinfo {author} {\bibfnamefont {F.}~\bibnamefont {Guinea}},\ }\bibfield  {title} {{\selectlanguage {english}\bibinfo {title} {Coulomb interaction, phonons, and superconductivity in twisted bilayer graphene}},\ }\href {https://doi.org/10.1073/pnas.2107874118} {\bibfield  {journal} {\bibinfo  {journal} {Proceedings of the National Academy of Sciences}\ }\textbf {\bibinfo {volume} {118}},\ \bibinfo {pages} {e2107874118} (\bibinfo {year} {2021})}\BibitemShut {NoStop}%
\bibitem [{\citenamefont {Sharma}\ \emph {et~al.}(2020)\citenamefont {Sharma}, \citenamefont {Trushin}, \citenamefont {Sushkov}, \citenamefont {Vignale},\ and\ \citenamefont {Adam}}]{sharma2020superconductivity}%
  \BibitemOpen
  \bibfield  {author} {\bibinfo {author} {\bibfnamefont {G.}~\bibnamefont {Sharma}}, \bibinfo {author} {\bibfnamefont {M.}~\bibnamefont {Trushin}}, \bibinfo {author} {\bibfnamefont {O.~P.}\ \bibnamefont {Sushkov}}, \bibinfo {author} {\bibfnamefont {G.}~\bibnamefont {Vignale}},\ and\ \bibinfo {author} {\bibfnamefont {S.}~\bibnamefont {Adam}},\ }\bibfield  {title} {\bibinfo {title} {Superconductivity from collective excitations in magic-angle twisted bilayer graphene},\ }\href {https://doi.org/10.1103/PhysRevResearch.2.022040} {\bibfield  {journal} {\bibinfo  {journal} {Phys. Rev. Res.}\ }\textbf {\bibinfo {volume} {2}},\ \bibinfo {pages} {022040} (\bibinfo {year} {2020})}\BibitemShut {NoStop}%
\bibitem [{\citenamefont {Peng}\ \emph {et~al.}(2024)\citenamefont {Peng}, \citenamefont {Yudhistira}, \citenamefont {Vignale},\ and\ \citenamefont {Adam}}]{peng2024theoretical}%
  \BibitemOpen
  \bibfield  {author} {\bibinfo {author} {\bibfnamefont {L.}~\bibnamefont {Peng}}, \bibinfo {author} {\bibfnamefont {I.}~\bibnamefont {Yudhistira}}, \bibinfo {author} {\bibfnamefont {G.}~\bibnamefont {Vignale}},\ and\ \bibinfo {author} {\bibfnamefont {S.}~\bibnamefont {Adam}},\ }\bibfield  {title} {\bibinfo {title} {Theoretical determination of the effect of a screening gate on plasmon-induced superconductivity in twisted bilayer graphene},\ }\href {https://doi.org/10.1103/PhysRevB.109.045404} {\bibfield  {journal} {\bibinfo  {journal} {Phys. Rev. B}\ }\textbf {\bibinfo {volume} {109}},\ \bibinfo {pages} {045404} (\bibinfo {year} {2024})}\BibitemShut {NoStop}%
\bibitem [{\citenamefont {Wei}\ \emph {et~al.}(2025)\citenamefont {Wei}, \citenamefont {von Oppen},\ and\ \citenamefont {Glazman}}]{wei2025dirac}%
  \BibitemOpen
  \bibfield  {author} {\bibinfo {author} {\bibfnamefont {N.}~\bibnamefont {Wei}}, \bibinfo {author} {\bibfnamefont {F.}~\bibnamefont {von Oppen}},\ and\ \bibinfo {author} {\bibfnamefont {L.~I.}\ \bibnamefont {Glazman}},\ }\bibfield  {title} {\bibinfo {title} {Dirac-point spectroscopy of flat-band systems with the quantum twisting microscope},\ }\href {https://doi.org/10.1103/PhysRevB.111.085128} {\bibfield  {journal} {\bibinfo  {journal} {Phys. Rev. B}\ }\textbf {\bibinfo {volume} {111}},\ \bibinfo {pages} {085128} (\bibinfo {year} {2025})}\BibitemShut {NoStop}%
\bibitem [{\citenamefont {Stauber}\ and\ \citenamefont {Kohler}(2016)}]{stauber2016quasi}%
  \BibitemOpen
  \bibfield  {author} {\bibinfo {author} {\bibfnamefont {T.}~\bibnamefont {Stauber}}\ and\ \bibinfo {author} {\bibfnamefont {H.}~\bibnamefont {Kohler}},\ }\bibfield  {title} {\bibinfo {title} {Quasi-flat plasmonic bands in twisted bilayer graphene},\ }\href@noop {} {\bibfield  {journal} {\bibinfo  {journal} {Nano letters}\ }\textbf {\bibinfo {volume} {16}},\ \bibinfo {pages} {6844} (\bibinfo {year} {2016})}\BibitemShut {NoStop}%
\bibitem [{\citenamefont {Cavicchi}\ \emph {et~al.}(2024)\citenamefont {Cavicchi}, \citenamefont {Torre}, \citenamefont {Jarillo-Herrero}, \citenamefont {Koppens},\ and\ \citenamefont {Polini}}]{cavicchi_theory_2024}%
  \BibitemOpen
  \bibfield  {author} {\bibinfo {author} {\bibfnamefont {L.}~\bibnamefont {Cavicchi}}, \bibinfo {author} {\bibfnamefont {I.}~\bibnamefont {Torre}}, \bibinfo {author} {\bibfnamefont {P.}~\bibnamefont {Jarillo-Herrero}}, \bibinfo {author} {\bibfnamefont {F.~H.~L.}\ \bibnamefont {Koppens}},\ and\ \bibinfo {author} {\bibfnamefont {M.}~\bibnamefont {Polini}},\ }\bibfield  {title} {{\selectlanguage {english}\bibinfo {title} {Theory of intrinsic acoustic plasmons in twisted bilayer graphene}},\ }\href {https://doi.org/10.1103/PhysRevB.110.045431} {\bibfield  {journal} {\bibinfo  {journal} {Physical Review B}\ }\textbf {\bibinfo {volume} {110}},\ \bibinfo {pages} {045431} (\bibinfo {year} {2024})}\BibitemShut {NoStop}%
\bibitem [{\citenamefont {Pines}\ and\ \citenamefont {Schrieffer}(1962)}]{pines1962approach}%
  \BibitemOpen
  \bibfield  {author} {\bibinfo {author} {\bibfnamefont {D.}~\bibnamefont {Pines}}\ and\ \bibinfo {author} {\bibfnamefont {J.~R.}\ \bibnamefont {Schrieffer}},\ }\bibfield  {title} {\bibinfo {title} {Approach to equilibrium of electrons, plasmons, and phonons in quantum and classical plasmas},\ }\href {https://doi.org/10.1103/PhysRev.125.804} {\bibfield  {journal} {\bibinfo  {journal} {Phys. Rev.}\ }\textbf {\bibinfo {volume} {125}},\ \bibinfo {pages} {804} (\bibinfo {year} {1962})}\BibitemShut {NoStop}%
\bibitem [{\citenamefont {Mahan}(2013)}]{mahan2013many}%
  \BibitemOpen
  \bibfield  {author} {\bibinfo {author} {\bibfnamefont {G.~D.}\ \bibnamefont {Mahan}},\ }\href@noop {} {\emph {\bibinfo {title} {Many-particle physics}}}\ (\bibinfo  {publisher} {Springer Science \& Business Media},\ \bibinfo {year} {2013})\BibitemShut {NoStop}%
\bibitem [{\citenamefont {Bistritzer}\ and\ \citenamefont {MacDonald}(2011)}]{bistritzer_moire_2011}%
  \BibitemOpen
  \bibfield  {author} {\bibinfo {author} {\bibfnamefont {R.}~\bibnamefont {Bistritzer}}\ and\ \bibinfo {author} {\bibfnamefont {A.~H.}\ \bibnamefont {MacDonald}},\ }\bibfield  {title} {{\selectlanguage {english}\bibinfo {title} {Moiré bands in twisted double-layer graphene}},\ }\href {https://doi.org/10.1073/pnas.1108174108} {\bibfield  {journal} {\bibinfo  {journal} {Proceedings of the National Academy of Sciences}\ }\textbf {\bibinfo {volume} {108}},\ \bibinfo {pages} {12233} (\bibinfo {year} {2011})}\BibitemShut {NoStop}%
\bibitem [{\citenamefont {Lopes~dos Santos}\ \emph {et~al.}(2007)\citenamefont {Lopes~dos Santos}, \citenamefont {Peres},\ and\ \citenamefont {Castro~Neto}}]{lopesdossantos_graphene_2007}%
  \BibitemOpen
  \bibfield  {author} {\bibinfo {author} {\bibfnamefont {J.~M.~B.}\ \bibnamefont {Lopes~dos Santos}}, \bibinfo {author} {\bibfnamefont {N.~M.~R.}\ \bibnamefont {Peres}},\ and\ \bibinfo {author} {\bibfnamefont {A.~H.}\ \bibnamefont {Castro~Neto}},\ }\bibfield  {title} {{\selectlanguage {english}\bibinfo {title} {Graphene {Bilayer} with a {Twist}: {Electronic} {Structure}}},\ }\href {https://doi.org/10.1103/PhysRevLett.99.256802} {\bibfield  {journal} {\bibinfo  {journal} {Physical Review Letters}\ }\textbf {\bibinfo {volume} {99}},\ \bibinfo {pages} {256802} (\bibinfo {year} {2007})}\BibitemShut {NoStop}%
\bibitem [{\citenamefont {Vafek}\ and\ \citenamefont {Kang}(2020)}]{vafek2020renormalization}%
  \BibitemOpen
  \bibfield  {author} {\bibinfo {author} {\bibfnamefont {O.}~\bibnamefont {Vafek}}\ and\ \bibinfo {author} {\bibfnamefont {J.}~\bibnamefont {Kang}},\ }\bibfield  {title} {\bibinfo {title} {Renormalization group study of hidden symmetry in twisted bilayer graphene with coulomb interactions},\ }\href {https://doi.org/10.1103/PhysRevLett.125.257602} {\bibfield  {journal} {\bibinfo  {journal} {Phys. Rev. Lett.}\ }\textbf {\bibinfo {volume} {125}},\ \bibinfo {pages} {257602} (\bibinfo {year} {2020})}\BibitemShut {NoStop}%
\bibitem [{Note1()}]{Note1}%
  \BibitemOpen
  \bibinfo {note} {In our notation, $\protect \hat {W}$ does not include the first order Coulomb interaction and is different from the conventional notation in the $GW$ approximation.}\BibitemShut {Stop}%
\bibitem [{Note2()}]{Note2}%
  \BibitemOpen
  \bibinfo {note} {Note that $G_{\protect \text {incoh}}$ is the scale for the features in $dI/dV$ resulting from elastic, momentum-conserving tunneling~\cite {xiao_theory_2024}.}\BibitemShut {Stop}%
\bibitem [{\citenamefont {Gao}\ \emph {et~al.}(2024)\citenamefont {Gao}, \citenamefont {Jimeno-Pozo}, \citenamefont {Pantaleon}, \citenamefont {Codecido}, \citenamefont {Sharifi}, \citenamefont {Zhang}, \citenamefont {Liu}, \citenamefont {Watanabe}, \citenamefont {Taniguchi}, \citenamefont {Bockrath}, \citenamefont {Guinea},\ and\ \citenamefont {Lau}}]{gao2024double}%
  \BibitemOpen
  \bibfield  {author} {\bibinfo {author} {\bibfnamefont {X.}~\bibnamefont {Gao}}, \bibinfo {author} {\bibfnamefont {A.}~\bibnamefont {Jimeno-Pozo}}, \bibinfo {author} {\bibfnamefont {P.~A.}\ \bibnamefont {Pantaleon}}, \bibinfo {author} {\bibfnamefont {E.}~\bibnamefont {Codecido}}, \bibinfo {author} {\bibfnamefont {D.~L.}\ \bibnamefont {Sharifi}}, \bibinfo {author} {\bibfnamefont {Z.}~\bibnamefont {Zhang}}, \bibinfo {author} {\bibfnamefont {Y.}~\bibnamefont {Liu}}, \bibinfo {author} {\bibfnamefont {K.}~\bibnamefont {Watanabe}}, \bibinfo {author} {\bibfnamefont {T.}~\bibnamefont {Taniguchi}}, \bibinfo {author} {\bibfnamefont {M.~W.}\ \bibnamefont {Bockrath}}, \bibinfo {author} {\bibfnamefont {F.}~\bibnamefont {Guinea}},\ and\ \bibinfo {author} {\bibfnamefont {C.~N.}\ \bibnamefont {Lau}},\ }\href {https://arxiv.org/abs/2412.01578} {\bibinfo {title} {Double-edged role of interactions in superconducting twisted bilayer graphene}} (\bibinfo {year} {2024}),\ \Eprint {https://arxiv.org/abs/2412.01578}
  {arXiv:2412.01578 [cond-mat.mes-hall]} \BibitemShut {NoStop}%
\bibitem [{\citenamefont {Wang}\ \emph {et~al.}(2025)\citenamefont {Wang}, \citenamefont {Zhu}, \citenamefont {Burg}, \citenamefont {Swain}, \citenamefont {Watanabe}, \citenamefont {Taniguchi}, \citenamefont {Zheng}, \citenamefont {MacDonald},\ and\ \citenamefont {Tutuc}}]{wang2025independently}%
  \BibitemOpen
  \bibfield  {author} {\bibinfo {author} {\bibfnamefont {Y.}~\bibnamefont {Wang}}, \bibinfo {author} {\bibfnamefont {J.}~\bibnamefont {Zhu}}, \bibinfo {author} {\bibfnamefont {G.~W.}\ \bibnamefont {Burg}}, \bibinfo {author} {\bibfnamefont {A.}~\bibnamefont {Swain}}, \bibinfo {author} {\bibfnamefont {K.}~\bibnamefont {Watanabe}}, \bibinfo {author} {\bibfnamefont {T.}~\bibnamefont {Taniguchi}}, \bibinfo {author} {\bibfnamefont {Y.}~\bibnamefont {Zheng}}, \bibinfo {author} {\bibfnamefont {A.~H.}\ \bibnamefont {MacDonald}},\ and\ \bibinfo {author} {\bibfnamefont {E.}~\bibnamefont {Tutuc}},\ }\bibfield  {title} {\bibinfo {title} {Independently tunable flat bands and correlations in a graphene double moir\'e system},\ }\href {https://doi.org/10.1103/PhysRevLett.134.096204} {\bibfield  {journal} {\bibinfo  {journal} {Phys. Rev. Lett.}\ }\textbf {\bibinfo {volume} {134}},\ \bibinfo {pages} {096204} (\bibinfo {year} {2025})}\BibitemShut {NoStop}%
\bibitem [{\citenamefont {Barrier}\ \emph {et~al.}(2024)\citenamefont {Barrier}, \citenamefont {Peng}, \citenamefont {Xu}, \citenamefont {Fal'ko}, \citenamefont {Watanabe}, \citenamefont {Tanigushi}, \citenamefont {Geim}, \citenamefont {Adam},\ and\ \citenamefont {Berdyugin}}]{barrier_coulomb_2024}%
  \BibitemOpen
  \bibfield  {author} {\bibinfo {author} {\bibfnamefont {J.}~\bibnamefont {Barrier}}, \bibinfo {author} {\bibfnamefont {L.}~\bibnamefont {Peng}}, \bibinfo {author} {\bibfnamefont {S.}~\bibnamefont {Xu}}, \bibinfo {author} {\bibfnamefont {V.~I.}\ \bibnamefont {Fal'ko}}, \bibinfo {author} {\bibfnamefont {K.}~\bibnamefont {Watanabe}}, \bibinfo {author} {\bibfnamefont {T.}~\bibnamefont {Tanigushi}}, \bibinfo {author} {\bibfnamefont {A.~K.}\ \bibnamefont {Geim}}, \bibinfo {author} {\bibfnamefont {S.}~\bibnamefont {Adam}},\ and\ \bibinfo {author} {\bibfnamefont {A.~I.}\ \bibnamefont {Berdyugin}},\ }\href {https://doi.org/10.48550/arXiv.2412.01577} {\bibinfo {title} {Coulomb screening of superconductivity in magic-angle twisted bilayer graphene}} (\bibinfo {year} {2024}),\ \bibinfo {note} {arXiv:2412.01577 [cond-mat]}\BibitemShut {NoStop}%
\bibitem [{\citenamefont {C{\u{a}}lug{\u{a}}ru}\ \emph {et~al.}(2023)\citenamefont {C{\u{a}}lug{\u{a}}ru}, \citenamefont {Borovkov}, \citenamefont {Lau}, \citenamefont {Coleman}, \citenamefont {Song},\ and\ \citenamefont {Bernevig}}]{cualuguaru2023twisted}%
  \BibitemOpen
  \bibfield  {author} {\bibinfo {author} {\bibfnamefont {D.}~\bibnamefont {C{\u{a}}lug{\u{a}}ru}}, \bibinfo {author} {\bibfnamefont {M.}~\bibnamefont {Borovkov}}, \bibinfo {author} {\bibfnamefont {L.~L.}\ \bibnamefont {Lau}}, \bibinfo {author} {\bibfnamefont {P.}~\bibnamefont {Coleman}}, \bibinfo {author} {\bibfnamefont {Z.-D.}\ \bibnamefont {Song}},\ and\ \bibinfo {author} {\bibfnamefont {B.~A.}\ \bibnamefont {Bernevig}},\ }\bibfield  {title} {\bibinfo {title} {Twisted bilayer graphene as topological heavy fermion: Ii. analytical approximations of the model parameters},\ }\href@noop {} {\bibfield  {journal} {\bibinfo  {journal} {Low Temperature Physics}\ }\textbf {\bibinfo {volume} {49}},\ \bibinfo {pages} {640} (\bibinfo {year} {2023})}\BibitemShut {NoStop}%
\bibitem [{\citenamefont {Martin}\ \emph {et~al.}(2016)\citenamefont {Martin}, \citenamefont {Reining},\ and\ \citenamefont {Ceperley}}]{martin2016interacting}%
  \BibitemOpen
  \bibfield  {author} {\bibinfo {author} {\bibfnamefont {R.~M.}\ \bibnamefont {Martin}}, \bibinfo {author} {\bibfnamefont {L.}~\bibnamefont {Reining}},\ and\ \bibinfo {author} {\bibfnamefont {D.~M.}\ \bibnamefont {Ceperley}},\ }\href@noop {} {\emph {\bibinfo {title} {Interacting electrons}}}\ (\bibinfo  {publisher} {Cambridge University Press},\ \bibinfo {year} {2016})\BibitemShut {NoStop}%
\bibitem [{\citenamefont {Caruso}\ and\ \citenamefont {Giustino}(2016)}]{caruso2016gw}%
  \BibitemOpen
  \bibfield  {author} {\bibinfo {author} {\bibfnamefont {F.}~\bibnamefont {Caruso}}\ and\ \bibinfo {author} {\bibfnamefont {F.}~\bibnamefont {Giustino}},\ }\bibfield  {title} {\bibinfo {title} {The gw plus cumulant method and plasmonic polarons: application to the homogeneous electron gas},\ }\href@noop {} {\bibfield  {journal} {\bibinfo  {journal} {The European Physical Journal B}\ }\textbf {\bibinfo {volume} {89}},\ \bibinfo {pages} {1} (\bibinfo {year} {2016})}\BibitemShut {NoStop}%
\bibitem [{\citenamefont {Ledwith}\ \emph {et~al.}(2025{\natexlab{a}})\citenamefont {Ledwith}, \citenamefont {Dong}, \citenamefont {Vishwanath},\ and\ \citenamefont {Khalaf}}]{ledwith_nonlocal_2025}%
  \BibitemOpen
  \bibfield  {author} {\bibinfo {author} {\bibfnamefont {P.~J.}\ \bibnamefont {Ledwith}}, \bibinfo {author} {\bibfnamefont {J.}~\bibnamefont {Dong}}, \bibinfo {author} {\bibfnamefont {A.}~\bibnamefont {Vishwanath}},\ and\ \bibinfo {author} {\bibfnamefont {E.}~\bibnamefont {Khalaf}},\ }\href {https://doi.org/10.48550/arXiv.2408.16761} {\bibinfo {title} {Nonlocal {Moments} in the {Chern} {Bands} of {Twisted} {Bilayer} {Graphene}}} (\bibinfo {year} {2025}{\natexlab{a}}),\ \bibinfo {note} {arXiv:2408.16761 [cond-mat]}\BibitemShut {NoStop}%
\bibitem [{\citenamefont {Guinea}\ and\ \citenamefont {Walet}(2018)}]{guinea_electrostatic_2018}%
  \BibitemOpen
  \bibfield  {author} {\bibinfo {author} {\bibfnamefont {F.}~\bibnamefont {Guinea}}\ and\ \bibinfo {author} {\bibfnamefont {N.~R.}\ \bibnamefont {Walet}},\ }\bibfield  {title} {{\selectlanguage {english}\bibinfo {title} {Electrostatic effects, band distortions, and superconductivity in twisted graphene bilayers}},\ }\href {https://doi.org/10.1073/pnas.1810947115} {\bibfield  {journal} {\bibinfo  {journal} {Proceedings of the National Academy of Sciences}\ }\textbf {\bibinfo {volume} {115}},\ \bibinfo {pages} {13174} (\bibinfo {year} {2018})}\BibitemShut {NoStop}%
\bibitem [{\citenamefont {Novelli}\ \emph {et~al.}(2020)\citenamefont {Novelli}, \citenamefont {Torre}, \citenamefont {Koppens}, \citenamefont {Taddei},\ and\ \citenamefont {Polini}}]{novelli2020optical}%
  \BibitemOpen
  \bibfield  {author} {\bibinfo {author} {\bibfnamefont {P.}~\bibnamefont {Novelli}}, \bibinfo {author} {\bibfnamefont {I.}~\bibnamefont {Torre}}, \bibinfo {author} {\bibfnamefont {F.~H.~L.}\ \bibnamefont {Koppens}}, \bibinfo {author} {\bibfnamefont {F.}~\bibnamefont {Taddei}},\ and\ \bibinfo {author} {\bibfnamefont {M.}~\bibnamefont {Polini}},\ }\bibfield  {title} {\bibinfo {title} {Optical and plasmonic properties of twisted bilayer graphene: Impact of interlayer tunneling asymmetry and ground-state charge inhomogeneity},\ }\href {https://doi.org/10.1103/PhysRevB.102.125403} {\bibfield  {journal} {\bibinfo  {journal} {Phys. Rev. B}\ }\textbf {\bibinfo {volume} {102}},\ \bibinfo {pages} {125403} (\bibinfo {year} {2020})}\BibitemShut {NoStop}%
\bibitem [{\citenamefont {Zhu}\ \emph {et~al.}(2024)\citenamefont {Zhu}, \citenamefont {Torre}, \citenamefont {Polini},\ and\ \citenamefont {MacDonald}}]{zhu2024weak}%
  \BibitemOpen
  \bibfield  {author} {\bibinfo {author} {\bibfnamefont {J.}~\bibnamefont {Zhu}}, \bibinfo {author} {\bibfnamefont {I.}~\bibnamefont {Torre}}, \bibinfo {author} {\bibfnamefont {M.}~\bibnamefont {Polini}},\ and\ \bibinfo {author} {\bibfnamefont {A.~H.}\ \bibnamefont {MacDonald}},\ }\bibfield  {title} {\bibinfo {title} {Weak-coupling theory of magic-angle twisted bilayer graphene},\ }\href {https://doi.org/10.1103/PhysRevB.110.L121117} {\bibfield  {journal} {\bibinfo  {journal} {Phys. Rev. B}\ }\textbf {\bibinfo {volume} {110}},\ \bibinfo {pages} {L121117} (\bibinfo {year} {2024})}\BibitemShut {NoStop}%
\bibitem [{\citenamefont {Peng}\ \emph {et~al.}(2025)\citenamefont {Peng}, \citenamefont {Vignale},\ and\ \citenamefont {Adam}}]{peng2025many}%
  \BibitemOpen
  \bibfield  {author} {\bibinfo {author} {\bibfnamefont {L.}~\bibnamefont {Peng}}, \bibinfo {author} {\bibfnamefont {G.}~\bibnamefont {Vignale}},\ and\ \bibinfo {author} {\bibfnamefont {S.}~\bibnamefont {Adam}},\ }\href {https://arxiv.org/abs/2502.06968} {\bibinfo {title} {Many-body perturbation theory for moir\'{e} systems}} (\bibinfo {year} {2025}),\ \Eprint {https://arxiv.org/abs/2502.06968} {arXiv:2502.06968 [cond-mat.str-el]} \BibitemShut {NoStop}%
\bibitem [{\citenamefont {Hofmann}\ \emph {et~al.}(2022)\citenamefont {Hofmann}, \citenamefont {Khalaf}, \citenamefont {Vishwanath}, \citenamefont {Berg},\ and\ \citenamefont {Lee}}]{hofmann2022fermionic}%
  \BibitemOpen
  \bibfield  {author} {\bibinfo {author} {\bibfnamefont {J.~S.}\ \bibnamefont {Hofmann}}, \bibinfo {author} {\bibfnamefont {E.}~\bibnamefont {Khalaf}}, \bibinfo {author} {\bibfnamefont {A.}~\bibnamefont {Vishwanath}}, \bibinfo {author} {\bibfnamefont {E.}~\bibnamefont {Berg}},\ and\ \bibinfo {author} {\bibfnamefont {J.~Y.}\ \bibnamefont {Lee}},\ }\bibfield  {title} {\bibinfo {title} {Fermionic monte carlo study of a realistic model of twisted bilayer graphene},\ }\href {https://doi.org/10.1103/PhysRevX.12.011061} {\bibfield  {journal} {\bibinfo  {journal} {Phys. Rev. X}\ }\textbf {\bibinfo {volume} {12}},\ \bibinfo {pages} {011061} (\bibinfo {year} {2022})}\BibitemShut {NoStop}%
\bibitem [{\citenamefont {Huang}\ \emph {et~al.}(2025)\citenamefont {Huang}, \citenamefont {Parthenios}, \citenamefont {Ulybyshev}, \citenamefont {Zhang}, \citenamefont {Assaad}, \citenamefont {Classen},\ and\ \citenamefont {Meng}}]{huang2025angle}%
  \BibitemOpen
  \bibfield  {author} {\bibinfo {author} {\bibfnamefont {C.}~\bibnamefont {Huang}}, \bibinfo {author} {\bibfnamefont {N.}~\bibnamefont {Parthenios}}, \bibinfo {author} {\bibfnamefont {M.}~\bibnamefont {Ulybyshev}}, \bibinfo {author} {\bibfnamefont {X.}~\bibnamefont {Zhang}}, \bibinfo {author} {\bibfnamefont {F.~F.}\ \bibnamefont {Assaad}}, \bibinfo {author} {\bibfnamefont {L.}~\bibnamefont {Classen}},\ and\ \bibinfo {author} {\bibfnamefont {Z.~Y.}\ \bibnamefont {Meng}},\ }\href {https://arxiv.org/abs/2412.11382} {\bibinfo {title} {Angle-tuned gross-neveu quantum criticality in twisted bilayer graphene: A quantum monte carlo study}} (\bibinfo {year} {2025}),\ \Eprint {https://arxiv.org/abs/2412.11382} {arXiv:2412.11382 [cond-mat.str-el]} \BibitemShut {NoStop}%
\bibitem [{\citenamefont {Datta}\ \emph {et~al.}(2023)\citenamefont {Datta}, \citenamefont {Calderón}, \citenamefont {Camjayi},\ and\ \citenamefont {Bascones}}]{datta_heavy_2023}%
  \BibitemOpen
  \bibfield  {author} {\bibinfo {author} {\bibfnamefont {A.}~\bibnamefont {Datta}}, \bibinfo {author} {\bibfnamefont {M.~J.}\ \bibnamefont {Calderón}}, \bibinfo {author} {\bibfnamefont {A.}~\bibnamefont {Camjayi}},\ and\ \bibinfo {author} {\bibfnamefont {E.}~\bibnamefont {Bascones}},\ }\bibfield  {title} {{\selectlanguage {english}\bibinfo {title} {Heavy quasiparticles and cascades without symmetry breaking in twisted bilayer graphene}},\ }\href {https://doi.org/10.1038/s41467-023-40754-4} {\bibfield  {journal} {\bibinfo  {journal} {Nature Communications}\ }\textbf {\bibinfo {volume} {14}},\ \bibinfo {pages} {5036} (\bibinfo {year} {2023})}\BibitemShut {NoStop}%
\bibitem [{\citenamefont {Rai}\ \emph {et~al.}(2024)\citenamefont {Rai}, \citenamefont {Crippa}, \citenamefont {Călugăru}, \citenamefont {Hu}, \citenamefont {Paoletti}, \citenamefont {De’~Medici}, \citenamefont {Georges}, \citenamefont {Bernevig}, \citenamefont {Valentí}, \citenamefont {Sangiovanni},\ and\ \citenamefont {Wehling}}]{rai_dynamical_2024}%
  \BibitemOpen
  \bibfield  {author} {\bibinfo {author} {\bibfnamefont {G.}~\bibnamefont {Rai}}, \bibinfo {author} {\bibfnamefont {L.}~\bibnamefont {Crippa}}, \bibinfo {author} {\bibfnamefont {D.}~\bibnamefont {Călugăru}}, \bibinfo {author} {\bibfnamefont {H.}~\bibnamefont {Hu}}, \bibinfo {author} {\bibfnamefont {F.}~\bibnamefont {Paoletti}}, \bibinfo {author} {\bibfnamefont {L.}~\bibnamefont {De’~Medici}}, \bibinfo {author} {\bibfnamefont {A.}~\bibnamefont {Georges}}, \bibinfo {author} {\bibfnamefont {B.~A.}\ \bibnamefont {Bernevig}}, \bibinfo {author} {\bibfnamefont {R.}~\bibnamefont {Valentí}}, \bibinfo {author} {\bibfnamefont {G.}~\bibnamefont {Sangiovanni}},\ and\ \bibinfo {author} {\bibfnamefont {T.}~\bibnamefont {Wehling}},\ }\bibfield  {title} {{\selectlanguage {english}\bibinfo {title} {Dynamical {Correlations} and {Order} in {Magic}-{Angle} {Twisted} {Bilayer} {Graphene}}},\ }\href {https://doi.org/10.1103/PhysRevX.14.031045} {\bibfield  {journal} {\bibinfo  {journal} {Physical Review X}\ }\textbf {\bibinfo
  {volume} {14}},\ \bibinfo {pages} {031045} (\bibinfo {year} {2024})}\BibitemShut {NoStop}%
\bibitem [{\citenamefont {Zhou}\ \emph {et~al.}(2024)\citenamefont {Zhou}, \citenamefont {Wang}, \citenamefont {Tong},\ and\ \citenamefont {Song}}]{zhou2024kondo}%
  \BibitemOpen
  \bibfield  {author} {\bibinfo {author} {\bibfnamefont {G.-D.}\ \bibnamefont {Zhou}}, \bibinfo {author} {\bibfnamefont {Y.-J.}\ \bibnamefont {Wang}}, \bibinfo {author} {\bibfnamefont {N.}~\bibnamefont {Tong}},\ and\ \bibinfo {author} {\bibfnamefont {Z.-D.}\ \bibnamefont {Song}},\ }\bibfield  {title} {\bibinfo {title} {Kondo phase in twisted bilayer graphene},\ }\href {https://doi.org/10.1103/PhysRevB.109.045419} {\bibfield  {journal} {\bibinfo  {journal} {Phys. Rev. B}\ }\textbf {\bibinfo {volume} {109}},\ \bibinfo {pages} {045419} (\bibinfo {year} {2024})}\BibitemShut {NoStop}%
\bibitem [{\citenamefont {Calderón}\ \emph {et~al.}(2025)\citenamefont {Calderón}, \citenamefont {Camjayi}, \citenamefont {Datta},\ and\ \citenamefont {Bascones}}]{calderón2025cascades}%
  \BibitemOpen
  \bibfield  {author} {\bibinfo {author} {\bibfnamefont {M.~J.}\ \bibnamefont {Calderón}}, \bibinfo {author} {\bibfnamefont {A.}~\bibnamefont {Camjayi}}, \bibinfo {author} {\bibfnamefont {A.}~\bibnamefont {Datta}},\ and\ \bibinfo {author} {\bibfnamefont {E.}~\bibnamefont {Bascones}},\ }\href {https://arxiv.org/abs/2412.20855} {\bibinfo {title} {Cascades in transport and optical conductivity of twisted bilayer graphene}} (\bibinfo {year} {2025}),\ \Eprint {https://arxiv.org/abs/2412.20855} {arXiv:2412.20855 [cond-mat.str-el]} \BibitemShut {NoStop}%
\bibitem [{\citenamefont {Georges}\ \emph {et~al.}(1996)\citenamefont {Georges}, \citenamefont {Kotliar}, \citenamefont {Krauth},\ and\ \citenamefont {Rozenberg}}]{georges_dynamical_1996}%
  \BibitemOpen
  \bibfield  {author} {\bibinfo {author} {\bibfnamefont {A.}~\bibnamefont {Georges}}, \bibinfo {author} {\bibfnamefont {G.}~\bibnamefont {Kotliar}}, \bibinfo {author} {\bibfnamefont {W.}~\bibnamefont {Krauth}},\ and\ \bibinfo {author} {\bibfnamefont {M.~J.}\ \bibnamefont {Rozenberg}},\ }\bibfield  {title} {{\selectlanguage {english}\bibinfo {title} {Dynamical mean-field theory of strongly correlated fermion systems and the limit of infinite dimensions}},\ }\href {https://doi.org/10.1103/RevModPhys.68.13} {\bibfield  {journal} {\bibinfo  {journal} {Reviews of Modern Physics}\ }\textbf {\bibinfo {volume} {68}},\ \bibinfo {pages} {13} (\bibinfo {year} {1996})}\BibitemShut {NoStop}%
\bibitem [{\citenamefont {Hu}\ \emph {et~al.}(2025)\citenamefont {Hu}, \citenamefont {Song},\ and\ \citenamefont {Bernevig}}]{hu_projected_2025}%
  \BibitemOpen
  \bibfield  {author} {\bibinfo {author} {\bibfnamefont {H.}~\bibnamefont {Hu}}, \bibinfo {author} {\bibfnamefont {Z.-D.}\ \bibnamefont {Song}},\ and\ \bibinfo {author} {\bibfnamefont {B.~A.}\ \bibnamefont {Bernevig}},\ }\href {https://doi.org/10.48550/arXiv.2502.14039} {\bibinfo {title} {Projected and {Solvable} {Topological} {Heavy} {Fermion} {Model} of {Twisted} {Bilayer} {Graphene}}} (\bibinfo {year} {2025}),\ \bibinfo {note} {arXiv:2502.14039 [cond-mat]}\BibitemShut {NoStop}%
\bibitem [{\citenamefont {Ledwith}\ \emph {et~al.}(2025{\natexlab{b}})\citenamefont {Ledwith}, \citenamefont {Vishwanath},\ and\ \citenamefont {Khalaf}}]{ledwith2025exotic}%
  \BibitemOpen
  \bibfield  {author} {\bibinfo {author} {\bibfnamefont {P.~J.}\ \bibnamefont {Ledwith}}, \bibinfo {author} {\bibfnamefont {A.}~\bibnamefont {Vishwanath}},\ and\ \bibinfo {author} {\bibfnamefont {E.}~\bibnamefont {Khalaf}},\ }\href {https://arxiv.org/abs/2505.08779} {\bibinfo {title} {Exotic carriers from concentrated topology: Dirac trions as the origin of the missing spectral weight in twisted bilayer graphene}} (\bibinfo {year} {2025}{\natexlab{b}}),\ \Eprint {https://arxiv.org/abs/2505.08779} {arXiv:2505.08779 [cond-mat.str-el]} \BibitemShut {NoStop}%
\bibitem [{\citenamefont {Song}\ and\ \citenamefont {Bernevig}(2022)}]{song_magic-angle_2022}%
  \BibitemOpen
  \bibfield  {author} {\bibinfo {author} {\bibfnamefont {Z.-D.}\ \bibnamefont {Song}}\ and\ \bibinfo {author} {\bibfnamefont {B.~A.}\ \bibnamefont {Bernevig}},\ }\bibfield  {title} {{\selectlanguage {english}\bibinfo {title} {Magic-{Angle} {Twisted} {Bilayer} {Graphene} as a {Topological} {Heavy} {Fermion} {Problem}}},\ }\href {https://doi.org/10.1103/PhysRevLett.129.047601} {\bibfield  {journal} {\bibinfo  {journal} {Physical Review Letters}\ }\textbf {\bibinfo {volume} {129}},\ \bibinfo {pages} {047601} (\bibinfo {year} {2022})}\BibitemShut {NoStop}%
\bibitem [{\citenamefont {van Loon}\ \emph {et~al.}(2014)\citenamefont {van Loon}, \citenamefont {Hafermann}, \citenamefont {Lichtenstein}, \citenamefont {Rubtsov},\ and\ \citenamefont {Katsnelson}}]{vanloon2014plasmons}%
  \BibitemOpen
  \bibfield  {author} {\bibinfo {author} {\bibfnamefont {E.~G. C.~P.}\ \bibnamefont {van Loon}}, \bibinfo {author} {\bibfnamefont {H.}~\bibnamefont {Hafermann}}, \bibinfo {author} {\bibfnamefont {A.~I.}\ \bibnamefont {Lichtenstein}}, \bibinfo {author} {\bibfnamefont {A.~N.}\ \bibnamefont {Rubtsov}},\ and\ \bibinfo {author} {\bibfnamefont {M.~I.}\ \bibnamefont {Katsnelson}},\ }\bibfield  {title} {\bibinfo {title} {Plasmons in strongly correlated systems: Spectral weight transfer and renormalized dispersion},\ }\href {https://doi.org/10.1103/PhysRevLett.113.246407} {\bibfield  {journal} {\bibinfo  {journal} {Phys. Rev. Lett.}\ }\textbf {\bibinfo {volume} {113}},\ \bibinfo {pages} {246407} (\bibinfo {year} {2014})}\BibitemShut {NoStop}%
\bibitem [{\citenamefont {Wu}\ and\ \citenamefont {Das~Sarma}(2020)}]{wu2020collective}%
  \BibitemOpen
  \bibfield  {author} {\bibinfo {author} {\bibfnamefont {F.}~\bibnamefont {Wu}}\ and\ \bibinfo {author} {\bibfnamefont {S.}~\bibnamefont {Das~Sarma}},\ }\bibfield  {title} {\bibinfo {title} {Collective excitations of quantum anomalous hall ferromagnets in twisted bilayer graphene},\ }\href {https://doi.org/10.1103/PhysRevLett.124.046403} {\bibfield  {journal} {\bibinfo  {journal} {Phys. Rev. Lett.}\ }\textbf {\bibinfo {volume} {124}},\ \bibinfo {pages} {046403} (\bibinfo {year} {2020})}\BibitemShut {NoStop}%
\bibitem [{\citenamefont {Kumar}\ \emph {et~al.}(2021)\citenamefont {Kumar}, \citenamefont {Xie},\ and\ \citenamefont {MacDonald}}]{kumar2021lattice}%
  \BibitemOpen
  \bibfield  {author} {\bibinfo {author} {\bibfnamefont {A.}~\bibnamefont {Kumar}}, \bibinfo {author} {\bibfnamefont {M.}~\bibnamefont {Xie}},\ and\ \bibinfo {author} {\bibfnamefont {A.~H.}\ \bibnamefont {MacDonald}},\ }\bibfield  {title} {\bibinfo {title} {Lattice collective modes from a continuum model of magic-angle twisted bilayer graphene},\ }\href {https://doi.org/10.1103/PhysRevB.104.035119} {\bibfield  {journal} {\bibinfo  {journal} {Phys. Rev. B}\ }\textbf {\bibinfo {volume} {104}},\ \bibinfo {pages} {035119} (\bibinfo {year} {2021})}\BibitemShut {NoStop}%
\bibitem [{\citenamefont {Bernevig}\ \emph {et~al.}(2021{\natexlab{a}})\citenamefont {Bernevig}, \citenamefont {Lian}, \citenamefont {Cowsik}, \citenamefont {Xie}, \citenamefont {Regnault},\ and\ \citenamefont {Song}}]{bernevig2021twisted_excitations}%
  \BibitemOpen
  \bibfield  {author} {\bibinfo {author} {\bibfnamefont {B.~A.}\ \bibnamefont {Bernevig}}, \bibinfo {author} {\bibfnamefont {B.}~\bibnamefont {Lian}}, \bibinfo {author} {\bibfnamefont {A.}~\bibnamefont {Cowsik}}, \bibinfo {author} {\bibfnamefont {F.}~\bibnamefont {Xie}}, \bibinfo {author} {\bibfnamefont {N.}~\bibnamefont {Regnault}},\ and\ \bibinfo {author} {\bibfnamefont {Z.-D.}\ \bibnamefont {Song}},\ }\bibfield  {title} {\bibinfo {title} {Twisted bilayer graphene. v. exact analytic many-body excitations in coulomb hamiltonians: Charge gap, goldstone modes, and absence of cooper pairing},\ }\href {https://doi.org/10.1103/PhysRevB.103.205415} {\bibfield  {journal} {\bibinfo  {journal} {Phys. Rev. B}\ }\textbf {\bibinfo {volume} {103}},\ \bibinfo {pages} {205415} (\bibinfo {year} {2021}{\natexlab{a}})}\BibitemShut {NoStop}%
\bibitem [{\citenamefont {Khalaf}\ \emph {et~al.}(2020)\citenamefont {Khalaf}, \citenamefont {Bultinck}, \citenamefont {Vishwanath},\ and\ \citenamefont {Zaletel}}]{khalaf2020soft}%
  \BibitemOpen
  \bibfield  {author} {\bibinfo {author} {\bibfnamefont {E.}~\bibnamefont {Khalaf}}, \bibinfo {author} {\bibfnamefont {N.}~\bibnamefont {Bultinck}}, \bibinfo {author} {\bibfnamefont {A.}~\bibnamefont {Vishwanath}},\ and\ \bibinfo {author} {\bibfnamefont {M.~P.}\ \bibnamefont {Zaletel}},\ }\href {https://arxiv.org/abs/2009.14827} {\bibinfo {title} {Soft modes in magic angle twisted bilayer graphene}} (\bibinfo {year} {2020}),\ \Eprint {https://arxiv.org/abs/2009.14827} {arXiv:2009.14827 [cond-mat.str-el]} \BibitemShut {NoStop}%
\bibitem [{\citenamefont {Bultinck}\ \emph {et~al.}(2020)\citenamefont {Bultinck}, \citenamefont {Khalaf}, \citenamefont {Liu}, \citenamefont {Chatterjee}, \citenamefont {Vishwanath},\ and\ \citenamefont {Zaletel}}]{bultinck2020ground}%
  \BibitemOpen
  \bibfield  {author} {\bibinfo {author} {\bibfnamefont {N.}~\bibnamefont {Bultinck}}, \bibinfo {author} {\bibfnamefont {E.}~\bibnamefont {Khalaf}}, \bibinfo {author} {\bibfnamefont {S.}~\bibnamefont {Liu}}, \bibinfo {author} {\bibfnamefont {S.}~\bibnamefont {Chatterjee}}, \bibinfo {author} {\bibfnamefont {A.}~\bibnamefont {Vishwanath}},\ and\ \bibinfo {author} {\bibfnamefont {M.~P.}\ \bibnamefont {Zaletel}},\ }\bibfield  {title} {\bibinfo {title} {Ground state and hidden symmetry of magic-angle graphene at even integer filling},\ }\href {https://doi.org/10.1103/PhysRevX.10.031034} {\bibfield  {journal} {\bibinfo  {journal} {Phys. Rev. X}\ }\textbf {\bibinfo {volume} {10}},\ \bibinfo {pages} {031034} (\bibinfo {year} {2020})}\BibitemShut {NoStop}%
\bibitem [{\citenamefont {Xie}\ and\ \citenamefont {MacDonald}(2020)}]{xie2020nature}%
  \BibitemOpen
  \bibfield  {author} {\bibinfo {author} {\bibfnamefont {M.}~\bibnamefont {Xie}}\ and\ \bibinfo {author} {\bibfnamefont {A.~H.}\ \bibnamefont {MacDonald}},\ }\bibfield  {title} {\bibinfo {title} {Nature of the correlated insulator states in twisted bilayer graphene},\ }\href {https://doi.org/10.1103/PhysRevLett.124.097601} {\bibfield  {journal} {\bibinfo  {journal} {Phys. Rev. Lett.}\ }\textbf {\bibinfo {volume} {124}},\ \bibinfo {pages} {097601} (\bibinfo {year} {2020})}\BibitemShut {NoStop}%
\bibitem [{\citenamefont {Wunsch}\ \emph {et~al.}(2006)\citenamefont {Wunsch}, \citenamefont {Stauber}, \citenamefont {Sols},\ and\ \citenamefont {Guinea}}]{wunsch2006dynamical}%
  \BibitemOpen
  \bibfield  {author} {\bibinfo {author} {\bibfnamefont {B.}~\bibnamefont {Wunsch}}, \bibinfo {author} {\bibfnamefont {T.}~\bibnamefont {Stauber}}, \bibinfo {author} {\bibfnamefont {F.}~\bibnamefont {Sols}},\ and\ \bibinfo {author} {\bibfnamefont {F.}~\bibnamefont {Guinea}},\ }\bibfield  {title} {\bibinfo {title} {Dynamical polarization of graphene at finite doping},\ }\href@noop {} {\bibfield  {journal} {\bibinfo  {journal} {New Journal of Physics}\ }\textbf {\bibinfo {volume} {8}},\ \bibinfo {pages} {318} (\bibinfo {year} {2006})}\BibitemShut {NoStop}%
\bibitem [{\citenamefont {Bistritzer}\ and\ \citenamefont {MacDonald}(2010)}]{bistritzer2010transport}%
  \BibitemOpen
  \bibfield  {author} {\bibinfo {author} {\bibfnamefont {R.}~\bibnamefont {Bistritzer}}\ and\ \bibinfo {author} {\bibfnamefont {A.~H.}\ \bibnamefont {MacDonald}},\ }\bibfield  {title} {\bibinfo {title} {Transport between twisted graphene layers},\ }\href {https://doi.org/10.1103/PhysRevB.81.245412} {\bibfield  {journal} {\bibinfo  {journal} {Phys. Rev. B}\ }\textbf {\bibinfo {volume} {81}},\ \bibinfo {pages} {245412} (\bibinfo {year} {2010})}\BibitemShut {NoStop}%
\bibitem [{\citenamefont {Liu}\ \emph {et~al.}(2021)\citenamefont {Liu}, \citenamefont {Khalaf}, \citenamefont {Lee},\ and\ \citenamefont {Vishwanath}}]{liu2021nematic}%
  \BibitemOpen
  \bibfield  {author} {\bibinfo {author} {\bibfnamefont {S.}~\bibnamefont {Liu}}, \bibinfo {author} {\bibfnamefont {E.}~\bibnamefont {Khalaf}}, \bibinfo {author} {\bibfnamefont {J.~Y.}\ \bibnamefont {Lee}},\ and\ \bibinfo {author} {\bibfnamefont {A.}~\bibnamefont {Vishwanath}},\ }\bibfield  {title} {\bibinfo {title} {Nematic topological semimetal and insulator in magic-angle bilayer graphene at charge neutrality},\ }\href {https://doi.org/10.1103/PhysRevResearch.3.013033} {\bibfield  {journal} {\bibinfo  {journal} {Phys. Rev. Res.}\ }\textbf {\bibinfo {volume} {3}},\ \bibinfo {pages} {013033} (\bibinfo {year} {2021})}\BibitemShut {NoStop}%
\bibitem [{\citenamefont {Bernevig}\ \emph {et~al.}(2021{\natexlab{b}})\citenamefont {Bernevig}, \citenamefont {Song}, \citenamefont {Regnault},\ and\ \citenamefont {Lian}}]{bernevig2021twisted_symmetries}%
  \BibitemOpen
  \bibfield  {author} {\bibinfo {author} {\bibfnamefont {B.~A.}\ \bibnamefont {Bernevig}}, \bibinfo {author} {\bibfnamefont {Z.-D.}\ \bibnamefont {Song}}, \bibinfo {author} {\bibfnamefont {N.}~\bibnamefont {Regnault}},\ and\ \bibinfo {author} {\bibfnamefont {B.}~\bibnamefont {Lian}},\ }\bibfield  {title} {\bibinfo {title} {Twisted bilayer graphene. iii. interacting hamiltonian and exact symmetries},\ }\href {https://doi.org/10.1103/PhysRevB.103.205413} {\bibfield  {journal} {\bibinfo  {journal} {Phys. Rev. B}\ }\textbf {\bibinfo {volume} {103}},\ \bibinfo {pages} {205413} (\bibinfo {year} {2021}{\natexlab{b}})}\BibitemShut {NoStop}%
\end{thebibliography}%
\end{document}